\documentclass{aa} 

\usepackage{amsmath}	
\usepackage{amssymb}	
\usepackage{CJK}        
\usepackage{pifont}
\usepackage{marvosym}
\usepackage{hyperref}   
\usepackage{graphicx}
\usepackage{txfonts}
\usepackage{natbib}
\usepackage{multirow}
\usepackage{enumitem}
\usepackage{url}
\usepackage{color}
\usepackage{xcolor}
\usepackage{orcidlink}
\hypersetup{
    colorlinks=true,
    linkcolor=blue,
    filecolor=blue,      
    urlcolor=blue,
    citecolor=blue
}
\newcommand{\TargetNum}{2,188,240}
\newcommand{\ReferenceNum}{424,164}

\newcommand{\TeffJS}{0.0589}
\newcommand{\LoggJS}{0.0410}
\newcommand{\FeHJS}{0.0278}

\newcommand{\NmatchLo}{100}
\newcommand{\NmatchUp}{1000}
\newcommand{\Ratio}{25}

\graphicspath{{./figures}}

\makeatletter
\renewcommand*\aa@pageof{, page \thepage{} of \pageref*{LastPage}}
\makeatother

\begin{document}
\begin{CJK*}{UTF8}{gbsn}

\title{Measuring the diffuse interstellar bands at 5780, 5797, and 6614 \AA \ in low-resolution spectra of cool stars from LAMOST}

\author{
Xiao-Xiao Ma (马晓骁)\inst{1, 2}\orcidlink{0000-0002-9279-2783} \and
Jian-Jun Chen (陈建军)\inst{3, 2}\orcidlink{0000-0003-4525-1287} \and
A-Li Luo (罗阿理)\inst{1, 2}\orcidlink{0000-0001-7865-2648} \and
He Zhao (赵赫)\inst{4}\orcidlink{0000-0003-2645-6869} \and
Ji-Wei Shi (时纪伟)\inst{3, 2}\orcidlink{0009-0001-4516-8486} \and
Jing Chen (陈静)\inst{5, 6}\orcidlink{0000-0001-8869-653X} \and
Jun-Chao Liang (梁俊超)\inst{1, 2}\orcidlink{0009-0001-9085-8718} \and
Shu-Guo Ma (马树国)\inst{1}\orcidlink{0000-0001-5066-5682} \and
Cai-Xia Qu (屈彩霞)\inst{1, 2}\orcidlink{0000-0002-5460-2205} \and
Bi-Wei Jiang (姜碧沩)\inst{7}\orcidlink{0000-0003-3168-2617}
}

\institute{
CAS Key Laboratory of Optical Astronomy, National Astronomical Observatories, Chinese Academy of Sciences, Beijing 100101, China \\
e-mail: \href{mailto:lal@nao.cas.cn}{lal@nao.cas.cn}
\and
School of Astronomy and Space Science, University of Chinese Academy of Sciences, Beijing 100049, China
\and
Key Laboratory of Space Astronomy and Technology, National Astronomical Observatories, Chinese Academy of Sciences, Beijing 100101, China \\
e-mail: \href{mailto:jjchen@nao.cas.cn}{jjchen@nao.cas.cn}
\and
Purple Mountain Observatory and Key Laboratory of Radio Astronomy, Chinese Academy of Sciences, 10 Yuanhua Road, Nanjing 210033, People's Republic of China
\and 
Nanjing Institute of Astronomical Optics \& Technology, Chinese Academy of Sciences, Nanjing 210042, China
\and
CAS Key Laboratory of Astronomical Optics \& Technology, Nanjing Institute of Astronomical Optics \& Technology, Nanjing 210042, China
\and
Department of Astronomy, Beijing Normal University, Beijing 100875, People's Republic of China
}

\date{Received 7 July 2024 / Accepted 28 September 2024}

\abstract  
{
The limited number of high-resolution spectra of hot stars is inadequate for statistical studies of diffuse interstellar bands (DIBs). In contrast, the vast quantity of low-resolution spectroscopic surveys on cool stars holds great potential for investigating the relationship between DIBs and the known interstellar medium (ISM), as well as the spatial distribution of their unidentified carriers.
}
{
We attempt to measure the DIBs $\lambda 5780$, $\lambda 5797$, and $\lambda 6614$ in over two million low-resolution spectra of cool stars from the Large Sky Area Multi-Object Fiber Spectroscopic Telescope (LAMOST). Based on these DIB measurements, we reviewed and investigated the correlation between DIBs and extinction; the kinematics of DIBs; and the Galactic distribution of DIBs from a statistical perspective.
}
{
We developed a pipeline to measure the DIBs $\lambda 5780$, $\lambda 5797$, and $\lambda 6614$ in the LAMOST low-resolution spectra. Four modules in the pipeline consist of building the target and reference dataset; extracting the ISM residual spectra from the target spectra; measuring the DIBs in the residual spectra; and quality control of the measurements.
}
{ 
We obtained DIB measurements of spectra of late-type stars from LAMOST, and selected 176,831, 13,473, and 110,152 high-quality (HQ) measurements of the DIBs $\lambda 5780$, $\lambda 5797$, and $\lambda 6614$, respectively, corresponding to 142,074, 11,480, and 85,301 unique sources. Using these HQ measurements, we present Galactic maps of the DIBs $\lambda 5780$ and $\lambda 6614$ in the northern sky for the first time. The central wavelengths of the DIBs $\lambda 5780$, $\lambda 5797$, and $\lambda 6614$ in air are determined to be $5780.48 \pm 0.01$, $5796.94 \pm 0.02$, and $6613.64 \pm 0.01$ \AA, respectively, based on their kinematics. A statistical fit of the equivalent widths of these three DIBs per unit extinction provides values of 0.565, 0.176, and 0.256 $\rm \AA \, mag^{-1}$. As a result of this work, three catalogs of the HQ measurements for the DIBs $\lambda 5780$, $\lambda 5797$, and $\lambda 6614$ are provided via \url{https://nadc.china-vo.org/res/r101404/}.
}
{
To the best of our knowledge, this is the largest number of measurements of these three DIBs to date. It is also the first time that Galactic maps of the DIBs $\lambda 5780$ and $\lambda 6614$ in the northern hemisphere are presented, and that the central wavelengths of the DIBs $\lambda 5780$, $\lambda 5797$, and $\lambda 6614$ are estimated from kinematics.
}

\titlerunning{LAMOST DIB}
\maketitle


\section{Introduction} \label{sec:intro}

Diffuse interstellar bands (DIBs) are a set of broad interstellar absorption features that are ubiquitously observed from the optical to the near-infrared (NIR) wavelength range (\citealt{2019ApJ...878..151F}; \citealt{2022A&A...662A..81E}; \citealt{2022ApJS..262....2H}; \citealt{2023MNRAS.521.3727V}; \citealt{2024MNRAS.532.2065C}). Since the first discovery of the DIBs $\lambda 5780$ and $\lambda 5797$ \footnote{We follow a naming convention based on the approximate central wavelength (\AA) of DIBs in air.} by \cite{1922LicOB..10..141H}, many efforts have been dedicated to identifying the carriers of DIBs (e.g., \citealt{2007ApJ...661L.167I}; \citealt{2011ApJ...735..124K}; \citealt{2023A&A...675L...9Z}), but with the exception of $\rm C_{60}^+$ (\citealt{2015Natur.523..322C}), which has been found to be responsible for five DIBs in the NIR wavelength range (see a review by \citealt{2020JMoSp.36711243L}), the carriers of DIBs remain unknown.

In attempts to identify the carriers of DIBs and guide laboratory searches for them, attention has been turned to the characterization of DIBs based on astronomical observations, such as their intrinsic profiles in high-resolution spectra and their correlations with other tracers of the interstellar medium (ISM). For example, \cite{2022A&A...662A..24M} constrained the sizes of possible carrier molecules by comparing the peak-to-peak separation in the DIBs $\lambda 5797$, $\lambda 6379$, and $\lambda 6614$ using the extensive spectra from the ESO diffuse interstellar bands large exploration survey (EDIBLES). To better understand the behavior of DIBs in different interstellar environments, \cite{2017ApJ...850..194F} analyzed how the pattern of the equivalent widths (EWs) of DIBs ---normalized by the corresponding extinction--- changes with the molecular fraction $f_{\rm H2}$.

The first observed DIBs, and those seen to have the strongest signal (\citealt{1930ApJ....72...98M}), the DIBs $\lambda 5780$, $\lambda 5797$, and $\lambda 6614$, as well as their interrelations, have been extensively studied and discussed in the literature. Owing to the proximity of the DIBs $\lambda 5780$ and $\lambda 5797$, they are usually investigated as a pair. The well-known ``$\sigma-\zeta$'' effect was found from the varying strength ratio of these two prominent DIBs (e.g., \citealt{1988A&A...190..339K}; \citealt{2011A&A...533A.129V}; \citealt{2013ApJ...774...72K}). These works indicated that the EWs of the DIB $\lambda 5780$ in the ``$\zeta-$type'' sight lines tend to be weaker than those in the ``$\sigma-$type'' sight lines, while the corresponding EWs of the DIB $\lambda 5797$ are similar. Upon discovering that the relationship between DIBs and reddening is linear, but with considerable dispersion (including the DIBs $\lambda 5780$, $\lambda 5797$, and $\lambda 6614$; \citealt{2012A&A...544A.136R}; \citealt{2013A&A...555A..25P}), and that there exists a ``skin effect'' (\citealt{1974ApJ...194..313S}; \citealt{1995ARA&A..33...19H}), the correlations with other ISM tracers, such as OH, CH, CO, H\,\scalebox{0.8}{I} and $\rm H_2$, have been explored (\citealt{2004A&A...414..949W}; \citealt{2011ApJ...727...33F}; \citealt{2015MNRAS.452.3629L}; \citealt{2017ApJ...850..194F}; \citealt{2019A&A...625A..55W}). It is now well established from those studies that the DIB $\lambda 5797$ is more closely related to the column density of $\rm H_2$ than DIB $\lambda 5780$, while this latter shows a stronger correlation with that of H\,\scalebox{0.8}{I}. In addition, through studies of the correlations between DIBs, an almost perfectly correlated DIB pair has been identified, namely DIB $\lambda 6196$-$\lambda 6614$ (\citealt{1997A&A...326..822C}; \citealt{1999A&A...351..680M}; \citealt{2002A&A...384..215G}; \citealt{2010ApJ...708.1628M}; \citealt{2016A&A...585A..12B}), as have some relatively well correlated DIB pairs, such as DIB $\lambda 5780$-$\lambda 6614$ and DIB $\lambda 5797$-$\lambda 5850$ (\citealt{2011ApJ...727...33F}; \citealt{2021MNRAS.507.5236S}; \citealt{2022MNRAS.510.3546F}), and even some anticorrelated DIB pairs, such as DIB $\lambda 4984$-$\lambda 7559$ and $\lambda 5418$-$\lambda 7562$ (\citealt{2022MNRAS.510.3546F}), which can help to group the DIBs into specific DIB families. Furthermore, in high-resolution spectra, the longward blue wing in the DIB $\lambda 5780$ profile (\citealt{1981ApJ...244..844S}; \citealt{1976MNRAS.174..571D}), the asymmetry profile in the DIB $\lambda 5797$ (\citealt{2022A&A...662A..24M}), and the triple-peak fine structure in the DIB $\lambda 6614$ profile (\citealt{1982ApJ...252..610H}; \citealt{2008MNRAS.386.2003G}), all suggest that their carriers are complex macromolecules.

However, the majority of the aforementioned studies are mainly based on the high-resolution spectra of early-type stars without contamination from the stellar lines. It is challenging to carry out larger-scale and more statistically significant analyses as a result of the limited sight lines toward the early-type stars. \cite{2013ApJ...778...86K} introduced a best neighbor matching method (BNM) to extract the DIB $\lambda 8621$ from approximately 500,000 spectra of late-type stars in the Radial Velocity Experiment survey (RAVE; \citealt{2006AJ....132.1645S}) and provided a statistical ratio between DIB $\lambda 8621$ and extinction. Although the objects observed in the Sloan Digital Sky Survey (SDSS; \citealt{2009AJ....137.4377Y}) do not concentrate on the Galactic plane where the reddening is high, \cite{2015MNRAS.452.3629L} still managed to measure over 20 DIBs from the spectra of cool stars by stacking the spectra with similar extinction, and mapped out the distribution of DIBs at high latitudes in the Milky Way. Later on, using the cool stars from the Apache Point Observatory Galactic Evolution Experiment (APOGEE; \citealt{2017AJ....154...94M}), \cite{2015ApJ...798...35Z} presented a projected Galactic map of the single DIBs at 1.5273 $\mu$m and were the first to estimate its precise central wavelength based on kinematics information. Similarly, \cite{2023A&A...674A..40G} and \cite{2023A&A...680A..38G} further elaborated the properties of the DIBs $\lambda 8621$ and $\lambda 8648$ in the $Gaia$ radial velocity spectrometer (RVS) spectra (\citealt{2018A&A...616A...5C}; \citealt{2021A&A...653A.160S}) of both cool and hot stars. Furthermore, \cite{2024A&A...689A..38C} constructed a 3D Galactic map tracing the DIB $\lambda 8621$ based on the measurements of \cite{2023A&A...674A..40G}. Using the information provided by the GALactic Archaeology with HERMES (GALAH; \citealt{2021MNRAS.506..150B}) for approximately 872,000 cool stars, brand new DIBs were found in the residuals of these stellar spectra (\citealt{2023MNRAS.521.3727V}). Moreover, \cite{2023ApJ...954..141S} and \cite{2024A&A...683A.199Z} recently employed the methods based on machine learning to improve measurements of DIBs from the spectra of cool stars. To measure DIBs from cool stellar spectra, the stellar lines must be removed from spectra in order to extract the ISM composition. A critical distinction exists between the methods used by authors to obtain the stellar lines. On the one hand, studies such as those of \cite{2013A&A...550A..62C}, \cite{2015A&A...573A..35P}, \cite{2021A&A...645A..14Z}, and \cite{2023A&A...674A..40G} use theoretical methods to synthesize the stellar lines. On the other hand, observed spectra with low or no extinction are used to pairwise match the stellar components (e.g., \citealt{2008A&A...480L..13C, 2008A&A...492L...5C}; \citealt{2013ApJ...774...72K}; \citealt{2014Sci...345..791K}; \citealt{2015ApJ...798...35Z} and \citealt{2023A&A...680A..38G}), which is also the approach adopted in the present study.

In addition to the works of \cite{2012MNRAS.425.1763Y} and \cite{2015MNRAS.452.3629L}, which are based on low-resolution spectra (R $\approx$ 2000) obtained by SDSS, certain teams have also made efforts to glean hidden information from low-resolution spectra. Using spectra with a resolving power of R = 3300, \cite{2008A&A...480L..13C} investigated the DIBs $\lambda 5780$, $\lambda 5797$, $\lambda 6203$, $\lambda 6283$, and $\lambda 6614$ in the ISM of M31. These authors then used the spectra with similar resolution to extend their work to the DIBs in the ISM of M33 (\citealt{2008A&A...492L...5C}), and carried out the first survey of DIBs observed in the spectra of B-type supergiants in M31 (\citealt{2011ApJ...726...39C}). By means of the DIBs $\lambda 5780$ and $\lambda 5797$ measured from spectra (\citealt{2015ApJS..216...33F}) with R = $\sim$ 2000 in the northern hemisphere, \cite{2015ApJ...800...64F} probed the Local Bubble and its surroundings. Later, combining the DIB measurements derived from the spectra with a resolution of 5500 in the southern observations, they presented a 3D map of the hot Local Bubble (\citealt{2019NatAs...3..922F}). Although the quantity of low-resolution spectra used in these studies has risen tremendously compared to high-resolution spectra, it is still insufficient to support a broader perspective for investigating the properties of DIBs, such as their spatial distribution and kinematics in the Milky Way.

In the present work, we take full advantage of the LAMOST low-resolution survey (LAMOST LRS; \citealt{2012RAA....12..723Z}; \citealt{2012RAA....12..735D}; \citealt{2015RAA....15.1095L}). The survey almost covers the entire northern sky from the declination of -10$^{\circ}$ to nearly 90$^{\circ}$, with particularly dense coverage of the ISM-rich Galactic plane. So far, a few million spectra have been obtained for cool stars across the optical wavelength range, which allow us to detect several DIBs simultaneously. We selected the DIBs $\lambda 5780$, $\lambda 5797$, and $\lambda 6614$ as our subjects, because their higher intensities compared to the other DIBs (\citealt{2019ApJ...878..151F}; \citealt{2023MNRAS.521.3727V}) mean that we are more likely to be able to measure them in the individual spectra of cool stars. Although the DIBs $\lambda 4430$ and $\lambda 6283$ are also strong, they are not included in this work because of the complexity of the spectral region in which they are found. For the DIB $\lambda 4430$, it is hard to determine the continuum level due to the adjacent strong hydrogen Balmer lines and its own broad profile. For the DIB $\lambda 6283$, the O$_2$ telluric band wraps around it (\citealt{1975ApJ...196..129H}; \citealt{1994AandAS..106...39J}) and is not cleanly removed in the LAMOST spectra. As a consequence, we focus on the DIBs $\lambda 5780$, $\lambda 5797$, and $\lambda 6614$, build the largest sample of measurements of these three DIBs to date, and conduct a comprehensive and statistically robust analysis of their properties.

The remainder of this paper is organized as follows. In Sect. \ref{sec:data}, we describe the data used in this work. The pipeline for measuring the DIBs $\lambda 5780$, $\lambda 5797$, and $\lambda 6614$ in the spectra of cool stars is detailed in Sect. \ref{sec:method}. In Sect. \ref{sec:vaildation}, we control the quality of our measurements and carry out a series of validation tests to ensure the reliability of the measurements. We present and discuss our results in Sect. \ref{sec:results}. Section \ref{sec:code} introduces the code and data availability. Finally, we summarize our findings in Sect. \ref{sec:summary}.


\section{Data} \label{sec:data}

LAMOST (the Large Sky Area Multi-Object-fiber Spectroscopic Telescope; \citealt{2012RAA....12..723Z}; \citealt{2012RAA....12..735D}; \citealt{2015RAA....15.1095L}) is a reflecting Schmidt telescope with a $5^{\circ}$ field-of-view and an effective aperture of 4 meters. Thanks to 4000 fibers averagely distributed among 16 spectrometers (i.e., each one accepts 250 fibers), LAMOST is able to simultaneously observe 4000 objects in theory. There are two resolving modes for LAMOST: one is the low resolution (R = 1800 at 5500 \AA) and the other is the medium resolution (R = 7500 at 5163 \AA \ and 6593 \AA). When switching to the low-resolution mode employed here, the designed optical band covers the range from 3700 \AA \ to 9000 \AA.

The data used in this work are from the data release 10 of LAMOST low-resolution spectroscopic survey (LAMOST LRS DR10) \footnote{\url{https://www.lamost.org/dr10}.}. A total of 11,817,430 spectra are published in this release, which contains 11,473,644 stellar spectra, 263,444 galaxy spectra, and 80,342 quasar spectra. All the published spectra are reduced by wavelength calibration, sky background subtraction, telluric correction, and relative flux calibration (\citealt{2012RAA....12..723Z}; \citealt{2015RAA....15.1095L}). Note that the telluric absorption is negligible in the regions of the DIBs $\lambda 5780$, $\lambda 5797$, and $\lambda 6614$ investigated in this work (see Fig. 16 in \citealt{2000AJ....120.1499M} and Fig. 2 in \citealt{2015EPJWC..8901001K} for more details). All the LAMOST targets are cross-matched with $Gaia$ DR3 sources (\citealt{2023A&A...674A...1G}) within a radius of 3 arcsec, and the $Gaia$ IDs, namely {\tt gaia\_source\_id}, are included in the official LAMOST DR10 catalog \footnote{\url{https://www.lamost.org/dr10/v1.0/catalogue}}. The stellar spectra we focus on are corresponding to $\sim$2,000,000 unique stars and the majority of them concentrate on the Galactic plane. For the cool stars, that is, the F, G, K, and late A-type stars, the LAMOST stellar parameter pipeline (LASP) provides the effective temperature ($T_{\rm eff}$), surface gravity ($\log g$), metallicity abundance ([Fe/H]), and radial velocity (RV) with the precision of 100 K, 0.19 dex, 0.14 dex, and 6 $\rm km\ s^{-1}$, respectively \footnote{Here we refer reader to the note of LAMOST DR10 via \url{https://www.lamost.org/dr10/v1.0/doc/release-note}.}. In addition, the continuum-normalized flux are calculated for these spectra with stellar parameters (\citealt{2012RAA....12..453S}).

Benefiting from the estimations of geometric distances from parallaxes in $Gaia$ DR3 (\citealt{2021AJ....161..147B}), the relatively reliable distance of an object in LAMOST can be simply derived. Combined with the three-dimensional dust reddening map (version name ``bayestar2019'', \citealt{2019ApJ...887...93G}), once the celestial coordinate and astrometric distance of an object is given, the fine-grained extinction $\rm E(B-V)$ \footnote{A recalibration factor of 0.884 (\citealt{2011ApJ...737..103S}) is applied for $\rm E(B-V)$.} can be calculated at the sight line of each object.



\section{Method} \label{sec:method}

\begin{figure*}[ht]
\includegraphics[width=1.0\textwidth]{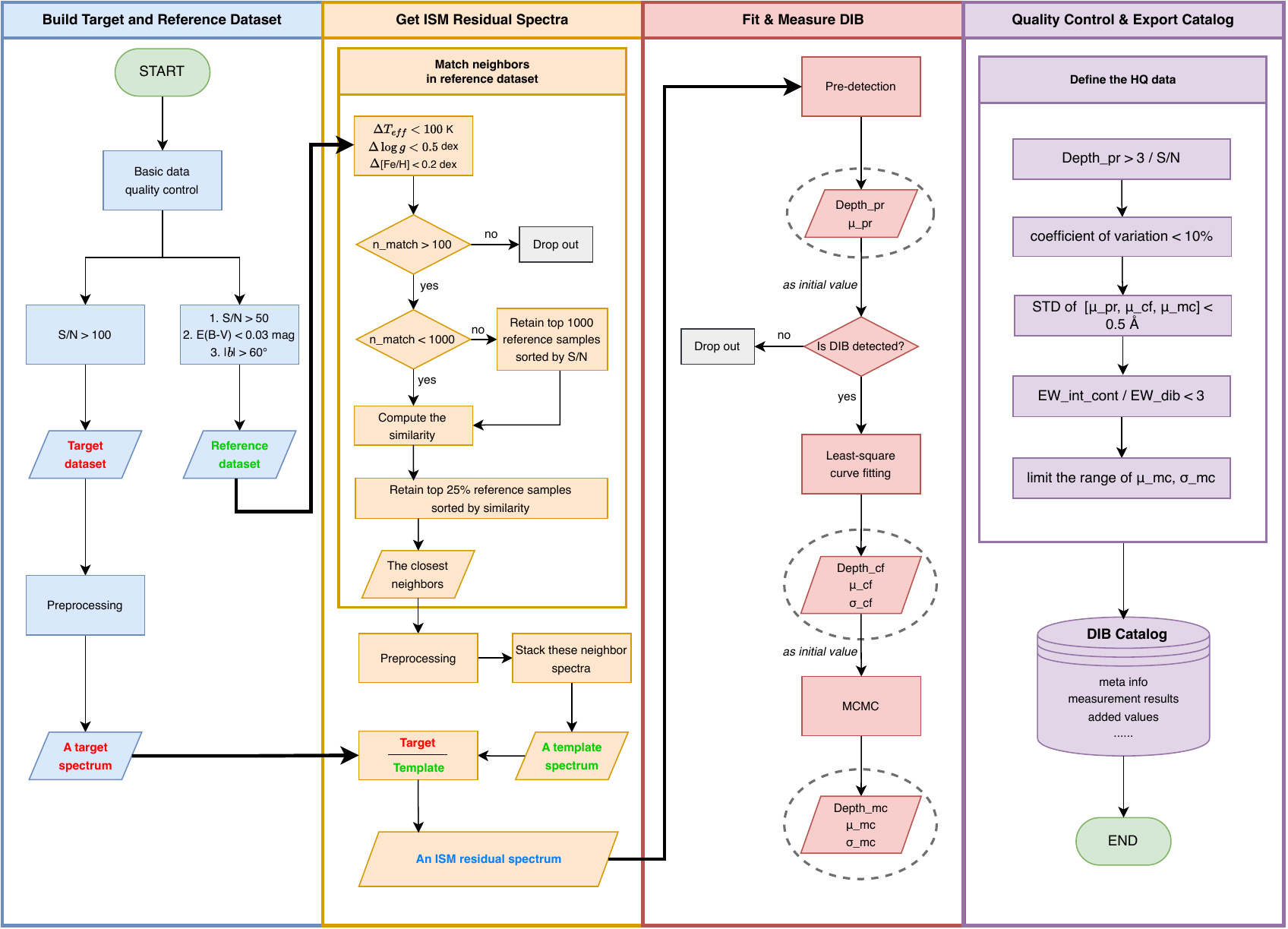}
\caption{
Schematic flowchart of the pipeline for measuring the DIBs $\lambda 5780$, $\lambda 5797$, and $\lambda 6614$ of cool stars in the LAMOST LRS DR10. Four colors encode the four parts of the pipeline, and the output of each modular function is used for the input of the subsequent one.
\label{fig:pipeline}
}
\end{figure*}

Figure \ref{fig:pipeline} shows an overview of the pipeline for measuring the DIBs $\lambda 5780$, $\lambda 5797$, and $\lambda 6614$ of cool stars in the LAMOST LRS DR10. Four swimlanes in the figure represent the four modules of the pipeline, which are responsible for building the target and reference dataset, deriving the ISM residual spectra of the target dataset, measuring the DIBs in the residual spectra, and quality control of the final measurements. The details of each module are described in the following subsections.

\subsection{Building the dataset} \label{subsec:dataset}

For the cool stars, the regions nearby DIBs are usually blended with the stellar absorption lines, which need to be thoroughly removed as much as possible. Inspired by the cases for the success in detecting the DIBs in the spectra of cool stars (\citealt{2013ApJ...778...86K}; \citealt{2012MNRAS.425.1763Y}; \citealt{2015MNRAS.452.3629L}; \citealt{2023MNRAS.521.3727V}), we build two datasets, that is, the target dataset in which the DIBs are likely to be embedded in the spectra, and the reference dataset in which the spectra are expected to contain the pure stellar features within the region of the DIB to be measured.

At first, both the target and reference datasets are filtered by the basic data quality control for the better stellar parameter matching in Sect. \ref{subsec:cis}. They are required to have valid stellar parameters within the error of 100 K, 0.2 dex, 0.1 dex, and 10 $\rm km\ s^{-1}$ for $T_{\rm eff}$, $\log g$, [Fe/H], and radial velocity (RV), respectively.

\begin{figure*}[ht]
\includegraphics[width=1.0\textwidth]{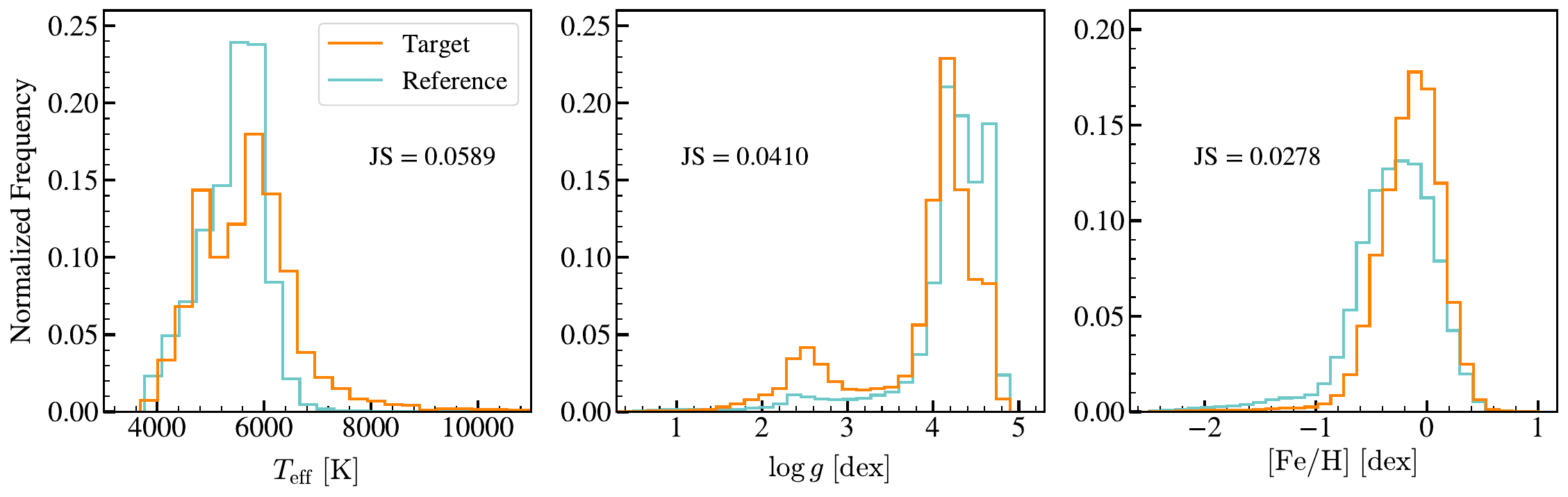}
\caption{
Distribution of the three atmospheric parameters, namely, $T_{\rm eff}$ (left panel), $\log g$ (middle panel), and [Fe/H] (right panel), for the target and reference datasets. The JS divergences of the three parameters between the two datasets are annotated in each panel.
\label{fig:hist_sp}
}
\end{figure*}

In general, given that the absorption depths of the DIBs $\lambda 5780$, $\lambda 5797$, and $\lambda 6614$ are commonly larger than one percent of continuum (\citealt{2023MNRAS.521.3727V}), we conservatively select the spectra with the signal-to-noise ratio in $r$ band (the keyword {\tt snrr} in the LAMOST catalog, hereafter S/N, which is derived from the average signal-to-noise ratio of the pixels within the $r$ band, i.e., from 5600 to 6800 \AA) greater than 100 as our target dataset. In that the reference dataset are used to subtract the components that do not contribute to the DIB in the target dataset, they are supposed to meet the following conditions to ensure that the interstellar clouds in the foreground are as thin as possible: (i) S/N $> 50$; (ii) $\rm E(B-V) < 0.03 \ mag$; (iii) Galactic latitude, $|b| > 60^{\circ}$. There are \TargetNum \ spectra in the target dataset and \ReferenceNum \ spectra in the reference dataset after the above filtering. The distribution of three atmospheric parameters for the two datasets are shown in Fig. \ref{fig:hist_sp}. The JS divergences (\citealt{MENENDEZ1997307}), which range from 0 to 1 and are often used to measure the similarity between two distributions, of $T_{\rm eff}$, $\log g$, and [Fe/H] between the target and reference dataset are \TeffJS, \LoggJS, and \FeHJS, respectively. These values closer to zero indicate it is of high similarity between the two datasets, which makes the reference dataset matchable for the subtraction of the stellar features in the target dataset.

\subsection{Getting the ISM residual spectra} \label{subsec:cis}

\begin{figure}
\centering
\includegraphics[width=0.48\textwidth]{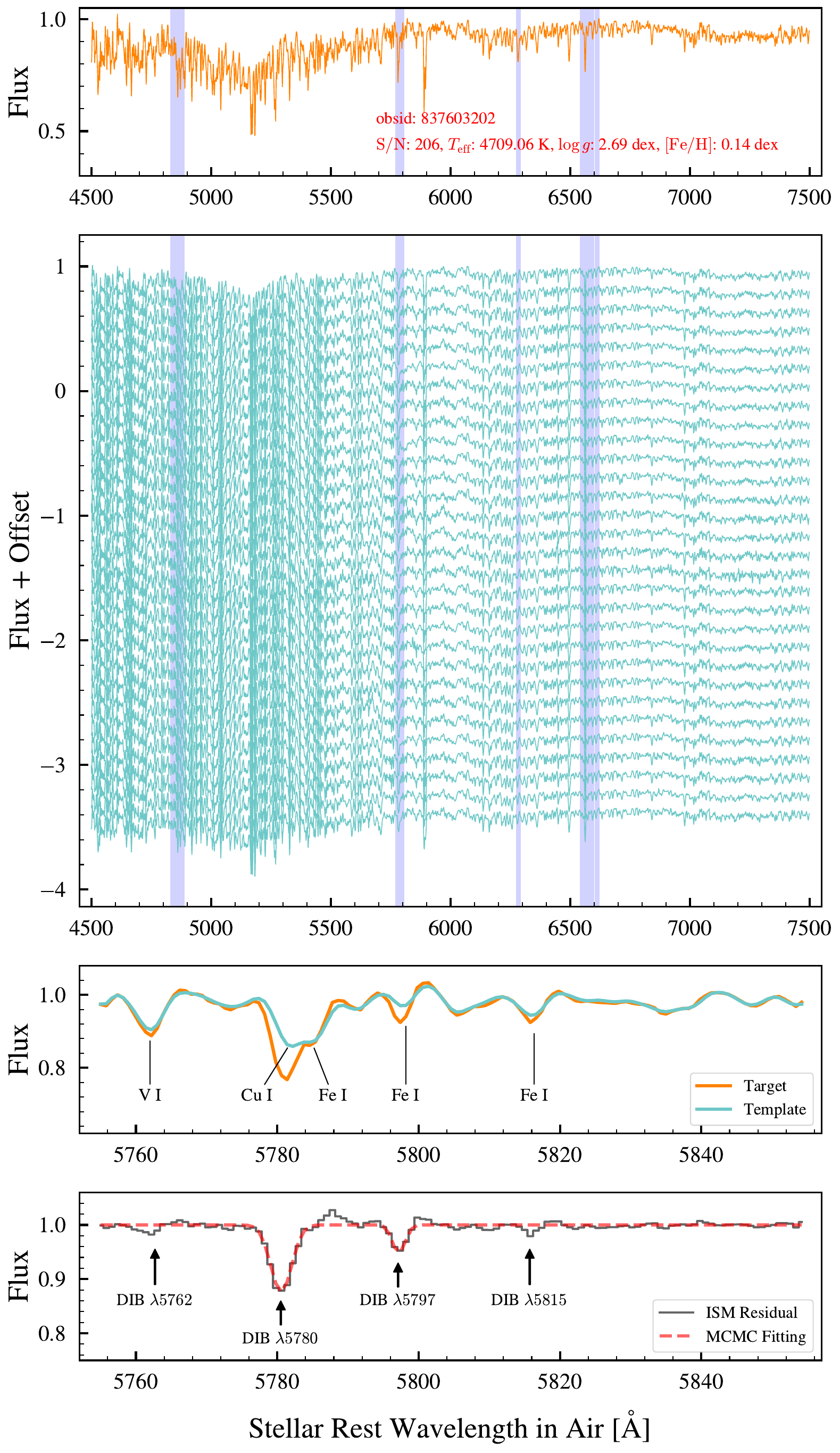}
\caption{
Example measurement of the DIBs $\lambda 5780$ and $\lambda 5797$ from a target spectrum. From top to bottom, (i) the first panel shows the target spectrum to be measured, with the red text indicating its {\tt obsid} (the unique ID of the spectrum in the LAMOST catalog) and atmospheric parameters. (ii) The second panel provides the first 30 best neighboring spectra of the target spectrum, which are used to build the template spectrum displayed in the third panel. The purple-shaded areas in the first and second panels are the masked regions that do not participate in the similarity computation. (iii) The third panel illustrates the target and template spectrum after local normalization. The most prominent stellar absorptions that may affect the DIB extraction are annotated in the panel. (iv) The fourth panel presents the ISM residual spectrum (in black) of the target and the fit (in red) of the DIBs $\lambda 5780$ and $\lambda 5797$ by MCMC. The shallow DIBs $\lambda 5762$ and $\lambda 5815$ in the ISM residual spectrum are also annotated but not fitted in the panel.
\label{fig:method}
}
\end{figure}

An ISM residual spectrum is the integral signals of DIB at the line-of-sight of a target object, so that an observed spectrum in the target dataset can be viewed as the product of the ISM residual and the stellar components. Although it is impossible to resolve the authentic stellar features out of a target spectrum, one of the workarounds is stacking the neighboring spectra of the target in the reference dataset to approximate the stellar ingredients we call the template spectrum (\citealt{2013ApJ...778...86K}). Then, the ISM residual spectrum can be obtained by the ratio of the target spectrum to the template spectrum from the stacked closest spectra, as can be seen in Fig. \ref{fig:method}.

\subsubsection{Matching neighbors} \label{subsubsec:neighbor}

The quality of the best neighbor matching determines whether the stellar components in the target spectrum can be cleanly deducted. It is essential to ensure that these neighbors are as close to the target possible. We impose constraints on the selection of the best neighboring samples based on the atmospheric parameter space, match number, S/N, and similarity in spectral morphology.

We make a preliminary selection within a certain range of atmospheric parameters. For the samples in the reference dataset that satisfy $\Delta T_{\rm eff} < 100$ K, $\Delta \log g < 0.5$ dex, and $\rm \Delta [Fe/H] < 0.2$ dex with respect to the target, if the number of those samples is no more than \NmatchLo, we drop them out. Then, to balance the computation efficiency and match accuracy, the top \NmatchUp \ samples (if enough) with the highest S/N in $r$ band as the neighboring candidates of the target are retained.

However, the proximity of atmospheric parameter space cannot guarantee the similarity of spectral morphology due to the measurement error of parameters. We further adopt the reciprocal of Euclidean distance between the target and the neighbor to represent the similarity in spectral morphology. Each continuum-normalized spectrum is processed for the alignment of similarity computation as follows: (i) mask the bad flux pixels according to the {\tt ormask} flag; (ii) shift the spectrum to the rest frame; (iii) convert the wavelength from vacuum to air; (iv) truncate the spectrum to [4500, 7500] \AA \ to avoid the bias of Balmer series at the blue end and the low response efficiency at the red end, and then rebin the spectrum with a step of 1 \AA; (v) mask the regions that do not contribute to the metal lines, including H\scalebox{0.9}{$\alpha$}, H\scalebox{0.8}{$\beta$}, strong DIBs $\lambda 5780$, $\lambda 5797$, $\lambda 6283$, and $\lambda 6614$ (see Table \ref{tab:exclusion} for their detailed wavelength ranges). Ultimately, the top \Ratio \% similar candidate samples, namely those with no fewer than 25 and at most 250, are selected as the closest neighbors prepared for the stacked template spectrum in Sect. \ref{subsubsec:ism}. For instance, the first 30 closest neighbors of a target spectrum of {\tt obsid = 837603202} can be seen in the second panel of Fig. \ref{fig:method}.

\renewcommand{\arraystretch}{1.1}
\begin{table}
\centering
\setlength\tabcolsep{15pt}
\caption{Masked region used to match neighboring spectra in Sect. \ref{subsubsec:neighbor} \label{tab:exclusion}}
\begin{tabular}{cc}
\hline\hline 
Feature name & Wavelength range [\AA] \\
\hline
H\scalebox{0.9}{$\alpha$} & [6540, 6600] \\
H\scalebox{0.8}{$\beta$} & [4830, 4890] \\
DIB $\lambda 5780$ and $\lambda 5797$ & [5770, 5807] \\
DIB $\lambda 6283$ & [6273, 6293] \\
DIB $\lambda 6614$ & [6604, 6624] \\
\hline
\end{tabular}
\end{table}

\subsubsection{Preprocessing} \label{subsubsec:preprocess}

Similar to Sect. \ref{subsubsec:neighbor}, a series of preprocessing steps are required to align the target and its neighbors before extracting the ISM residual spectrum of the target. These preprocessing steps still include removing the bad flux pixels, shifting the spectrum to the rest frame, and converting the wavelength from vacuum to air. Unlike Sect. \ref{subsubsec:neighbor}, the renormalized flux within the local region instead of the continuum-normalized flux is utilized. Although the continuum-normalized flux provided by LASP meets the requirement of spectral morphology matching in Sect. \ref{subsubsec:neighbor}, it is too coarse to measure the weak DIBs when anchoring the local region of the DIBs. Therefore, a more careful local normalization for the flux within the local DIB region needs to be performed.

\begin{figure}
\centering
\includegraphics[width=0.48\textwidth]{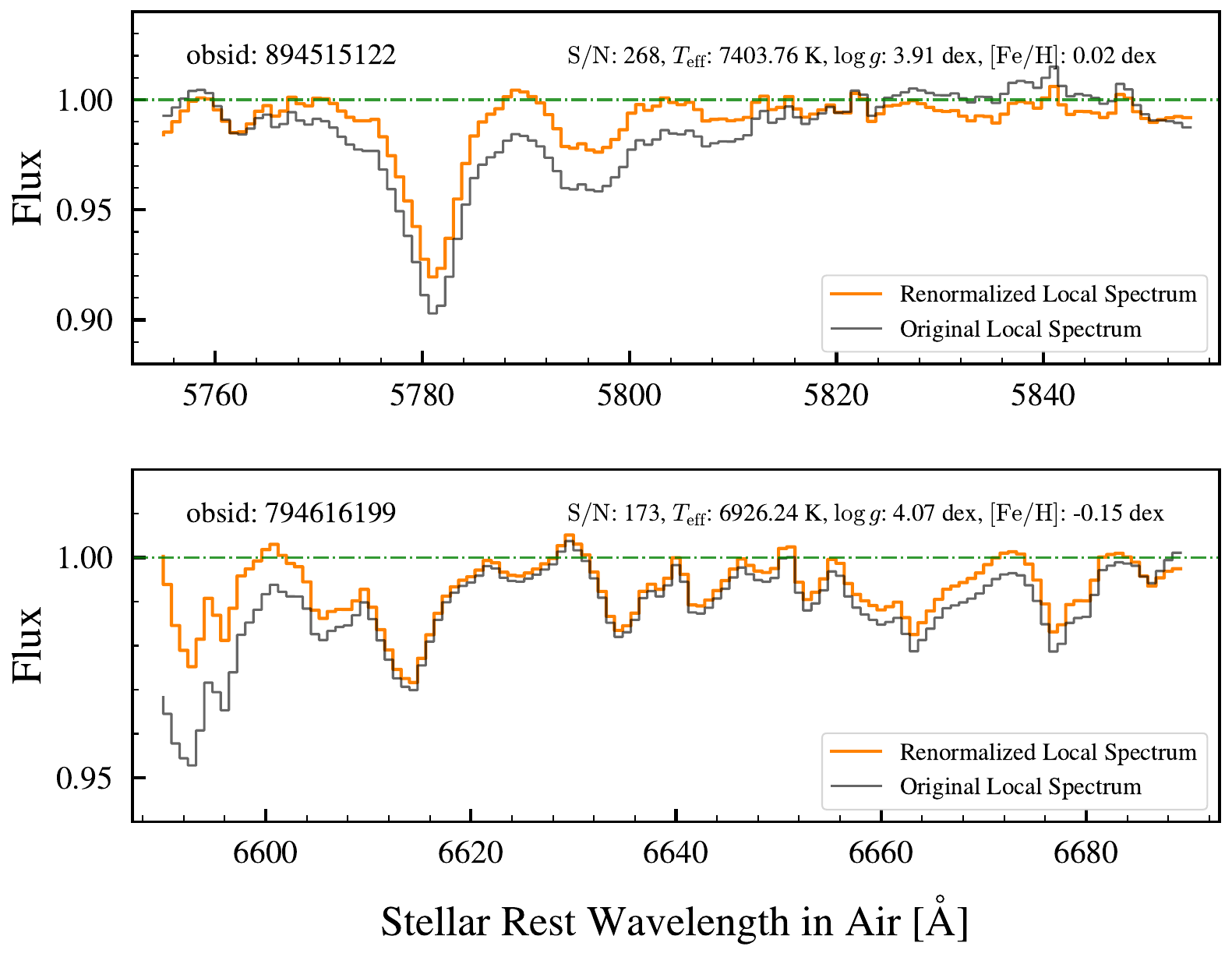}
\caption{
Examples of the local renormalization for two target spectra. The upper panel is the target spectrum within the local region covering the DIBs $\lambda 5780$ and $\lambda 5797$, and the lower panel is the target spectrum within the local region including the DIB $\lambda 6614$. In each panel, the black line represents the original spectrum, and the orange line is the renormalized spectrum. The {\tt obsid} of the target spectrum and its atmospheric parameters are also annotated in each panel.
\label{fig:local_renorm}
}
\end{figure}

Initially, the wavelength range of the local DIB region is determined. For the DIB $\lambda 5780$ and $\lambda 5797$, they are merged into a single local region, namely [5755, 5855] \AA \, owing to their close central wavelengths. For the DIB $\lambda 6614$, the blue end keeps away from the H\scalebox{0.9}{$\alpha$} line to alleviate the influence of the quite strong H\scalebox{0.9}{$\alpha$} feature on the blue wing of the DIB $\lambda 6614$. In addition, the red end of the local region is moderately extended to compensate for the accuracy of the local continuum normalization. The final selection for the DIB $\lambda 6614$ is [6590, 6690] \AA. Subsequently, the flux in these two local regions is linearly interpolated and rebinned to an interval of 0.8 \AA, resulting in a total of 125 pixels. We refer to the method by \cite{2021A&A...645A..14Z} as the local normalization of the flux within the selected local DIB region. The only difference is that we use a fifth-order polynomial curve instead of a second-order one to fit the local spectrum because of the cool stars in this work. A 20-iteration continuum fitting is performed as follows: (i) fit the local spectrum by a fifth-order polynomial curve; (ii) compute the differences of the local spectrum to the fitting curve, as well as their standard deviation; (iii) replace the flux values with the fitted polynomial ones if the differences between them are larger than five times of the standard deviation when the pixels are located above the polynomial curve, or 0.5 times when the pixels are below the polynomial curve; (iv) use the remaining and replaced pixels as the new local spectrum to repeat the above steps. The final normalized spectrum is obtained by dividing the last local spectrum by the last fitted continuum. Figure \ref{fig:local_renorm} provides two examples of the local renormalization for the target spectra with {\tt obsid = 894515122} and {\tt obsid = 794616199}. It is clear that the local renormalization can effectively correct the continuum level of some absorption features.

The flux error is estimated while performing the local normalization because the flux error after local normalization is not only a key input parameter for the subsequent MCMC fitting in Sect. \ref{subsec:measure}, but also an important indicator for evaluating the quality of the ISM residual spectra in Sect. \ref{sec:vaildation}. Although the LAMOST provides the error of the flux in a format of the inverse variance (1/error$^2$), many works, such as \cite{2015MNRAS.448..822X} and \cite{2017ApJ...836....5H} have reported this error is overestimated. \cite{2017ApJ...836....5H} and \cite{2020ApJS..246....9Z} have given the same approximations of the inverse variance of the flux after normalization, which can be expressed as Eq. (\ref{eq:ivar}),
\begin{equation} \label{eq:ivar}
    \rm ivar_{norm} = F_{cont}^2 \times ivar_{rel} \ ,
\end{equation}
where $\rm F_{cont}$ is the fitted continuum and $\rm ivar_{rel}$ is the inverse variance of the flux. The final renormalized flux error equals the inverse of the square root of $\rm ivar_{norm}$.

\subsubsection{Isolating the ISM residual spectra} \label{subsubsec:ism}

Once the locally normalized spectra of the target and its best neighbors are ready, the high-quality template spectrum for the target can be obtained by averaging those spectra of the best neighbors weighted by their S/N in $r$ band, which characterizes the stellar features in the target spectrum. Then, the ISM residual spectrum of the target is isolated by dividing the target spectrum by the template spectrum, as shown in the third and fourth panels of Fig. \ref{fig:method}. By comparing the template spectrum to the spectral standard star \textbf{Arcturus} (a K1.5\,III star with the luminosity of 170 $L_{\odot}$) using an interactive database (\citealt{2011CaJPh..89..395L}) called SpectroWeb (\url{http://spectra.freeshell.org/spectroweb.html}), we also list the most prominent stellar lines that may contaminate the DIBs in the template spectrum in the third panel of Fig. \ref{fig:method}. 

The error of the ISM spectrum is deduced by the error propagation as Eq. (\ref{eq:err_cis}),
\begin{equation} \label{eq:err_cis}
    \rm E_{res} = \frac{F_{tar}}{F_{temp}} \times \sqrt{\rm (\frac{E_{tar}}{F_{tar}})^2 + (\frac{E_{temp}}{F_{temp}})^2} \ ,
\end{equation}
where $\rm F_{tar}$ and $\rm F_{temp}$ are the normalized flux of the target and template spectra, respectively, and $\rm E_{tar}$ and $\rm E_{temp}$ are their corresponding errors. For the $\rm E_{temp}$, it is calculated by the same operation as that of $\rm F_{temp}$, that is, the weighted average of the flux errors of the best neighbors.

\subsection{Measuring diffuse interstellar bands} \label{subsec:measure}

Most of the DIB profiles are fitted by the symmetric Gaussian function, such as \cite{2015MNRAS.452.3629L} and \cite{2021A&A...645A..14Z}, but some studies based on the early-type stars in high resolution have found that the profiles of some DIBs are asymmetric and multiple components. For instance, \cite{2008MNRAS.386.2003G} reported the DIB $\lambda 6614$ displays a triple-peak fine structure. Considering the low resolution of LAMOST, the instrumental broadening effect (see Appendix \ref{appd:ib}) dominates the shape of the DIBs $\lambda 5780$, $\lambda 5797$, and $\lambda 6614$ and their fine structures are blurred. In such a resolution, \cite{2012MNRAS.425.1763Y}, \cite{2015MNRAS.452.3629L}, and \cite{2019NatAs...3..922F} have proven the Gaussian function is still a good approximation for the profile of those DIBs. Thus, we adopt the Gaussian function to fit the DIBs in the ISM residual spectra. The process of measuring the DIBs consists of three steps: (i) pre-detect the central wavelength and depth of the DIB to determine whether there is a DIB signal; (ii) use the parameters obtained in the first step and the fixed Gaussian widths (3.85, 2, and 2 \AA \ for the DIBs $\lambda 5780$, $\lambda 5797$, and $\lambda 6614$, respectively) as the initial values of the least-square curve fitting to get the three parameters of the Gaussian profile, namely, the central wavelength, depth, and Gaussian width; (iii) further optimize the parameters obtained in the second step using the MCMC method to get the final fitting results. An MCMC fitting demo of the DIBs $\lambda 5780$ and $\lambda 5797$ is presented in the fourth panel of Fig. \ref{fig:method}.

\subsubsection{Pre-detection} \label{subsubsec:predetect}

In the context of big data mining, pre-detection not only allows the rapid identification of samples that probably contain DIBs, but also provides reasonable initial values for the curve fitting in Sect. \ref{subsubsec:curvefit}. We take the rest-frame central wavelength and FWHM of the DIBs reported by \cite{2019ApJ...878..151F} into account, and set the wavelength range of the pre-detection segment to be about $\pm 300 \ \rm km\ s^{-1}$ around the central wavelength of the DIB. For the DIBs $\lambda 5780$, $\lambda 5797$, and $\lambda 6614$, their pre-detection segments are [5775, 5785], [5793, 5801], and [6608, 6620] \AA, respectively. Such a wide wavelength range is chosen to deal with those potential DIBs with large RVs or broad profiles. If the maximum depth of the ISM residual spectrum within the pre-detection segment exceeds $\rm \frac{1}{S/N}$ of the target spectrum, where S/N is the average signal-to-noise ratio within the $r$ band that covers the DIBs $\lambda 5780$, $\lambda 5797$, and $\lambda 6614$, we proceed to the next step of curve fitting. The maximum depth and the wavelength corresponding to it are taken as the initial depth and central wavelength of the DIB profile. Otherwise, we drop out the target.

\subsubsection{Curve fitting} \label{subsubsec:curvefit}

Both curve fitting and MCMC fitting in Sect. \ref{subsubsec:mcmc} are done in a narrower region called the ``focus region'', which is dynamically determined according to the central wavelength of the DIB pre-detected in Sect. \ref{subsubsec:predetect}. The flux values beyond the focus region are set to the continuum level to prevent the other potential features, such as certain weaker DIBs (see the bottom panel of Fig. \ref{fig:method}), from disturbing the fitting. The wavelength interval of the focus region is designed based on the 2$\sigma$ principle, where $\sigma$ refers to the Gaussian width. For instance, if a DIB has a $\sigma$ of 2 \AA, its entire profile will span approximately $2 \times 3 \sigma$, namely 12 \AA. Therefore, a wavelength interval of at least $2 \times 2 \sigma = 8$ \AA \ is necessary for a reasonable fitting. Specifically, for the DIBs $\lambda 5797$ and $\lambda 6614$, the width of the focus region is set to 6 \AA \ on both sides of the pre-detected central wavelength. For the DIB $\lambda 5780$, the width is set to 10 \AA \ because the typical FWHM of the DIB $\lambda 5780$ is larger than those of the DIBs $\lambda 5797$ and $\lambda 6614$ (\citealt{2008ApJ...680.1256H}; \citealt{2019ApJ...878..151F}). On top of the above cases, if the DIBs $\lambda 5780$ and $\lambda 5797$ are both pre-detected, the blue end of the focus region is set to 6 \AA \ on the left side of the pre-detected central wavelength of the DIB $\lambda 5780$, and the red end is set to 6 \AA \ on the right side of the pre-detected central wavelength of the DIB $\lambda 5797$. Similar to the wavelength interval of the pre-detection in Sect. \ref{subsubsec:predetect}, such a focus region is wide enough to handle the broadened DIBs, which may be caused by multiple DIB components or the instrumental broadening effect.

We employ the method of least-square curve fitting (the function {\tt curve\_fit} in the Python module {\tt scipy.optimize}) to obtain the central wavelength, depth and FWHM of the single Gaussian profile of the DIB as Eq. (\ref{eq:gaussian}).
\begin{equation} \label{eq:gaussian}
    f_{\theta}(x;\text{D},\mu,\sigma) = \text{D} \times \text{exp}\left(-\frac{(x-\mu)^2}{2\sigma^2}\right) + 1 \ ,
\end{equation}
where $x$ means the wavelength, and $\rm D$, $\rm \mu$, $\rm \sigma$ denote the depth, central wavelength, and Gaussian width, respectively. In the case where the DIBs $\lambda 5780$ and $\lambda 5797$ are both detected, a double Gaussian profile is adopted to simultaneously fit the two DIBs, which can overcome the blend of the two close DIBs when the DIBs are greatly broadened. The initial values of the depth and central wavelength are taken from the pre-detection results in Sect. \ref{subsubsec:predetect}, and the initial values of the Gaussian width are set to 3.85, 2, and 2 \AA \ for the DIBs $\lambda 5780$, $\lambda 5797$, and $\lambda 6614$, respectively. Such initial values are larger than the typical values of the DIBs owing to the instrumental broadening effect in the low resolution (\citealt{2012MNRAS.425.1763Y}; \citealt{2015MNRAS.452.3629L}). Actually, there is little influence on the final fitting results while using the canonical Gaussian width, for instance, the FWHM from \cite{2019ApJ...878..151F}, as the initial value.

\subsubsection{MCMC fitting} \label{subsubsec:mcmc}

The prior in the Bayesian inference can rule out some cases that obviously violate the physics, for example, the central wavelength of DIB is far away from the well-known value. And the MCMC method is able to efficiently sample the multi-parameter distribution from the posterior constructed by the product of the likelihood and prior based on the Bayesian theory (Eq. (\ref{eq:posterior})).
\begin{equation} \label{eq:posterior}
    \rm P(\theta|Y) \propto P(Y|\theta) \times P(\theta) \ ,
\end{equation}
where $\rm \theta$ represents the parameters of the DIB profile, that is, $\rm \theta=\left(D, \mu, \sigma\right)$, and $\rm Y$ denotes the observed data. Specifically, the likelihood and prior are expressed as Eqs. (\ref{eq:likelihood}) and (\ref{eq:prior}) in logarithmic form, respectively. 

\begin{equation} \label{eq:likelihood}
\begin{aligned}
    \ln \text{P}(\text{Y}|\theta) = -\frac{1}{2}  \sum \left[ \frac{(y - f_{\theta}(x))^2}{y_{\text{err}}^2} + \ln(y_{\text{err}}^2) \right] \ ,
\end{aligned}
\end{equation}
where $x$, $y$, and $y_{\text{err}}$ as the observed data $\rm Y$ are the wavelength, flux, and flux error of the ISM residual spectrum, respectively, and $f_{\theta}(x)$ is the model of DIB profile which is set out in Eq. (\ref{eq:gaussian}).

\begin{equation} \label{eq:prior}
\rm \ln P(\theta) = \ln P(D) + \ln P(\mu) + \ln P(\sigma) \ ,
\end{equation}
where $\rm P(D)$, $\rm P(\mu)$, and $\rm P(\sigma)$ are the priors of the depth, central wavelength, and Gaussian width, respectively. The prior of $\rm D$ satisfies the uniform distribution, and the priors of $\rm \mu$ and $\rm \sigma$ are both set to be the Gaussian distributions with the means of the curve fitting results in Sect. \ref{subsubsec:curvefit} and the standard deviations of 0.5 \AA. Table \ref{tab:prior} summarizes the detailed upper and lower limits of the priors of $\rm D$, $\rm \mu$, and $\rm \sigma$ for the DIBs $\lambda 5780$, $\lambda 5797$, and $\lambda 6614$.

\renewcommand{\arraystretch}{1.1}
\begin{table}
\caption{Upper and lower limits of the priors \label{tab:prior}}
\centering
\setlength\tabcolsep{14pt}
\begin{tabular}{cccc}

\hline\hline
DIB & $\rm D$ \tablefootmark{(1)} & $\rm \mu$ [\AA] & $\rm \sigma$ [\AA] \\
\hline
$\lambda 5780$ & [0, 0.4] & [5773, 5787] & [0.1, 5] \\
$\lambda 5797$ & [0, 0.3] & [5790, 5804] & [0.1, 5] \\
$\lambda 6614$ & [0, 0.4] & [6608, 6620] & [0.1, 5] \\
\hline

\end{tabular}
\tablefoot{ \\
\tablefoottext{1}{The upper limit of the prior of $\rm D$ is set according to the maximum depth of all the targets measured by pre-detection in Sect. \ref{subsubsec:predetect}.}
}
\end{table}

The Python package {\tt emcee} is called to carry out the MCMC fitting. We set 100 walkers and 250 steps for each walker, where the first 50 steps are used as the burn-in stage and the last 200 steps are used to sample the posterior distribution. The initial guesses of the walkers are randomly seeded around the curve fitting results in Sect. \ref{subsubsec:curvefit} with a standard deviation of 0.01, which can greatly reduce the length of the chain as a result of the fast convergence of the MCMC fitting. We take the 50th percentile and half of the difference between the 16th and the 84th percentiles of the posterior distribution sampled by the MCMC as our best estimate and statistical uncertainty. The equivalent width (EW) of the DIB is calculated by $\rm EW = \sqrt{2\pi} \times D \times \sigma$. The EW error is estimated by the error propagation from the errors of $\rm D$ and $\rm \sigma$.


\section{Quality control and validation} \label{sec:vaildation}

\subsection{Quality control} \label{subsec:qc}

\subsubsection{Defining the high-quality samples} \label{subsubsec:hq}

A total of \TargetNum \ target spectra are processed by the pipeline described in Sect. \ref{sec:method}, and the statistics of all the valid measurements for the DIBs $\lambda 5780$, $\lambda 5797$, and $\lambda 6614$ are summarized in Table \ref{tab:stat}. To analyze the results based on the reliable DIB measurements, we define the high-quality (HQ) samples. For each HQ sample, it must meet the following criteria:

\renewcommand{\arraystretch}{1.1}
\begin{table}
\caption{Statistics of DIB measurement \label{tab:stat}}
\centering
\setlength\tabcolsep{8pt}
\begin{tabular}{crrr}

\hline\hline
DIB & Number of all & Number of HQ & HQ ratio \\
\hline
$\lambda 5780$ & 854,357 & 176,831 & 20.7\% \\
$\lambda 5797$ & 355,665 & 13,473 & 3.8\% \\
$\lambda 6614$ & 266,870 & 110,152 & 41.3\% \\
\hline

\end{tabular}
\end{table}

\begin{enumerate}
\item The depth fitted by the MCMC $\rm D_{mc} >  \frac{3}{S/N}$. The threshold of $\rm \frac{3}{S/N}$ is set to ensure the DIB signal is significantly above the noise level.
\item The coefficient of variation (CV) of the depths $\rm \{ D_{pr}, \ D_{cf}, \ D_{mc} \} < $ 10\%, where the CV is defined as the standard deviation divided by the mean value, and the subscripts $\rm pr$, $\rm cf$ and $\rm mc$ represent the pre-detection, curve fitting and MCMC fitting in Sect. \ref{sec:method}, respectively.
\item The standard deviation of the central wavelengths $\rm \{ \mu_{pr}, \ \mu_{cf}, \ \mu_{mc} \} < $ 0.5 \AA.
\item $\left| \rm \mu_{mc} - \lambda_0 \right| < $ 3 \AA, where $\rm \lambda_0$ is the typical central wavelength of DIB (\citealt{1975ApJ...196..129H}; \citealt{2019ApJ...878..151F}; \citealt{2023MNRAS.521.3727V}), that is, $\rm \lambda_0 = 5780.6, \ 5797.1, \ 6613.6$ \AA \ for the DIBs $\lambda 5780$, $\lambda 5797$, and $\lambda 6614$, respectively. This criterion is mainly used to confine the RV of the DIB carrier to within around $\pm 200 \ \rm km \ s^{-1}$, which is a reasonable assumption if the DIB carrier primarily tracks the ISM at a distance of several kiloparsecs from the Sun (\citealt{2021A&A...645A..14Z}).
\item The Gaussian width $\rm \sigma_{mc} > $ 1 \AA \ and $\rm \sigma_{mc} < $ 3 \AA, where the lower limit of 1 \AA \ is set due to the instrumental broadening effect (see Appendix \ref{appd:ib}), and the upper limit of 3 \AA \ is set based on the typical FWHM of DIB (\citealt{2023MNRAS.521.3727V}), which will be elaborated in Sect. \ref{subsubsec:lambda_sigma}.
\item $\rm EW_{DIB} / EW_{cont} > $ 33\% for the DIB $\lambda 5780$ and $>$ 25\% for the DIBs $\lambda 5797$ and $\lambda 6614$, where $\rm EW_{DIB}$ is the equivalent width of the DIB, and $\rm EW_{cont}$ is the integral equivalent width of the continuum outside the DIB region, namely $\rm [\mu_{mc} - 3 \sigma_{mc}, \ \mu_{mc} + 3 \sigma_{mc}]$. The constraints can ensure the DIBs rather than the stellar residuals or noises significantly contribute to the EWs.
\end{enumerate}

Besides the above criteria, we also tighten the constraints on the CV and central wavelength for the M- and K-type ($T_{\rm eff} < 5200 \ \rm K$) stars to get rid of the fake DIB signals caused by the stronger metal absorptions in cooler stars. Specifically, the CV must be less than 5\% and the standard deviation of $\rm \{ \mu_{pr}, \ \mu_{cf}, \ \mu_{mc} \}$ must be less than 0.25 \AA. After applying the above criteria, we obtain the HQ samples, and the summary of the HQ samples are also listed in Table \ref{tab:stat}. There are a total of 7,681 spectra simultaneously contain the HQ measurements of the DIBs $\lambda 5780$, $\lambda 5797$, and $\lambda 6614$. Besides the public access to three catalogs of the HQ samples for the DIBs $\lambda 5780$, $\lambda 5797$, and $\lambda 6614$, here we also release the results of these 7,681 HQ measurements \footnote{We emphasize that the ``HQ sample'' or ``HQ measurement'' refers to the individual DIB measurement of each cool stellar spectrum, and the ``HQ source'' refers to the mean value of the DIB measurements of the multiple-epoch spectra for the same star. See more details in Sects. \ref{subsubsec:multi_epoch} and \ref{sec:results}.} via \url{https://nadc.china-vo.org/res/r101404/}.

\subsubsection{Defining the marginal-quality samples} \label{subsubsec:mq}

Aside from the HQ samples, the marginal-quality (MQ) samples are also available online. Although the MQ samples cannot provide DIB measurements as reliable as the HQ samples, these low or null values can offer important boundary constraints for other studies, for example, 3D mapping. For each MQ DIB measurement, it is required to meet the following two criteria.

\begin{enumerate}
\item The EW fitted by the MCMC, $\rm EW_{DIB} \leq 0.03$ \AA, where the upper limit is set to 0.03 \AA \ according to the typical uncertainty of 0.04 \AA \ for the DIB measurements validated in Sect. \ref{subsubsec:ew}, and the lowest EW of around 0.03 \AA \ measured in the HQ samples.

\item $\rm EW_{cont} \leq 0.5$ \AA, where $\rm EW_{cont}$ is the integral equivalent width of the continuum outside the DIB region, namely $\rm [\mu_{mc} - 3 \sigma_{mc}, \ \mu_{mc} + 3 \sigma_{mc}]$. The upper limit is set to 0.5 \AA \ to ensure that the stellar residuals are cleaned up to the continuum level.
\end{enumerate}

After applying the above criteria, there are a total of 27,598, 42,735, and 21,635 MQ measurements of the DIBs $\lambda 5780$, $\lambda 5797$, and $\lambda 6614$, respectively, corresponding to 23,920, 36,825, and 18,145 unique stars. We note that we do not set a specific threshold to further distinguish between marginal and null values, as the uncertainties estimated by MCMC are provided in the catalogs. Therefore, users can determine the threshold based on their needs. However, we suggest using a ratio of $\rm err(EW_{DIB}) / \rm EW_{DIB}$ of 1 as the dividing line: values below 1 can be considered marginal, while values above 1 can be classified as null.

\subsection{Validation} \label{subsec:validation}

\subsubsection{Central wavelength versus Gaussian width} \label{subsubsec:lambda_sigma}

The $\lambda - \sigma$ 2D histogram has been a powerful diagnostic tool for visually distinguishing real DIBs from false positives in the context of automatic measurement of DIBs from a large spectroscopic survey (\citealt{2023ApJ...954..141S}; \citealt{2024A&A...683A.199Z}). The false-positive DIBs, which may be stellar absorption, are usually far away from the typical central wavelength of the DIBs, and their Gaussian widths are narrower than the typical values of the real DIBs as the FWHM of a DIB is generally broader than that of metal absorption line. In the 2D histogram, the true DIB measurements are located around the rest-frame central wavelength and typical Gaussian width of the DIB, but with a certain scatter, whereas the stellar features exhibit smaller Gaussian widths and are highly concentrated in a region.

\begin{figure*}[ht]
\includegraphics[width=1.0\textwidth]{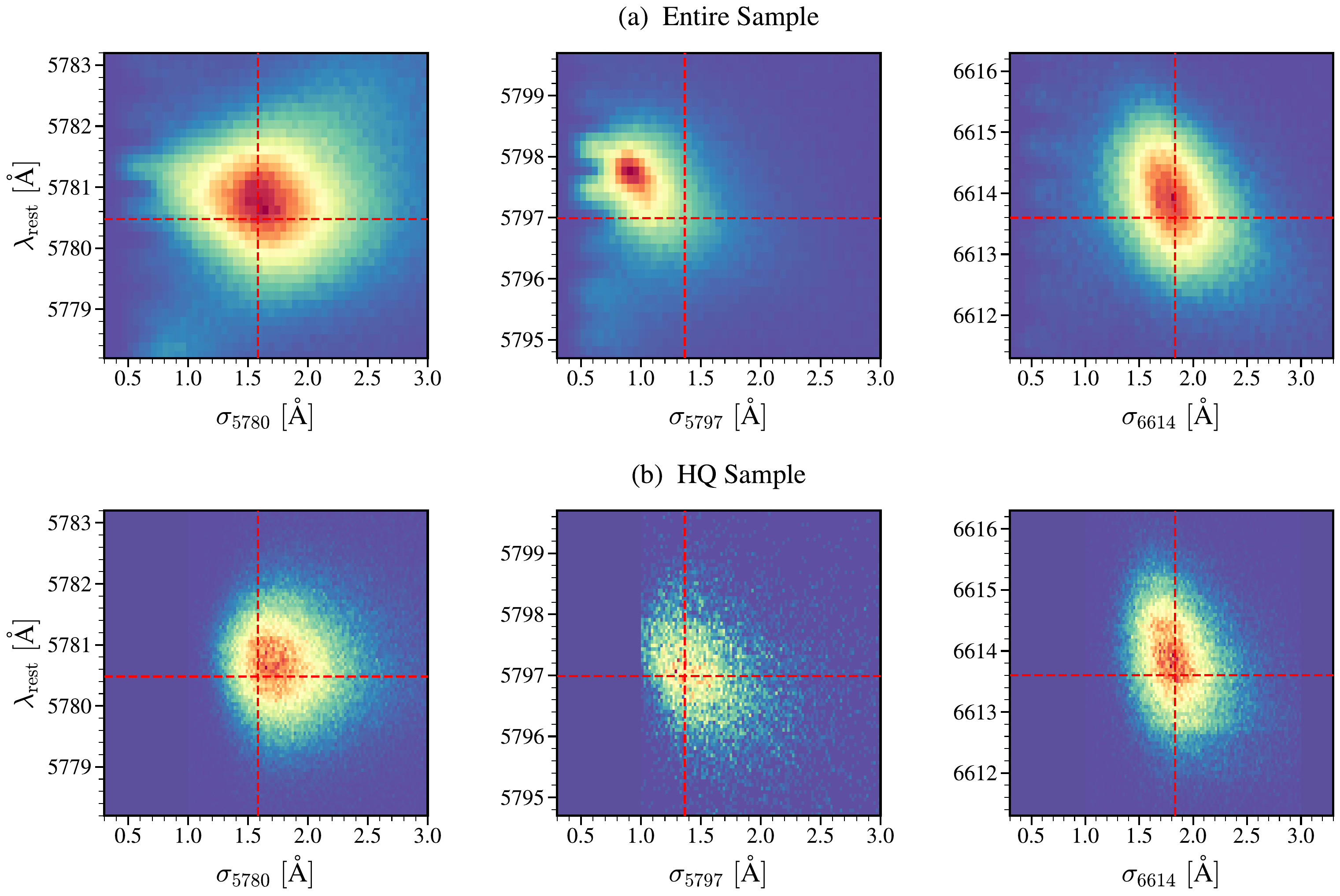}
\caption{
Two-dimensional histograms of the measured DIBs $\lambda 5780$, $\lambda 5797$, and $\lambda 6614$ as a function of the Gaussian width $\sigma$ and the central wavelength $\lambda$ in the rest frame. (a) The three upper panels present the density of all the samples that may have DIBs, (b) and the other three lower panels show the density of the HQ samples for which we have reliable DIB measurements. All of the histograms have the same bin size of 100. The red horizontal dashed lines denote the central wavelengths of the DIBs, which are fitted using the method described in Sect. \ref{subsec:cw}. The red vertical dashed lines indicate the typical Gaussian widths of the DIBs considering the instrumental broadening of LAMOST LRS (see Appendix \ref{appd:ib}), which are taken from \cite{2023MNRAS.521.3727V}.
\label{fig:lambda_sigma}
}
\end{figure*}

Figure \ref{fig:lambda_sigma} illustrates the $\lambda - \sigma$ distribution of the DIBs $\lambda 5780$, $\lambda 5797$, and $\lambda 6614$ for the entire samples and the HQ samples. It is clear that for either the entire samples or the HQ samples, the DIBs $\lambda 5780$ and $\lambda 6614$ are mainly distributed nearby the crossed red dashed lines that represent the statistical central wavelength fitted by the method in Sect. \ref{subsec:cw} and the broadened Gaussian width (see Appendix \ref{appd:ib}) of the DIB measured by \cite{2023MNRAS.521.3727V}. However, the distribution of the DIB $\lambda 5797$ for the entire samples displays an abnormal high-density region centered at about $(\sigma, \ \lambda) = (0.9,\ 5797.8)$ \AA, which deviates from the expected position marked by the red dashed lines. \cite{1975ApJ...196..129H} has reported a Si I line at 5797.859 \AA \ can obliterate the DIB $\lambda 5797$ in the spectra of the type F and later stars, which explains such an abnormal high-density region. Nevertheless, after the strict quality control, particularly with the criterion of $1 < \rm \sigma_{mc} < 3$ \AA, this anomalous area is almost completely eliminated, and the distribution of the DIB $\lambda 5797$ for the HQ samples is consistent with the expected position.

\subsubsection{Integral EW versus fitted EW} \label{subsubsec:ew}

\begin{figure*}[ht]
\includegraphics[width=1.0\textwidth]{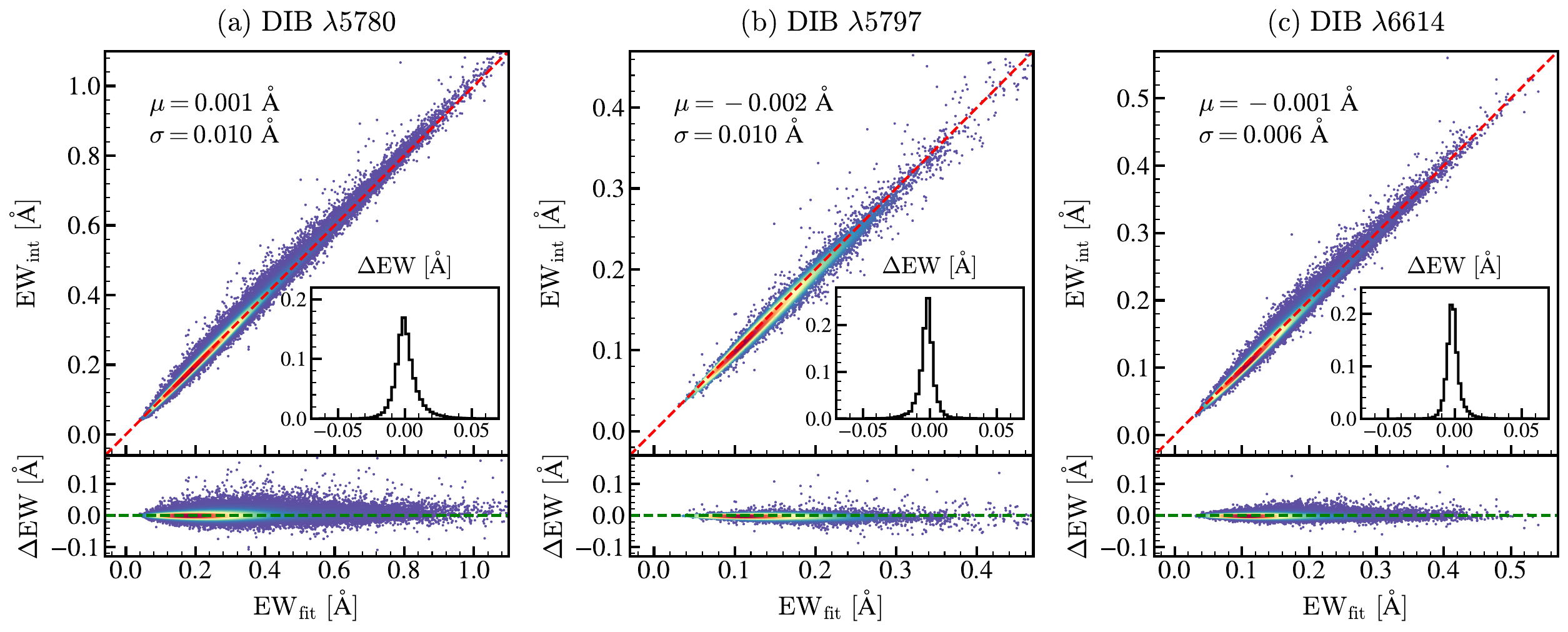}
\caption{
Comparison between the integral EW and the fitted EW of the DIBs $\lambda 5780$, $\lambda 5797$, and $\lambda 6614$ for their HQ samples, from the subfigures (a) to (c). For each subfigure, the upper panel shows the Gaussian KDE number density plot of the integral EW as a function of the fitted EW, with the red dashed line representing the one-to-one relation. The lower panel displays the Gaussian KDE number density plot of the difference between the integral EW and the fitted EW, namely $\rm \Delta EW = EW_{int}-EW_{fit}$ as a function of the fitted EW, with the green dashed line representing the zero benchmark. The distribution of $\rm \Delta EW$ is also overplotted in the upper panel with its mean and standard deviation annotated.
\label{fig:ew}
}
\end{figure*}

To quantify the reliability of the DIB measurements, we compute the integral EW ($\rm EW_{int}$) of the three DIBs in the HQ samples within the same wavelength range, that is, $\rm \mu_{mc} \pm 3\sigma_{mc}$ as the fitted EW ($\rm EW_{fit}$). As shown in Fig. \ref{fig:ew}, the $\rm EW_{int}$ is highly consistent with the $\rm EW_{fit}$, and the difference between them ($\rm \Delta EW$) is mainly concentrated in a tiny region that is less than 0.04 \AA \ for over 99\% of the samples among three DIBs. For all three DIBs, however, there are certain samples whose $\rm EW_{int}$ are far away from their $\rm EW_{fit}$, which stem from the fact that the continuum level is not well determined (\citealt{2015ApJ...800...64F}; \citealt{2019NatAs...3..922F}). For the DIB $\lambda 5780$, the blend with a very broad DIB $\lambda 5778$ is a confounding factor in the determination of the continuum level (\citealt{1975ApJ...196..129H}; \citealt{1987ApJ...312..860K}; \citealt{1993ApJ...407..142H}). Similar with the DIB $\lambda 5780$, the blue wing of the DIB $\lambda 5797$ is also overlapped with a quite broad DIB $\lambda 5795$ (\citealt{1987ApJ...312..860K}; \citealt{2011ApJ...727...33F}). Nevertheless, \cite{2011A&A...533A.129V} has reported that the broader DIBs $\lambda 5778$ and $\lambda 5795$ are largely eliminated from the ISM residual spectra in the local normalization, and they are too weak to significantly contaminate the EW measurements of the DIBs $\lambda 5780$ and $\lambda 5797$, even for high extinction. Such broader DIB blend does not exist for the DIB $\lambda 6614$ (\citealt{1982ApJ...252..610H}; \citealt{2008MNRAS.386.2003G}), which gives rise to better agreement (see Fig. \ref{fig:ew}(c)) between the $\rm EW_{int}$ and the $\rm EW_{fit}$ than the other two DIBs.

A sequential eye check for certain ISM residual spectra with large $\rm \Delta EW$ reveals that the blend with the broader DIB does destroy the continuum level and makes the fitted Gaussian width wider than the specified DIB. Assuming the continuum level is well defined, a multi-Gaussian profile may be a better choice to fit the DIB with a DIB blend. For example, \cite{2015MNRAS.452.3629L} has adopted a double Gaussian profile to simultaneously fit the DIB $\lambda 5780$ and $\lambda 5778$. In addition, the extra structural features are found among those ISM residual spectra, which are likely to be the stellar residuals (\citealt{2024A&A...683A.199Z}). One probable reason is that the stellar features are not completely erased by the subtraction of the stellar template from the target spectrum, which can be further attributed to the mismatch between the stellar template and the target spectrum, or the incorrect alignment caused by the false radial velocity. Nonetheless, the overall agreement between the $\rm EW_{int}$ and the $\rm EW_{fit}$ can support the reliability of the DIB measurements and reinforce the following analysis in statistical properties.

\subsubsection{Multiple epochs} \label{subsubsec:multi_epoch}

\begin{figure*}[ht]
\includegraphics[width=1.0\textwidth]{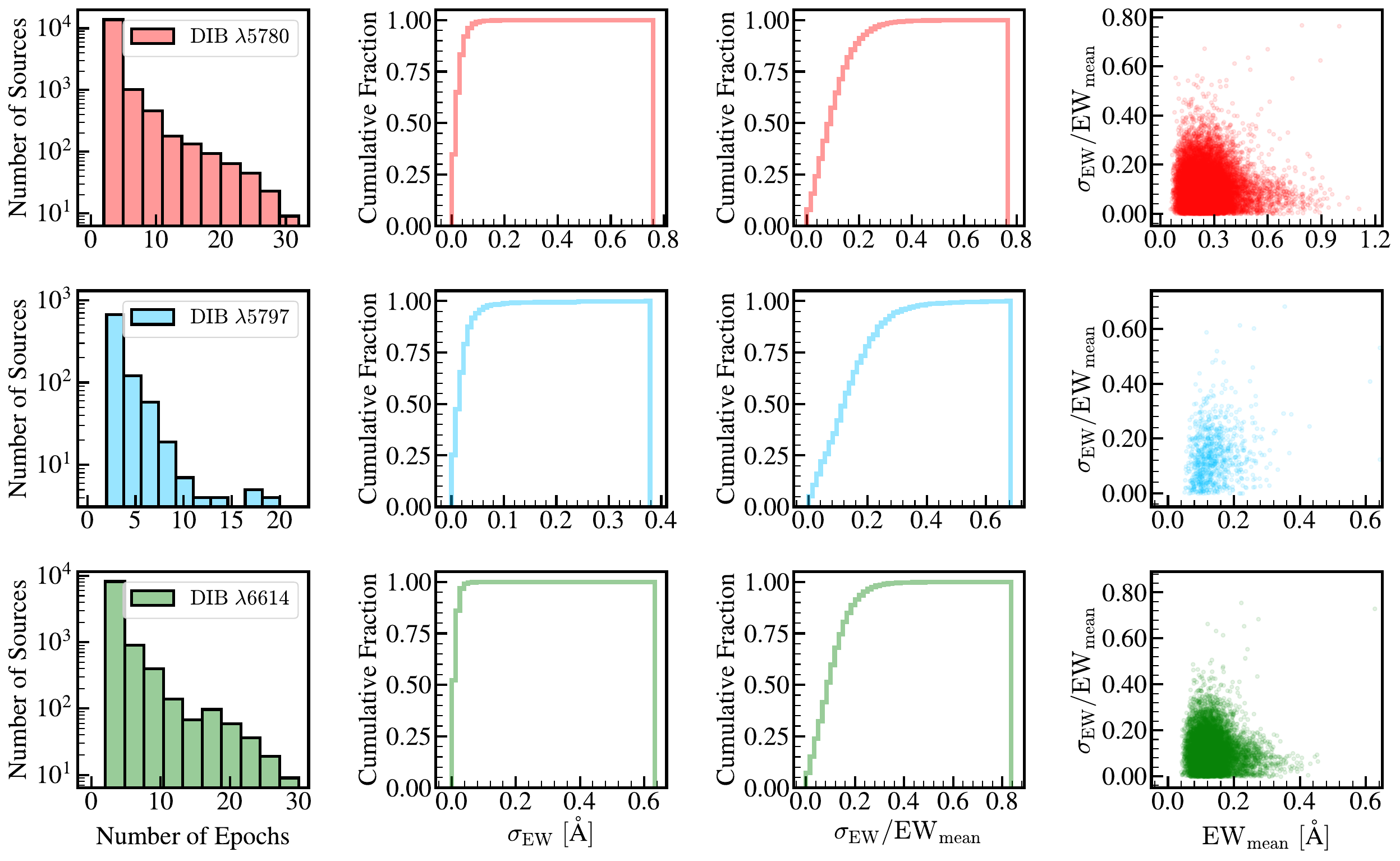}
\caption{
Three distinct colors encode the three DIBs $\lambda 5780$, $\lambda 5797$, and $\lambda 6614$ for the HQ samples with over two epochs. The panels in the first column present the distribution of the number of epochs. The panels in the second column depict the cumulative distribution of the standard deviation of the EWs ($\rm \sigma_{EW}$). The panels in the third column display the cumulative distribution of the coefficient of variation ($\rm \sigma_{EW} / EW_{mean}$). The panels in the fourth column show the scatter of the $\rm \sigma_{EW} / EW_{mean}$ as a function of the $\rm EW_{mean}$.
\label{fig:epochs}
}
\end{figure*}

Benefiting from the high efficiency in the spectrum acquisition, LAMOST has accumulated numerous sources with multiple epochs. These sources that have been observed many times can provide a good opportunity to validate the variation of DIB measurement. We select the sources with over two epochs from the HQ samples by grouping the sources with the same {\tt uid} (unique source identifier in LAMOST) and present the distribution of the number of epochs in the panels in the first column of Fig. \ref{fig:epochs}. There are 15,839, 891, and 9,939 sources with over two epochs for the DIBs $\lambda 5780$, $\lambda 5797$, and $\lambda 6614$, respectively. These sources have four epochs on average and the maximum number of epochs is up to 33. The panels in the second column of Fig. \ref{fig:epochs} illustrate the cumulative distribution of the standard deviation of the EWs ($\rm \sigma_{EW}$) for multiple-epoch sources. Almost 95\% of the sources have $\rm \sigma_{EW} < 0.05$ \AA \ for three DIBs, which is comparable to the typical value, namely 0.04 \AA \ of $\rm \Delta EW$ investigated in Sect. \ref{subsubsec:ew}. Such similarity between $\rm \sigma_{EW}$ and $\rm \Delta EW$ implies that the variation of the DIB measurements among multiple epochs of the same source is resulted from the measurement error rather than the intrinsic property of the DIB. Without the influence of blend with a broader DIB on the DIB $\lambda 6614$ (see Sect. \ref{subsubsec:ew}), it is not surprising that the cumulative distribution of the DIB $\lambda 6614$ accumulates toward 1 more rapidly than that of the DIBs $\lambda 5780$ and $\lambda 5797$. 

In order to present the relative variation of the EWs, we also supply the cumulative distribution of the coefficient of variation (CV = $\rm \sigma_{EW} / EW_{mean}$) in the panels in the third column of Fig. \ref{fig:epochs}. Well over three quarters of the sources have CV $< 0.2$ for the DIB $\lambda 5797$, and this proportion is more than 90\% for the DIBs $\lambda 5780$ and $\lambda 6614$. The large CV of the multiple-epoch sources generally comes from the small EW, which appears to be more sensitive to the measurement error, as can be seen in the panels in the fourth column of Fig. \ref{fig:epochs}. It is noted that both the cumulative distribution of the $\rm \sigma_{EW}$ and the CV demonstrate the long tail. Besides the measurement error mentioned above, such an imbalance is driven by the differences among the atmospheric parameters of multiple spectra for the same source. Specifically, although these spectra originate from the same source, the differences in observation conditions, processing procedures, instrument running status, etc., result in variations in atmospheric parameter measurements among these spectra. Consequently, their matched closest neighboring template spectra also differ, which ultimately leads to a noticeable discrepancy in the ISM residual spectra. In the visual inspection, we found that the maximum difference in $T_{\rm eff}$ among multiple-epoch spectra of the same source can reach 500 K, far exceeding the match threshold of 100 K set in Sect. \ref{subsubsec:neighbor}. \cite{2024A&A...683A.199Z} and \cite{2023ApJ...954..141S} have circumvented the issues of atmospheric parameters and directly extracted the DIB $\lambda 8621$ signal from the parameter-free spectra using machine learning methods. The applicability of this method, however, requires further validation, which is also one of our future directions.

\subsubsection{Residuals} \label{subsubsec:residuals}

\begin{figure*}[ht]
\includegraphics[width=1.0\textwidth]{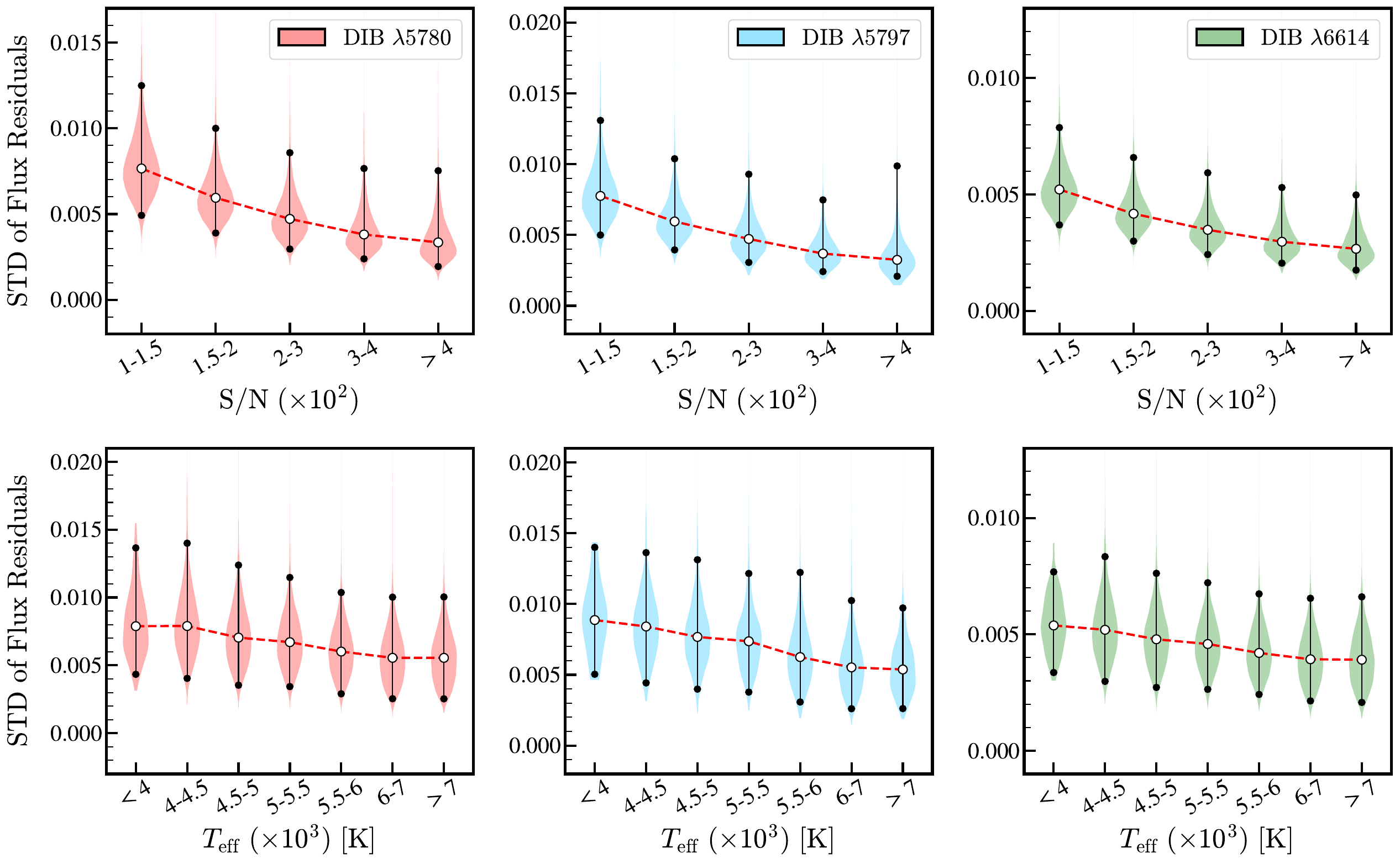}
\caption{
Trend of the standard deviation of the residuals between the flux of the ISM residual spectrum beyond the DIB region and that of the continuum as a function of S/N in the upper panels, and $T_{\rm eff}$ in the lower panels. Three distinct colors encode the three DIBs $\lambda 5780$, $\lambda 5797$, and $\lambda 6614$. In each panel, a violin plot is used to present the distribution of the standard deviation of the residuals for the binned HQ samples. The median of each bin is annotated with a white dot, and the 2.5 and 97.5 percentiles are marked with black dots. The red dashed line connects the median of each bin to show the trend of the standard deviation of the residuals.
\label{fig:res}
}
\end{figure*}

In addition to validating the measurements of DIBs, the quality of ISM residual spectra is also verified. Ideally, after subtracting the stellar features using the template spectrum, the flux of the ISM residual spectrum outside the DIB region which is from $\rm \mu_{mc}-3\sigma_{mc}$ to $\rm \mu_{mc}+3\sigma_{mc}$ should be at the continuum level with no stellar components. Thus, we calculate the standard deviation of the residuals between the flux of ISM residual spectrum outside the DIB region and that of the continuum, and then provide the trend of STD with S/N and $T_{\rm eff}$ in Fig. \ref{fig:res}. As can be seen in the figure, the STD decreases as S/N increases, which indicates that higher S/N leads to smaller residuals in the ISM residual spectrum or cleaner subtraction of stellar components. The STD also decreases with increasing $T_{\rm eff}$ as a consequence of the weakening of metal lines in hotter stars. However, this decreasing trend of $T_{\rm eff}$ is more gradual compared to that of S/N, reflecting the robustness of measurement, which is less disturbed by absorption lines in cool stars. Furthermore, the magnitude of STD is comparable to the level of normalized flux error estimated in \ref{subsubsec:preprocess}, which indicates that the pipeline does not introduce additional errors and that the ISM residual spectrum is quite clean, with no significant residual stellar components. It is noteworthy that although there probably exist the DIB components other than the interesting ones within the range of the ISM spectrum, they are usually weak absorptions, such as the DIBs $\lambda 5762$ and $\lambda 5815$ in Fig. \ref{fig:method}, documented in the previous DIB catalogs (\citealt{2019ApJ...878..151F}; \citealt{2023MNRAS.521.3727V}). Moreover, after instrumental broadening by the LAMOST LRS, the depths of those weak DIBs and the flux errors share the close values, which cannot cause any significant bias in the final results.

\section{Results and discussion} \label{sec:results}

In this section, we review and explore the spatial distribution, correlation, central wavelength and kinematics of the DIBs $\lambda 5780$, $\lambda 5797$, and $\lambda 6614$ in the context of statistics based on the HQ samples. Due to the multiple-epoch sources mentioned in Sect. \ref{subsubsec:multi_epoch}, we combine the measurements of DIB from the same source and take the mean value as its final result. After combination, the number of the HQ sources for the DIBs $\lambda 5780$, $\lambda 5797$, and $\lambda 6614$ is 142,074, 11,480, and 85,301 respectively.

\subsection{Spatial distribution} \label{subsec:spatial}

\begin{figure*}[ht]
\includegraphics[width=1.0\textwidth]{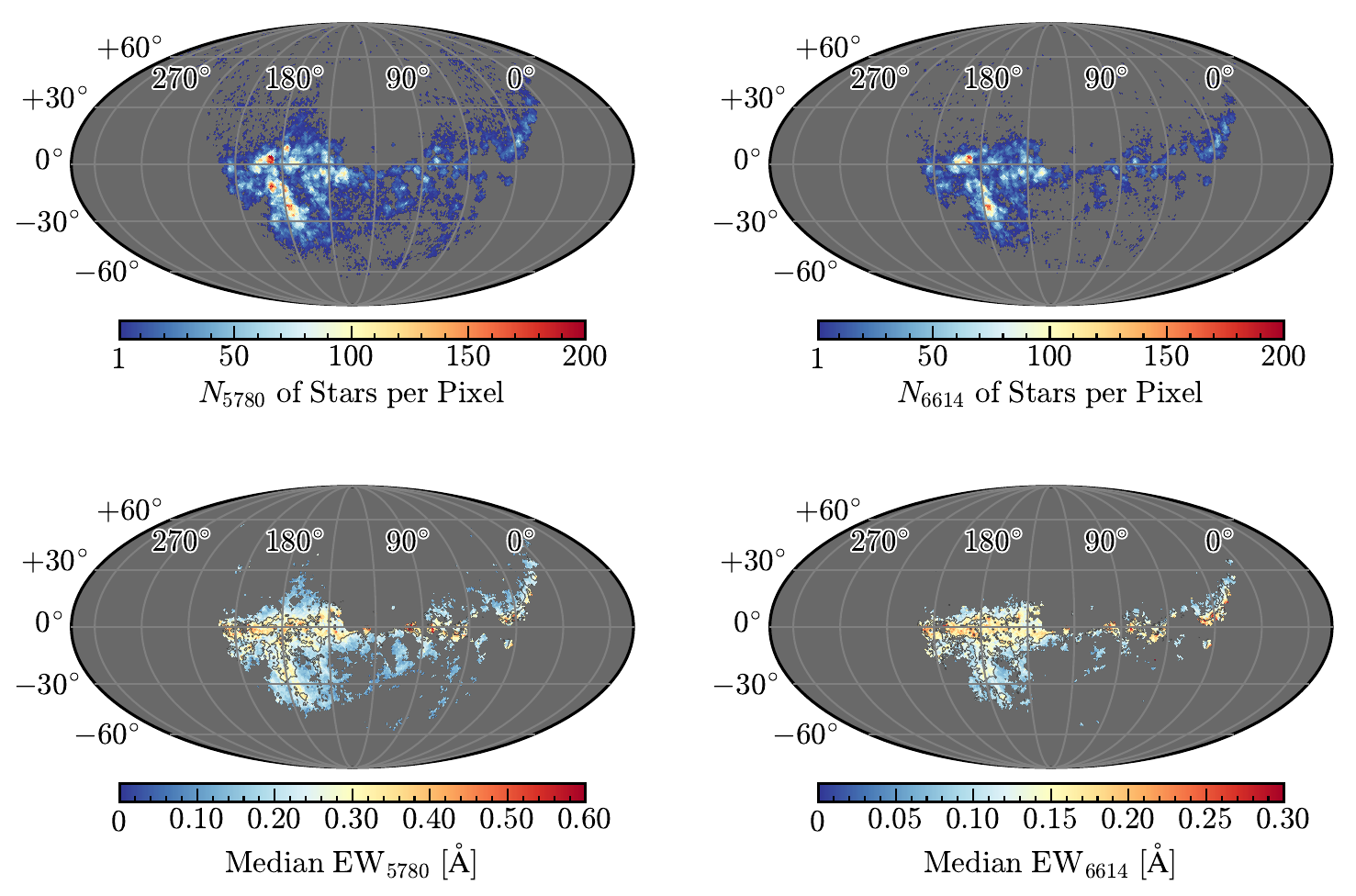}
\caption{
Galactic distribution of the DIBs $\lambda 5780$ and $\lambda 6614$ for the HQ samples. The upper panels show the number density of the background stars. The lower panels display the sky distribution of the strength of the DIBs with the color bar representing the median EW in each pixel. The black contours of the corresponding median $\rm E(B-V)$ at the level of 0.35 and 0.7 magnitude are overplotted in the median EW map. The resolution of the map is about 0.92$^{\circ}$ ($N_{\rm side}=64$) calculated by the HEALPix package. 
\label{fig:spatial}
}
\end{figure*}

Figure \ref{fig:spatial} presents the number density of the background stars and the sky distribution of the median EW of the DIBs $\lambda 5780$ and $\lambda 6614$. The absence of the DIB $\lambda 5797$ in the figure is due to the scarcity of the HQ sources caused by the low detection rate (see Table \ref{tab:stat}). In order to present a more generalized sky distribution of DIB in the lower panels of Fig. \ref{fig:spatial}, we remove the pixels in which the number density of the background stars is less than 3. Among the removed pixels, some exhibit significantly larger EWs than the values of DIB in the surrounding areas. Upon inspecting the spectra of the DIBs with large EW, we find that their background stars are located at considerable distances, which can lead to the large extinctions and EWs. However, there are some large EWs of DIB at the sight line of the short-distance background stars as well, which are likely to be the fake DIB signals investigated in Sect. \ref{sec:vaildation} or the incomplete correlation between the DIB and extinction discussed in Sect. \ref{subsubsec:dib_ebv}.

The DIBs $\lambda 5780$ and $\lambda 6614$ share a similar sky distribution, that is, the Galactic plane is the prominent region with the strong DIBs and the strength of the DIBs decreases with the increasing Galactic latitude. The lower number of samples for DIB $\lambda 6614$ compared to DIB $\lambda 5780$ at high latitudes is caused by the weaker intensity of DIB $\lambda 6614$ per unit extinction (\citealt{2023MNRAS.521.3727V}), making it less detectable in the regions of low extinction at high galactic latitudes. The high-EW regions of the DIBs $\lambda 5780$ and $\lambda 6614$, under the Galactic view, are both spatially correlated with the well-known molecular clouds, such as Taurus and Orion. This behavior is similar to that of the DIB $\lambda 8621$ investigated in $Gaia$ RVS spectra (\citealt{2023A&A...674A..40G}; \citealt{2023ApJ...954..141S}). In addition, there is a notable extent to high latitudes ($b \approx -45^{\circ}$) in the directions of Galactic anti-center, which does not entirely align with the distribution of molecular clouds. This stretch from the plane to high latitudes is similar to that of the DIB $\lambda 8621$ depicted by \cite{2023A&A...674A..40G}, \cite{2023ApJ...954..141S}, and \cite{2024A&A...683A.199Z}. The causes of such a phenomenon remain unclear, but it is likely attributable to the selection effect of the observation, as can be seen in the sky distribution of the number density in the upper panels of Fig. \ref{fig:spatial}. Another possible explanation is that the resolution of map is not high enough to resolve the sky distribution of the DIBs, which makes it appear as though the region is interconnected. It is anticipated that with the inclusion of DIB measurements from hot stars of LAMOST in the future, more detailed and accurate northern sky distribution of the DIBs will be obtained. Additionally, a three-dimensional spatial distribution of the DIBs, such as \cite{2014Sci...345..791K} and \cite{2019NatAs...3..922F}, will be constructed, which can provide a more comprehensive understanding of the DIBs and their positional and physical relationship with the ISM.

\subsection{Correlation} \label{subsec:correlation}

There have been numerous studies on the correlation of DIB with DIB and DIB with other ISM tracers, such as extinction, CO gas, atomic hydrogen, and molecular hydrogen (e.g., \citealt{1975ApJ...196..129H}; \citealt{2004A&A...414..949W}; \citealt{2011ApJ...727...33F}; \citealt{2015MNRAS.452.3629L}). However, these works are more or less limited by the number of sight lines or combined the spectra along the same sight line for high S/N, which sacrifice the statistical generalization. For the sake of statistical and homogeneous comparison, a total of 7,681 common samples for the DIBs $\lambda 5780$, $\lambda 5797$, and $\lambda 6614$ from the HQ samples are picked out to investigate the correlation of DIB V.S. extinction and DIB V.S. DIB. 

\subsubsection{Correlation between DIBs and E(B-V)} \label{subsubsec:dib_ebv}

\begin{figure}
\centering
\includegraphics[width=0.48\textwidth]{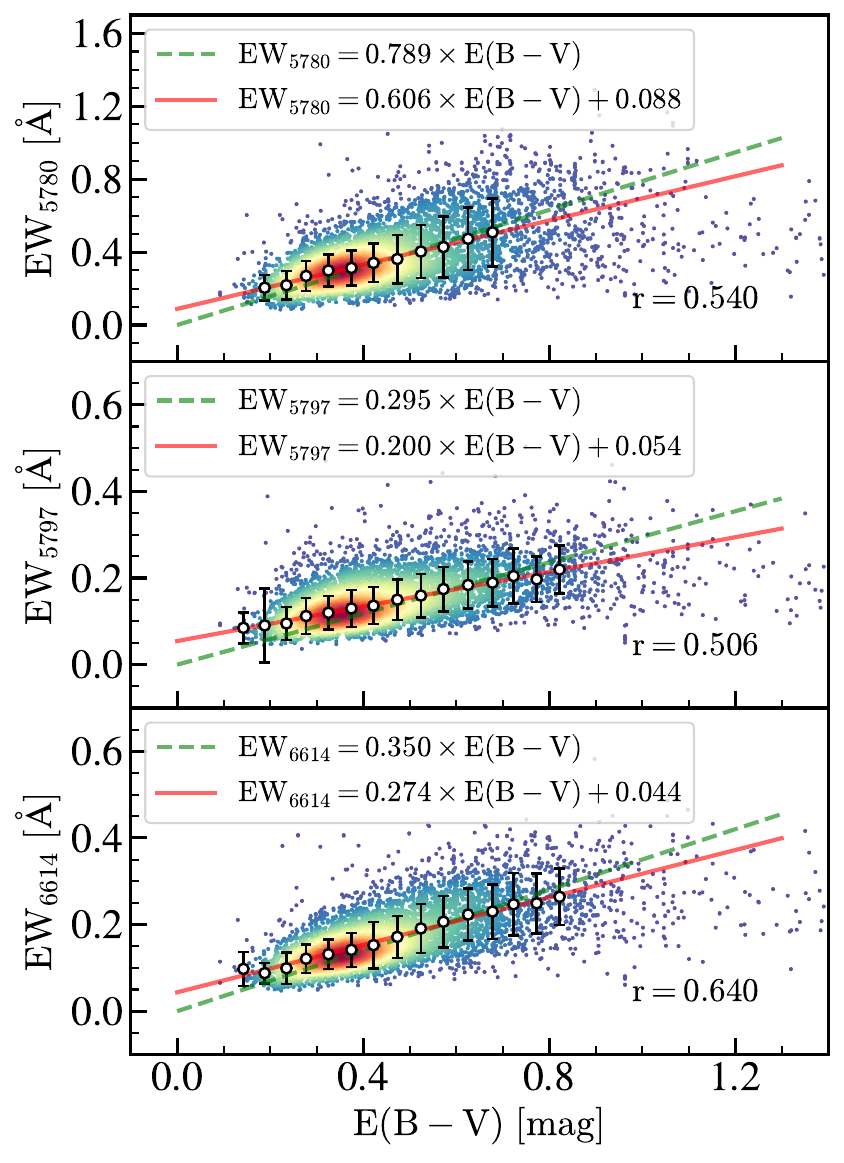}
\caption{
Gaussian KDE number density plot of the EWs of the DIBs $\lambda 5780$, $\lambda 5797$, and $\lambda 6614$ (from top to bottom) as a function of the extinction $\rm E(B-V)$ derived from \cite{2019ApJ...887...93G}. For each panel, the red solid line represents a linear fit to the median EW (which is marked with a white dot, and its standard deviation marked is with a black error bar) in each $\rm \Delta E(B-V) = 0.05 \ mag$ bin. The green dashed line is the same as the red solid line, but the intercept is forced to be zero. The coefficients of the linear fit and the Pearson correlation coefficient between the EW and $\rm E(B-V)$ are also annotated in each panel.
\label{fig:dib_ebv}
}
\end{figure}

To avoid zero or inaccurate values of $\rm E(B-V)$ at close distances, the heliocentric distance $d$ is limited to greater than 500 pc and a total of 6,514 common HQ sources are selected. Figure \ref{fig:dib_ebv} compares the EW of DIB with the extinction $\rm E(B-V)$ derived from \cite{2019ApJ...887...93G} by the linear fit to the median EW in each $\rm \Delta E(B-V) = 0.05 \ mag$ bin. All three DIBs agree with a linear positive correlation with $\rm E(B-V)$ in Table \ref{tab:dib_ebv}, but not perfect, as known for a few decades (\citealt{1995ARA&A..33...19H}). Nonetheless, the areas with high number density (in red) show smaller scatters and have better linear relationship (red solid lines). Even in sparsely sampled regions, such as those near low extinction of 0.15 mag and high extinction of over 0.6 mag, the medians do not deviate from the fitted curve. We do not include the samples with the higher extinction ($>$ 1 mag) in the fitting process due to the well-known ``skin effect'' (\citealt{1995ARA&A..33...19H}), namely the strength of DIB cannot indefinitely increase with the extinction as expected. Besides, we also attempt a linear fit forced through the origin (green dashed lines) to facilitate the comparison with the previous works, for instance, \cite{1975ApJ...196..129H}. It is clear that the green dashed lines are not suitable to represent the correlation of the DIBs with $\rm E(B-V)$, as they have larger fitting errors and deviate from the red solid lines.

\renewcommand{\arraystretch}{1.1}
\begin{table}
\caption{Correlation between DIB and $\rm E(B-V)$ \label{tab:dib_ebv} \tablefootmark{(1)}}
\centering
\setlength\tabcolsep{5pt}
\begin{tabular}{lcrrrrr}

\hline\hline
 & \multicolumn{3}{c}{Common sources \tablefootmark{(2)}} & \multicolumn{3}{c}{All HQ sources \tablefootmark{(3)}} \\ [0.1ex]
DIB & Number & $m$ & $n$ & Number & $m$ & $n$ \\
\hline
$\lambda 5780$ &       & 0.606 & 0.088 & 118,157 & 0.565 & 0.087 \\
$\lambda 5797$ & 6,514 & 0.200 & 0.054 &   9,760 & 0.176 & 0.069 \\
$\lambda 6614$ &       & 0.274 & 0.044 &  73,234 & 0.256 & 0.042 \\
\hline

\end{tabular}
\tablefoot{ \\
\tablefoottext{1}{Coefficients for $\rm EW \ [\AA] = \textit{m} \times E(B-V) [mag] + \textit{n}$.} \\
\tablefoottext{2}{Common sources at $d > 0.5 \ \rm kpc$.} \\
\tablefoottext{3}{HQ sources at $d > 0.5 \ \rm kpc$.} \\
}
\end{table}


As one of the strongest and earliest discovered DIBs, the correlation of DIB $\lambda 5780$ with $\rm E(B-V)$ is the most studied. Our slope of 0.606 $\rm \AA \, mag^{-1}$ is close to the slope of 0.64 $\rm \AA \, mag^{-1}$ reported by \cite{2011A&A...533A.129V} \footnote{\cite{2011A&A...533A.129V} utilized the data along the specified sight lines called ``$\sigma$-type'' to fit.}. The slopes of 0.549 $\rm \AA \, mag^{-1}$ computed by \cite{1975ApJ...196..129H} \footnote{The slope was fitted based on the data excluding the sight lines towards the CYG and SCO-OPH that have skin effect.}, 0.505 $\rm \AA \, mag^{-1}$ fitted by \cite{2011ApJ...727...33F} \footnote{The slope is actually computed by the inverse of the slope of 1.98, because \cite{2011ApJ...727...33F} used $\rm E(B-V) = \textit{a} \times EW + \textit{b}$ as their fitting function.}, 0.525 $\rm \AA \, mag^{-1}$ reported in \cite{2012A&A...544A.136R}, and 0.43 $\rm \AA \, mag^{-1}$ derived by \cite{2015MNRAS.452.3629L} \footnote{\cite{2015MNRAS.452.3629L} employed a power-law function, that is, $\rm EW = \textit{a} \times E(B-V)^{\gamma}$, as their fitting reference. Nonetheless, for the DIBs $\lambda 5780$, $\lambda 5797$, and $\lambda 6614$, the fitted values of $\gamma$ are 1.0, 0.96, and 0.97, which can approximate the linear function.} are all smaller than our slope. Some other works have even obtained the slopes lower than those found in the above works, such as the slope of 0.401 $\rm \AA \, mag^{-1}$ from \cite{2013A&A...555A..25P} and the slope of 0.419 $\rm \AA \, mag^{-1}$ from \cite{2011A&A...533A.129V} \footnote{The slope was fitted based on the data along the specified sight lines called ``$\zeta$-type''.}. There are three aspects that can cause the discrepancy of these slopes. One is the measurement error of the DIBs and the extinction. Apart from \cite{2015MNRAS.452.3629L}, the other studies focused on their data within an absolutely narrow range of EW or $\rm E(B-V)$. For instance, the mean EW of the DIB $\lambda 5780$ in \cite{2013A&A...555A..25P} is about 60 m\AA, with the measurement errors exceeding the EWs themselves (see Fig. 6 in their work). Similarly, the samples with large EW in \cite{2011ApJ...727...33F} hardly influence the fitting and significantly deviate from the fitted curve. Consequently, the fitting results are predominantly driven by the samples with $\rm E(B-V) < 0.2$ mag (refer to the enlarged region in Fig. 4 of their work). Another reason is the different profile definition of the DIB $\lambda 5780$ (see Sect. \ref{subsubsec:lambda_sigma} for more details). \cite{2015MNRAS.452.3629L} separated the broader DIB $\lambda 5778$ from the DIB $\lambda 5780$ and simultaneously fitted them with a double Gaussian profile, whereas the other works, including ours, treated them as a single DIB. Despite the discovery of the DIB $\lambda 5778$ as early as 1975 (\citealt{1975ApJ...196..129H}), the subsequent research on the $\lambda 5780$ often overlooked this blended broad DIB. If the contribution of the DIB $\lambda 5778$ by \cite{2015MNRAS.452.3629L} is attributed to the DIB $\lambda 5780$, the co-added result will align with the slope obtained in this work. The last key factor is the selection of the sight lines. \cite{2011A&A...533A.129V} elaborated the influence of $\sigma-$ and $\zeta-$type sight lines on DIB carrier according to the stars from the upper Scorpius. They concluded that the DIB $\lambda 5780$ in the $\sigma-$type sight line is stronger than that in the $\zeta-$type (\citealt{1988A&A...190..339K}; \citealt{2013ApJ...774...72K}), which may be caused by the different ultraviolet (UV) radiation fields, that is, the external ionized regions for the $\sigma-$type and the UV-shielded could cores for the $\zeta-$type. It is also worth noting that, compared to the DIBs $\lambda 5797$ and $\lambda 6614$, the dispersion of EWs in each extinction bin for the DIB $\lambda 5780$ increases with $\rm E(B-V)$, which further validates the $\sigma-\zeta$ effect on the DIB $\lambda 5780$. Such a $\sigma-\zeta$ effect on our slope is beyond the scope of this work, but our correlations of the DIB $\lambda 5780$ with $\rm E(B-V)$ offer more generalizable conclusions, and circumvent the sample selection effect (the limited number of sight lines) and the measurements that are sensitive to their own errors. Furthermore, the large number of samples in this work is likely to encompass more $\sigma-$ and $\zeta-$type sight lines, which will help us to understand the DIB $\lambda 5780$ along the different sight lines in a more comprehensive way. 

For the DIBs $\lambda 5797$ and $\lambda 6614$, their correlations with $\rm E(B-V)$ resembles those of previous works. Our slope of 0.2 $\rm \AA \, mag^{-1}$ for the DIB $\lambda 5797$ is consistent with 0.219 $\rm \AA \, mag^{-1}$ derived by \cite{1975ApJ...196..129H}, 0.174 $\rm \AA \, mag^{-1}$ computed by \cite{2011ApJ...727...33F}, and 0.18 $\rm \AA \, mag^{-1}$ fitted by \cite{2015MNRAS.452.3629L}. By similar token, the slopes of 0.262 $\rm \AA \, mag^{-1}$ from \cite{1975ApJ...196..129H}, 0.216 $\rm \AA \, mag^{-1}$ reported in \cite{2011ApJ...727...33F} and 0.22 $\rm \AA \, mag^{-1}$ documented by \cite{2015MNRAS.452.3629L} for the DIB $\lambda 6614$ are in good agreement with our slope of 0.274 $\rm \AA \, mag^{-1}$. Such consistency of the slopes among different works are unsurprising, as \cite{2013ApJ...774...72K} indicated that the DIBs $\lambda 5797$ and $\lambda 6614$ are unaffected by the $\sigma-\zeta$ effect. Additionally, although the DIB $\lambda 5797$ is influenced by the nearby DIB $\lambda 5795$ (\citealt{1987ApJ...312..860K}; \citealt{2011ApJ...727...33F}), this impact is confined to the blue wing of the DIB $\lambda 5797$ and not as significant as the complete overlap of the DIB $\lambda 5778$ with the broad DIB $\lambda 5780$. As to the DIB $\lambda 6614$, it is not blended with any other DIB (\citealt{1982ApJ...252..610H}; \citealt{2008MNRAS.386.2003G}), which makes the correlation with $\rm E(B-V)$ more straightforward. The probable cause of the slight difference in slope still stems from the measurement error of the DIBs and the extinction considering weaker EW per unit extinction of the DIBs $\lambda 5797$ and $\lambda 6614$ compared to the DIB $\lambda 5780$.

We also take the same data used to linear fittings to calculate the Pearson correlation coefficient between EW and $\rm E(B-V)$ for the three DIBs, which are 0.54, 0.51, and 0.64 for the DIBs $\lambda 5780$, $\lambda 5797$, and $\lambda 6614$, respectively. None of these DIBs reaches a significant level of correlation with extinction, consistent with the conclusions of many recent studies (\citealt{2011ApJ...727...33F}; \citealt{2017ApJ...850..194F}), which is caused by the variable interstellar environments as mentioned before. Furthermore, the differences of the number of data points that are used to compute correlation coefficients can also lead to the discrepancy of the interpretation of correlation. To generalize our conclusions further, besides the 6,514 common HQ sources, we perform the same fitting procedure on all HQ sources at heliocentric distances greater than 500 pc. The results are also summarized in Table \ref{tab:dib_ebv}, which are comparable to those of the common HQ sources. 

\subsubsection{Correlations between diffuse interstellar bands} \label{subsubsec:dib_dib}

\begin{figure}
\centering
\includegraphics[width=0.48\textwidth]{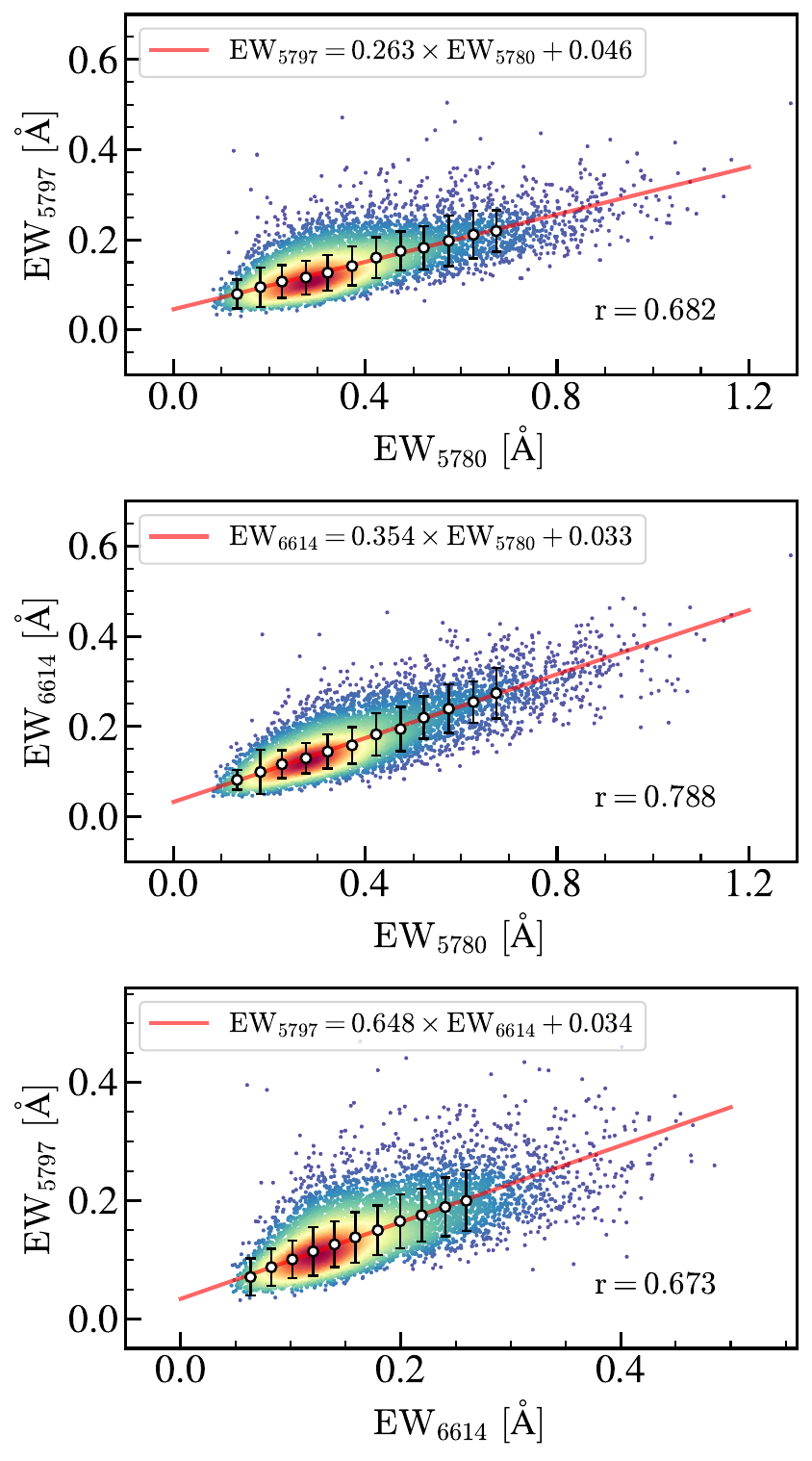}
\caption{
As in Fig. \ref{fig:dib_ebv}, but for the Gaussian KDE number density plot of EW V.S. EW. From the top panel to the bottom panel, the DIBs $\lambda 5780$, $\lambda 5797$, and $\lambda 6614$ are compared with each other. For each panel, the red solid line represents a linear fit to the median EW in each $\rm \Delta EW = 0.05 \ \AA$ bin. Except for the bottom panel, the bin size of the EW is $\rm \Delta EW = 0.02 \ \AA$. The coefficients of the linear fit and the Pearson correlation coefficient between the EWs are also annotated in each panel.
\label{fig:dib_dib}
}
\end{figure}

The correlation between DIBs can be used to infer whether their carriers come from the same DIB family or even share a common origin (\citealt{1999A&A...351..680M}; \citealt{2011ApJ...727...33F}; \citealt{2021MNRAS.507.5236S}). Figure \ref{fig:dib_dib} illustrates the mutual relation of DIB pairs and their Pearson correlation coefficients.

DIB $\lambda 5780 - \lambda 5797$: This pair of DIBs has been widely studied in the literature since the first discovery of the DIBs $\lambda 5780$ and $\lambda 5797$ (\citealt{1922LicOB..10..141H}). The general ratio of $\rm EW_{5797} / EW_{5780}$ is 0.263, which is lower than the value of 0.384 reported by \cite{2011ApJ...727...33F}. The high scatter of the data points used for fitting in \cite{2011ApJ...727...33F} is the main cause of this discrepancy. In Fig. 10 of \cite{2011ApJ...727...33F}, namely the figure showing the fitting results of $\rm EW_{5797}$ and $\rm EW_{5780}$, the fitting is predominantly driven by their samples with $\rm EW_{5780}$ of less than 140 m\AA \ (see the inset of their Fig. 10), but approximately one-third of the data points lie below the fitted line. In particular, there is a strong deviation trend for the samples with $\rm EW_{5780}$ between 150 and 300 m\AA. Additionally, our fitting is performed on the medians within each bin, rather than on all data points together as done by \cite{2011ApJ...727...33F}. Results of the later approach is often more susceptible to outliers. The famous $\sigma-\zeta$ effect (see the discussion in Sect. \ref{subsubsec:dib_ebv}) can be a probable cause of such a high scatter in \cite{2011ApJ...727...33F}, as seen, for instance, in the sight line towards Herschel 36 and 6 Cas annotated in their Fig. 10. It is also worth noting that while \cite{2011ApJ...727...33F} used the OB stars within the Galactic plane, our sample includes the sources at higher Galactic latitudes, which can also contribute to the statistical differences between the two studies. The ratio of $\rm EW_{5797} / EW_{5780}$, however, are more complex than hitherto thought, and cannot be adequately represented by a simple linear relationship. In addition to the $\sigma-\zeta$ effect, \cite{2019NatAs...3..922F} proposed that the carrier of the DIB $\lambda 5780$ can resist X-ray photodissociation and sputtering by fast ions, whereas the carrier of the DIB $\lambda 5797$ breaks down. A more sophisticated method, such as a 3D projection map (\citealt{2020Natur.578..237A}; \citealt{2023ApJ...954..141S}), is anticipated to depict it accurately.


DIB $\lambda 5780 - \lambda 6614$: Among all three DIB pairs, this pair shows the most significant correlation ranked by the Pearson correlation coefficient of 0.788. Interestingly, and somewhat paradoxically, many studies have provided conflicting conclusions regarding their classification into specific families or groups according to the different perspectives. For example, based on their intensity behavior through different types of clouds ($\sigma-$ and $\zeta-$types), \cite{2013ApJ...774...72K} contended that the DIB $\lambda 6614$ belongs to type I (where the linear relations with extinction are similar) and the DIB $\lambda 5780$ is type II (where they changed substantially). In contrast, \cite{2022MNRAS.510.3546F} categorized the DIBs $\lambda 5780$ and $\lambda 6614$ into the same $\sigma-$DIB group using a hierarchical agglomerative clustering method based on EW of DIB normalized to extinction. Additionally, in the studies conducted by \cite{1999A&A...351..680M}, \cite{2011ApJ...727...33F}, and \cite{2021MNRAS.507.5236S}, including ours, the correlation of this pair ranked among the top across all DIB pairs. One reasonable explanation for the conflict is that the carriers of the DIBs $\lambda 5780$ and $\lambda 6614$ display different properties in various physical environments, as \cite{2022MNRAS.510.3546F} argued that the DIBs form a rather continuous sequence. It is worth emulating a multidimensional approach by \cite{2015MNRAS.452.3629L} to decouple different effects on DIB. Through combining extra information such as the sky distribution and kinematics with the ISM tracers, a more exhaustive understanding of DIB will be achieved.

DIB $\lambda 6614 - \lambda 5797$: There is no evidence that supports the close relationship between the DIBs $\lambda 6614$ and $\lambda 5797$ (\citealt{1999A&A...351..680M}; \citealt{2011ApJ...727...33F}; \citealt{2021MNRAS.507.5236S}), as the Pearson correlation coefficient is 0.673, ranking the lowest among the three DIB pairs. Although the DIBs $\lambda 5780$ and $\lambda 5797$ unaffected by the $\sigma-\zeta$ effect (\citealt{2011A&A...533A.129V}), \cite{2022MNRAS.510.3546F} suggested that they belong to distinct DIB families.

\subsection{Central wavelength in the rest frame} \label{subsec:cw}

The central wavelength of DIB in the rest frame \footnote{In this work, the central wavelengths of the DIBs are all expressed in air and in the heliocentric rest frame without any correction of the RVs of the background stars except for validation of the stellar absorptions in Sect. \ref{subsec:validation}.} is one of the most important properties, which can identify the DIB carrier through the match of the band wavelengths and the strength ratios between the observational and laboratory measurements (\citealt{2015Natur.523..322C}; \citealt{2018ApJ...858...36C}; \citealt{2018A&A...614A..28L}; \citealt{2019ApJ...875L..28C}). Table \ref{tab:cw} provides the central wavelength of the DIBs $\lambda 5780$, $\lambda 5797$, and $\lambda 6614$ from five canonical DIB catalogs, which are measured using the Doppler shift of the well-identified interstellar atomic or molecular lines, such as Na\,\scalebox{0.8}{I} D doublet, K\,\scalebox{0.8}{I} line, and CH line (see Table 2 in \citealt{1994AandAS..106...39J}). This traditional method is based on the strong assumption that the DIB carriers co-move with the atoms or molecules so that they share the same Doppler shift. However, all the five catalogs focused on a single or a dozen of sight lines, and there are probably multiple ISM components along the sight line, which can lead to the confusion of the ISM Doppler shift (\citealt{1982ApJ...252..610H}). Another method to measure the central wavelength of DIB is based on an empirical assumption that, in the Local Standard of Rest (LSR), the matter moving towards the Galactic center (GC) or the Galactic anti-center (GAC) tends to be at rest, meaning the statistical radial velocity of the matter approaches zero (\citealt{2012ApJ...759..131B}). Without the interstellar reference lines, \cite{2015ApJ...798...35Z} were the first to successfully measure the central wavelength of the DIB at 1.5273 $\mu$m using this statistical method, so did \cite{2021A&A...645A..14Z}, \cite{2023A&A...674A..40G}, and \cite{2023ApJ...954..141S} for the DIB $\lambda 8621$.

\renewcommand{\arraystretch}{1.1}
\begin{table*}
\caption{Compilation of central wavelength [\AA] \label{tab:cw}}
\centering
\setlength\tabcolsep{12pt}
\begin{tabular}{ccc}

\hline\hline
DIB & This work & Cross-reference \\
\hline
$\lambda 5780$ & $5780.48 \pm 0.01$ & $5780.41 \pm 0.01 \, ^a$ \quad $5780.59 \pm 0.05 \, ^b$ \quad $5780.37 \pm 0.01 \, ^c$ \quad $5780.64 \, ^d$ \quad $5780.59 \pm 0.01 \, ^e$ \\
$\lambda 5797$ & $5796.94 \pm 0.02$ & $5797.03 \pm 0.02 \, ^a$ \quad $5797.11 \pm 0.05 \, ^b$ \quad $5796.96 \pm 0.10 \, ^c$ \quad $5797.18 \, ^d$ \quad $5797.19 \pm 0.03 \, ^e$ \\
$\lambda 6614$ & $6613.64 \pm 0.01$ & $6613.63 \pm 0.02 \, ^a$ \quad $6613.72 \pm 0.12 \, ^b$ \quad $6613.56 \pm 0.10 \, ^c$ \quad $6613.74 \, ^d$ \quad $6613.66 \pm 0.01 \, ^e$ \\
\hline

\end{tabular}
\tablefoot{
The letter superscript of the cross-reference denotes the following works: (a) \citealt{1975ApJ...196..129H}, (b) \citealt{1994AandAS..106...39J}, (c) \citealt{2000PASP..112..648G}, (d) \citealt{2019ApJ...878..151F}, and (e) \citealt{2023MNRAS.521.3727V}.
}
\end{table*}

\begin{figure}
\centering
\includegraphics[width=0.48\textwidth]{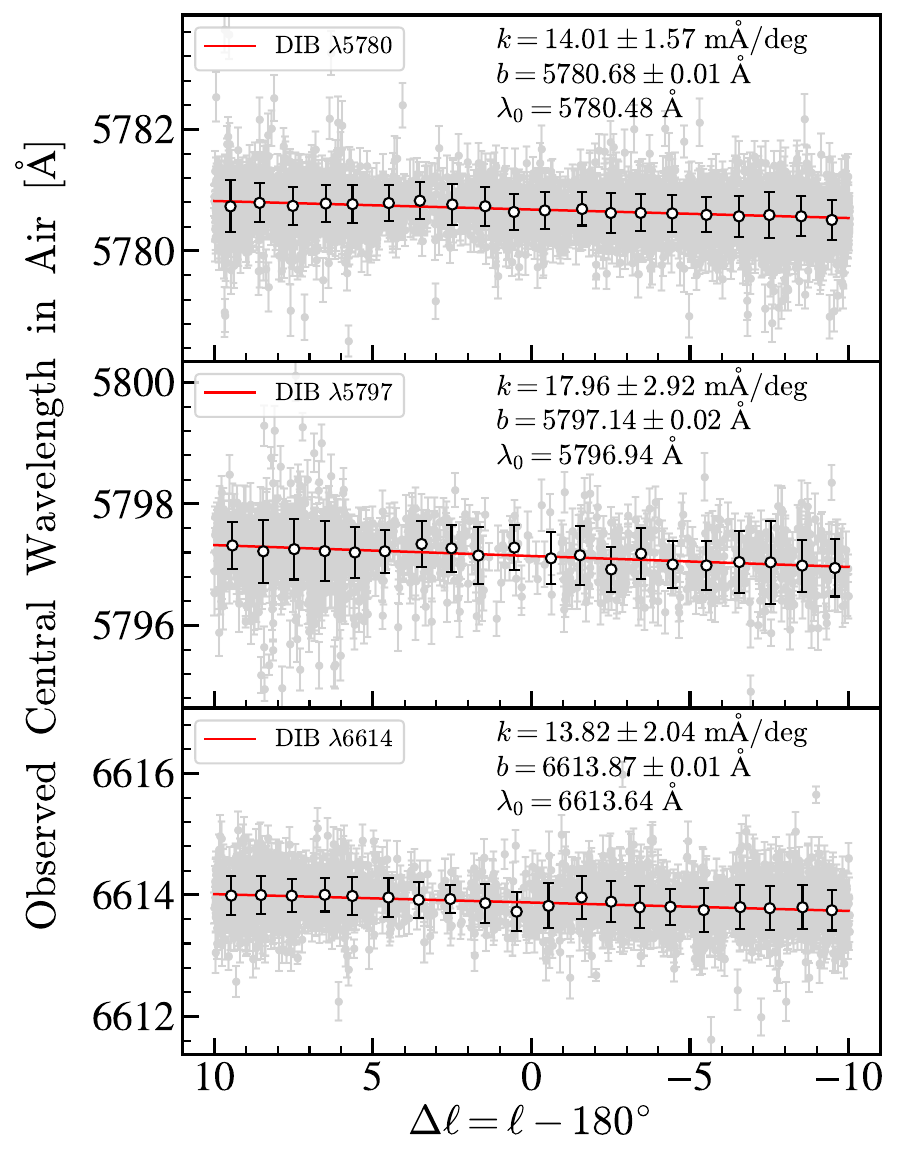}
\caption{
Observed central wavelength ($\lambda_{\rm obs}$) of the DIBs $\lambda 5780$, $\lambda 5797$, and $\lambda 6614$ as a function of the angular distance ($\Delta \ell$) from the GAC, from top to bottom. The gray dots and their error bars represent the measurements of the central wavelength of DIB and the corresponding uncertainties. The white dots and their black error bars denote the medians and the standard deviations in each $\Delta \ell = 1^{\circ}$ bin. The red lines are the linear fits to the medians of each bin using the least squares method. The fitting results, that is, the slopes $k$, the intercepts $b$, and the final central wavelengths $\lambda_0$ considering the solar motion, are annotated in each panel.  
\label{fig:cw}
}
\end{figure}

Although the spectral wavelength range of LAMOST LRS includes the interstellar absorption lines such as the Ca\,\scalebox{0.8}{II} K line at 3933.66 \AA \ and Na\,\scalebox{0.8}{I} D doublet at 5889.95 and 5895.92 \AA, these lines are prone to blending with stellar absorption lines at the same wavelengths due to the cool stars. Additionally, owing to the increasing modernization around the observation site, the light emitted by high-pressure sodium lamps used for street lighting has contaminated the Na\,\scalebox{0.8}{I} D doublet in the spectra (private communication, Chao Liu, 2023). \cite{2011MNRAS.415L..81P} also concluded that Na\,\scalebox{0.8}{I} D absorption at low resolution is a bad proxy for extinction. As a consequence, the statistical method instead of the ISM Doppler shift method is employed in this work, but we plan to compare the differences in the central wavelength of DIB between the two methods when the measurements of DIB from hot stars are available in the future. The samples with $|b| < 2^{\circ}$, $170^{\circ} < \ell < 190^{\circ}$, $d \leq 4 \ \rm kpc$, $\rm err(\mu_{mc}) < 0.5 \ \AA$, and $\rm err(RV) < 5 \ km\,s^{-1}$ are selected to fit the central wavelength of the DIBs. For the DIB $\lambda 5797$, we loosen the constraint of $|b| < 2^{\circ}$ to $|b| < 10^{\circ}$ because of its limited number of the HQ sources. There are 4934, 1542, and 3186 HQ sources in the GAC for the DIBs $\lambda 5780$, $\lambda 5797$, and $\lambda 6614$, respectively. Figure \ref{fig:cw} shows the observed central wavelength of DIB as a function of the angular distance from the GAC. By the linear fit to the medians of each bin ($\Delta \ell = 1^{\circ}$) using the least squares method, the observed central wavelengths of the DIBs $\lambda 5780$, $\lambda 5797$, and $\lambda 6614$ at $\ell = 180^{\circ}$ are $5780.68 \pm 0.01$, $5797.14 \pm 0.02$, and $6613.87 \pm 0.01$ \AA, respectively. We further take the effect of the solar motion into account, and the final central wavelength $\lambda_0$ can be expressed as $\lambda_{\rm obs} \times c/(c+U_{\odot})$, where $\lambda_{\rm obs}$ is the observed central wavelength, $c$ is the speed of light, and $U_{\odot} = 10.6 \ \rm km\,s^{-1}$ is the radial solar motion (\citealt{2019ApJ...885..131R}). As listed in Table \ref{tab:cw}, the final central wavelengths $\lambda_0$ of the DIBs $\lambda 5780$, $\lambda 5797$, and $\lambda 6614$ are $5780.48 \pm 0.01$, $5796.94 \pm 0.02$, and $6613.64 \pm 0.01$ \AA, respectively, which are all consistent with the central wavelengths from the cross-reference. Nonetheless, the central wavelength of the DIB $\lambda 5797$ is slightly lower than those from the cross-reference. The obtained slopes are also kind of smaller than the value of $23 \pm 3.4 \ \rm m\AA\,deg^{-1}$ derived by \cite{2023A&A...674A..40G} and the value of $19 \ \rm m\AA\,deg^{-1}$ estimated by \cite{2023ApJ...954..141S} for the DIB $\lambda 8621$ in the GAC, but are quite lower than the value of $57 \pm 8 \ \rm m\AA\,deg^{-1}$ reported by \cite{2015ApJ...798...35Z} for the DIB at 1.5273 $\mu$m.

\subsection{Kinematics} \label{subsec:kinematics}

\begin{figure}
\centering
\includegraphics[width=0.48\textwidth]{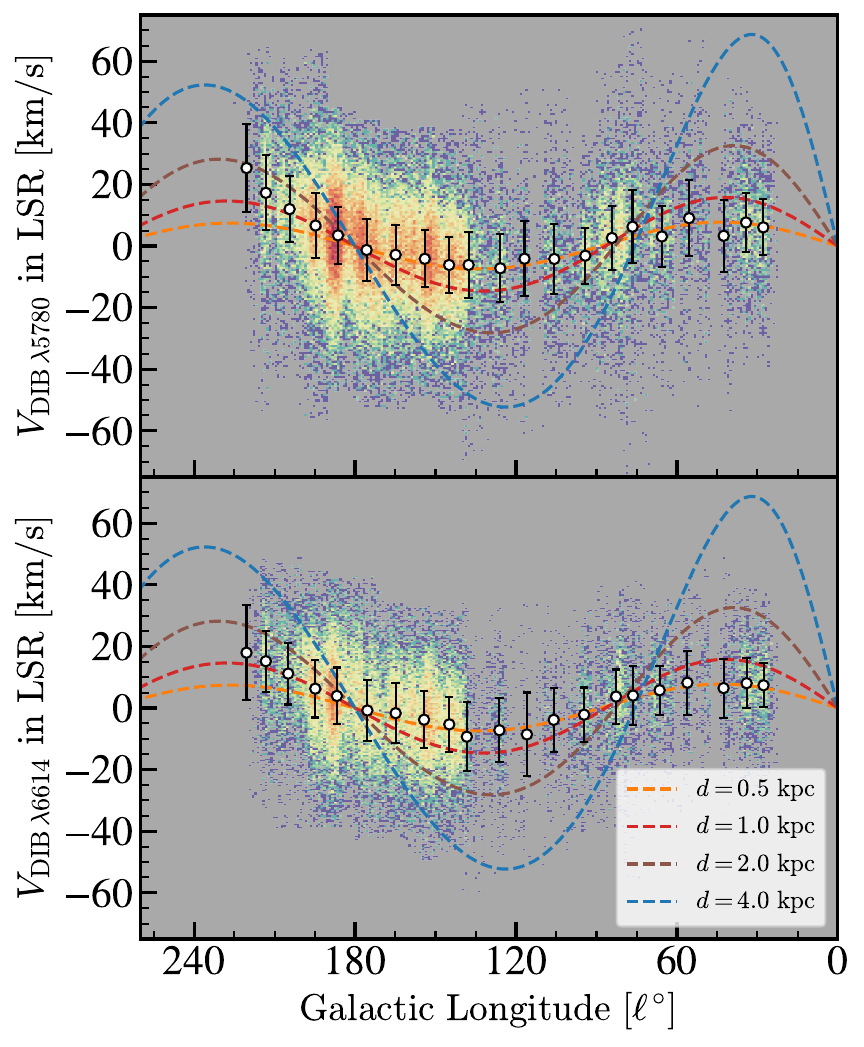}
\caption{
Tow-dimensional histogram of the LSR velocity of the DIBs $\lambda 5780$ (upper panel) and $\lambda 6614$ (lower panel) as a function of the Galactic longitude. The medians and the standard deviations in each $\Delta \ell = 10^{\circ}$ bin are denoted by the white dots and their black error bars. The Galactic velocity curves with different heliocentric distances $d$  calculated by the Model A5 of \cite{2019ApJ...885..131R} are overlaid in the plots.
\label{fig:lsr}
}
\end{figure}

Despite the unknown DIB carriers, many works have attempted to use DIB as a tracer to study the kinematics of the ISM since the DIB carriers are proven to be interstellar. \cite{2015ApJ...798...35Z} provided what 3D distribution the carriers of the DIB at 1.5273 $\mu$m have, and were the first to elaborately depict how these DIB carriers move with the Milky Way by the mean Galactic velocity curve. Following this work, \cite{2021A&A...654A.116Z} and \cite{2023A&A...674A..40G} displayed the kinematics of the DIB $\lambda 8621$ in the same way, while the average galactocentric distance of the DIB $\lambda 8621$ estimated by the velocity curve is about 7.5 kpc, smaller than 9 kpc derived from \cite{2015ApJ...798...35Z} due to the differences of the distances of the observed stars. In addition, a comparison of the LSR velocity of the DIB $\lambda 8621$ with that of CO gas (\citealt{2001ApJ...547..792D}) shows they own the same kinematics pattern, which was quantified by the linear correlation of a slope of 0.95 using the algorithm called ``peak-finding'' (\citealt{2023ApJ...954..141S}). Inspired by these works, we investigate the LSR velocity of the DIBs $\lambda 5780$ and $\lambda 6614$, $V_{\rm DIB}^{\rm LSR}$, as a function of the Galactic longitude $\ell$, as presented in Fig. \ref{fig:lsr}. The comparison of $V_{\rm DIB}^{\rm LSR}$ with the LSR velocity of the CO gas and the other tracers of the ISM will be explored in the forthcoming work.

We select the HQ sources with $|b| < 12^{\circ}$, $\rm err(\mu_{mc}) < 0.5 \ \AA$, and $\rm err(RV) < 5 \ km\,s^{-1}$ to study the kinematics of DIB and there are a total of 71,061 and 47,011 HQ sources for the DIBs $\lambda 5780$ and $\lambda 6614$. To convert the observed central wavelength of DIB to the LSR velocity, we take the effect of the solar motion $\boldsymbol{V_{\odot}} = (U_{\odot}, V_{\odot}, W_{\odot}) = (10.6,\ 10.7,\ 7.6) \ \rm km\,s^{-1}$ (\citealt{2019ApJ...885..131R}) relative to the LSR into account, and $V_{\rm DIB}^{\rm LSR}$ can be expressed as Eq. (\ref{eq:lsr}).

\begin{equation} \label{eq:lsr}
    V_{\rm DIB}^{\rm LSR} = c \times \left( \lambda_{\rm obs} / \lambda_0 - 1 \right) + \boldsymbol{V_{\odot}} / \hat{r} \ ,
\end{equation}
where $c$ is the speed of light, $\lambda_{\rm obs}$ is the observed central wavelength of DIB, $\lambda_0$ is the final central wavelength of DIB fitted by the statistical method in Sect. \ref{subsec:cw}, and $\hat{r}$ is the unit direction vector from the Sun to the source. The Galactic velocity curves with diverse heliocentric distances $d$ are computed using the Model A5 of \cite{2019ApJ...885..131R}. It can be seen from Fig. \ref{fig:lsr} that the fluctuation of $V_{\rm DIB}^{\rm LSR}$ for the whole selected samples reflects the motion of the DIB carrier with the Galactic rotation. The medians of $V_{\rm DIB}^{\rm LSR}$ in each bin ($\Delta \ell = 10^{\circ}$) are both consistent with the velocity curve with the heliocentric distance $d = 0.5 \ \rm kpc$ at the $30^{\circ} < \ell < 180^{\circ}$ and the $d = 1.5 \ \rm kpc$ at the $\ell > 180^{\circ}$. The results agree with the distance of the DIB $\lambda 8621$ estimated \cite{2023A&A...674A..40G}, but may be related to the fact that most of the background stars are located within 3 kpc.

\section{Data and code availability} \label{sec:code}

The final catalog of the DIB measurements of the 7,681 common HQ samples, and the catalogs of the HQ and the MQ samples for the DIBs $\lambda 5780$, $\lambda 5797$, and $\lambda 6614$ are publicly available at \url{https://nadc.china-vo.org/res/r101404/} in FITS tables. The codes used to perform the pipeline of measuring DIB are also provided via \url{https://github.com/iScottMark/LAMOST_DIB}.

The other data used in this work and the codes that generate the figures and support the findings of this study are available from the corresponding author upon reasonable request.

\section{Summary and conclusions} \label{sec:summary}

We successfully extracted 2,188,240 ISM residual spectra from the late-type stellar spectra in the LAMOST LRS using the method of subtracting the stellar templates. By fitting the ISM residual spectrum with the Gaussian profile, we obtained the largest sample of reliable measurements of the DIBs $\lambda 5780$, $\lambda 5797$, and $\lambda 6614$ in the northern sky to date. The selected 176,831, 13,473, and 110,152 HQ samples of the DIBs $\lambda 5780$, $\lambda 5797$, and $\lambda 6614$, respectively, corresponding to 142,074, 11,480, and 85,301 unique sources, not only allow for a more comprehensive review of DIB correlations from a statistical perspective, but also provide the first characterization of the sky distribution and kinematic properties of these three DIBs. In addition, we release a total of 27,598, 42,735, and 21,635 MQ measurements of the DIBs $\lambda 5780$, $\lambda 5797$, and $\lambda 6614$, respectively, corresponding to 23,920, 36,825, and 18,145 unique stars. These marginal or null measurements allow further studies, such as 3D mapping or the investigation of additional correlations. Our primary conclusions are as follows:

\vspace{5mm}
(i) We release to the community the largest sample of HQ measurements of the DIBs $\lambda 5780$ and $\lambda 6614$ to date, and by virtue of these data, we also present their Galactic distribution for the first time, at a resolution of 0.92$^{\circ}$/pixel.

(ii) In our statistical analysis with a sample size on the order of tens of thousands, we provide more consistent and robust EWs of DIBs per unit $\rm E(B-V)$. For the DIB $\lambda 5780$, we determine a value of 0.565 $\rm \AA \, mag^{-1}$, which is larger than those found in previous works, while for the DIBs $\lambda 5797$ and $\lambda 6614$, we obtain values of 0.176 and 0.256 $\rm \AA \, mag^{-1}$, respectively, which agree with previous findings.

(iii) Without assuming the co-motion of the DIB carriers and the ISM, we use a kinematic fitting method to obtain precise measurements of the central wavelengths of the DIBs $\lambda 5780$, $\lambda 5797$, and $\lambda 6614$, which are $5780.48 \pm 0.01$, $5796.94 \pm 0.02$, and $6613.64 \pm 0.01$ \AA, respectively.

(iv) For our HQ sources of the DIBs $\lambda 5780$ and $\lambda 6614$, the kinematic distances of their carriers, characterized by their velocity curves, are both mainly between 0.5 and 1.5 kpc from the Sun.



\begin{acknowledgements}
This work is supported by National Key R\&D Program of China (Grant NO. 2019YFA0405102) and National Natural Science Foundation of China (grant Nos. 12261141689, 11973060, and 12090044). Guoshoujing Telescope (the Large Sky Area Multi-Object Fiber Spectroscopic Telescope, LAMOST) is a National Major Scientific Project built by the Chinese Academy of Sciences. Funding for the Project has been provided by the National Development and Reform Commission. LAMOST is operated and managed by the National Astronomical Observatories, Chinese Academy of Sciences. He Zhao acknowledges the National Natural Science Foundation of China (grant No. 12203099), the China Postdoctoral Science Foundation (No. 2022M723373), and the Jiangsu Funding Program for Excellent Postdoctoral Talent.
\\
\\
\textit{Software}: Astropy (\citealt{2013A&A...558A..33A}; \citealt{2018AJ....156..123A}; \citealt{2022ApJ...935..167A}), TOPCAT \citep{2005ASPC..347...29T}, VizieR \citep{2000A&AS..143...23O}, Simbad \citep{2000A&AS..143....9W}, {\tt emcee} (\url{https://emcee.readthedocs.io/en/stable/}), {\tt uncertainties} (\url{https://uncertainties-python-package.readthedocs.io/en/latest/index.html})
\end{acknowledgements}


\bibliographystyle{aa}
\bibliography{ref.bib}

\begin{thebibliography}{98}
\expandafter\ifx\csname natexlab\endcsname\relax\def\natexlab#1{#1}\fi

\bibitem[{{Alves} {et~al.}(2020){Alves}, {Zucker}, {Goodman}, {Speagle},
  {Meingast}, {Robitaille}, {Finkbeiner}, {Schlafly}, \&
  {Green}}]{2020Natur.578..237A}
{Alves}, J., {Zucker}, C., {Goodman}, A.~A., {et~al.} 2020,
  \href{http://dx.doi.org/10.1038/s41586-019-1874-z}{\color{magenta}\nat},
  \href{https://ui.adsabs.harvard.edu/abs/2020Natur.578..237A}{578, 237}

\bibitem[{{Astropy Collaboration} {et~al.}(2022){Astropy Collaboration},
  {Price-Whelan}, {Lim}, {Earl}, {Starkman}, {Bradley}, {Shupe}, {Patil},
  {Corrales}, {Brasseur}, {N{\"o}the}, {Donath}, {Tollerud}, {Morris},
  {Ginsburg}, {Vaher}, {Weaver}, {Tocknell}, {Jamieson}, {van Kerkwijk},
  {Robitaille}, {Merry}, {Bachetti}, {G{\"u}nther}, {Aldcroft},
  {Alvarado-Montes}, {Archibald}, {B{\'o}di}, {Bapat}, {Barentsen},
  {Baz{\'a}n}, {Biswas}, {Boquien}, {Burke}, {Cara}, {Cara}, {Conroy},
  {Conseil}, {Craig}, {Cross}, {Cruz}, {D'Eugenio}, {Dencheva}, {Devillepoix},
  {Dietrich}, {Eigenbrot}, {Erben}, {Ferreira}, {Foreman-Mackey}, {Fox},
  {Freij}, {Garg}, {Geda}, {Glattly}, {Gondhalekar}, {Gordon}, {Grant},
  {Greenfield}, {Groener}, {Guest}, {Gurovich}, {Handberg}, {Hart},
  {Hatfield-Dodds}, {Homeier}, {Hosseinzadeh}, {Jenness}, {Jones}, {Joseph},
  {Kalmbach}, {Karamehmetoglu}, {Ka{\l}uszy{\'n}ski}, {Kelley}, {Kern},
  {Kerzendorf}, {Koch}, {Kulumani}, {Lee}, {Ly}, {Ma}, {MacBride}, {Maljaars},
  {Muna}, {Murphy}, {Norman}, {O'Steen}, {Oman}, {Pacifici}, {Pascual},
  {Pascual-Granado}, {Patil}, {Perren}, {Pickering}, {Rastogi}, {Roulston},
  {Ryan}, {Rykoff}, {Sabater}, {Sakurikar}, {Salgado}, {Sanghi}, {Saunders},
  {Savchenko}, {Schwardt}, {Seifert-Eckert}, {Shih}, {Jain}, {Shukla}, {Sick},
  {Simpson}, {Singanamalla}, {Singer}, {Singhal}, {Sinha}, {Sip{\H{o}}cz},
  {Spitler}, {Stansby}, {Streicher}, {{\v{S}}umak}, {Swinbank}, {Taranu},
  {Tewary}, {Tremblay}, {de Val-Borro}, {Van Kooten}, {Vasovi{\'c}}, {Verma},
  {de Miranda Cardoso}, {Williams}, {Wilson}, {Winkel}, {Wood-Vasey}, {Xue},
  {Yoachim}, {Zhang}, {Zonca}, \& {Astropy Project
  Contributors}}]{2022ApJ...935..167A}
{Astropy Collaboration}, {Price-Whelan}, A.~M., {Lim}, P.~L., {et~al.} 2022,
  \href{http://dx.doi.org/10.3847/1538-4357/ac7c74}{\color{magenta}\apj},
  \href{https://ui.adsabs.harvard.edu/abs/2022ApJ...935..167A}{935, 167}

\bibitem[{{Astropy Collaboration} {et~al.}(2018){Astropy Collaboration},
  {Price-Whelan}, {Sip{\H{o}}cz}, {G{\"u}nther}, {Lim}, {Crawford}, {Conseil},
  {Shupe}, {Craig}, {Dencheva}, {Ginsburg}, {VanderPlas}, {Bradley},
  {P{\'e}rez-Su{\'a}rez}, {de Val-Borro}, {Aldcroft}, {Cruz}, {Robitaille},
  {Tollerud}, {Ardelean}, {Babej}, {Bach}, {Bachetti}, {Bakanov}, {Bamford},
  {Barentsen}, {Barmby}, {Baumbach}, {Berry}, {Biscani}, {Boquien}, {Bostroem},
  {Bouma}, {Brammer}, {Bray}, {Breytenbach}, {Buddelmeijer}, {Burke},
  {Calderone}, {Cano Rodr{\'\i}guez}, {Cara}, {Cardoso}, {Cheedella}, {Copin},
  {Corrales}, {Crichton}, {D'Avella}, {Deil}, {Depagne}, {Dietrich}, {Donath},
  {Droettboom}, {Earl}, {Erben}, {Fabbro}, {Ferreira}, {Finethy}, {Fox},
  {Garrison}, {Gibbons}, {Goldstein}, {Gommers}, {Greco}, {Greenfield},
  {Groener}, {Grollier}, {Hagen}, {Hirst}, {Homeier}, {Horton}, {Hosseinzadeh},
  {Hu}, {Hunkeler}, {Ivezi{\'c}}, {Jain}, {Jenness}, {Kanarek}, {Kendrew},
  {Kern}, {Kerzendorf}, {Khvalko}, {King}, {Kirkby}, {Kulkarni}, {Kumar},
  {Lee}, {Lenz}, {Littlefair}, {Ma}, {Macleod}, {Mastropietro}, {McCully},
  {Montagnac}, {Morris}, {Mueller}, {Mumford}, {Muna}, {Murphy}, {Nelson},
  {Nguyen}, {Ninan}, {N{\"o}the}, {Ogaz}, {Oh}, {Parejko}, {Parley}, {Pascual},
  {Patil}, {Patil}, {Plunkett}, {Prochaska}, {Rastogi}, {Reddy Janga},
  {Sabater}, {Sakurikar}, {Seifert}, {Sherbert}, {Sherwood-Taylor}, {Shih},
  {Sick}, {Silbiger}, {Singanamalla}, {Singer}, {Sladen}, {Sooley},
  {Sornarajah}, {Streicher}, {Teuben}, {Thomas}, {Tremblay}, {Turner},
  {Terr{\'o}n}, {van Kerkwijk}, {de la Vega}, {Watkins}, {Weaver}, {Whitmore},
  {Woillez}, {Zabalza}, \& {Astropy Contributors}}]{2018AJ....156..123A}
{Astropy Collaboration}, {Price-Whelan}, A.~M., {Sip{\H{o}}cz}, B.~M., {et~al.}
  2018, \href{http://dx.doi.org/10.3847/1538-3881/aabc4f}{\color{magenta}\aj},
  \href{https://ui.adsabs.harvard.edu/abs/2018AJ....156..123A}{156, 123}

\bibitem[{{Astropy Collaboration} {et~al.}(2013){Astropy Collaboration},
  {Robitaille}, {Tollerud}, {Greenfield}, {Droettboom}, {Bray}, {Aldcroft},
  {Davis}, {Ginsburg}, {Price-Whelan}, {Kerzendorf}, {Conley}, {Crighton},
  {Barbary}, {Muna}, {Ferguson}, {Grollier}, {Parikh}, {Nair}, {Unther},
  {Deil}, {Woillez}, {Conseil}, {Kramer}, {Turner}, {Singer}, {Fox}, {Weaver},
  {Zabalza}, {Edwards}, {Azalee Bostroem}, {Burke}, {Casey}, {Crawford},
  {Dencheva}, {Ely}, {Jenness}, {Labrie}, {Lim}, {Pierfederici}, {Pontzen},
  {Ptak}, {Refsdal}, {Servillat}, \& {Streicher}}]{2013A&A...558A..33A}
{Astropy Collaboration}, {Robitaille}, T.~P., {Tollerud}, E.~J., {et~al.} 2013,
  \href{http://dx.doi.org/10.1051/0004-6361/201322068}{\color{magenta}\aap},
  \href{https://ui.adsabs.harvard.edu/abs/2013A&A...558A..33A}{558, A33}

\bibitem[{{Bailer-Jones} {et~al.}(2021){Bailer-Jones}, {Rybizki}, {Fouesneau},
  {Demleitner}, \& {Andrae}}]{2021AJ....161..147B}
{Bailer-Jones}, C.~A.~L., {Rybizki}, J., {Fouesneau}, M., {Demleitner}, M., \&
  {Andrae}, R. 2021,
  \href{http://dx.doi.org/10.3847/1538-3881/abd806}{\color{magenta}\aj},
  \href{https://ui.adsabs.harvard.edu/abs/2021AJ....161..147B}{161, 147}

\bibitem[{{Bailey} {et~al.}(2016){Bailey}, {van Loon}, {Farhang}, {Javadi},
  {Khosroshahi}, {Sarre}, \& {Smith}}]{2016A&A...585A..12B}
{Bailey}, M., {van Loon}, J.~T., {Farhang}, A., {et~al.} 2016,
  \href{http://dx.doi.org/10.1051/0004-6361/201526656}{\color{magenta}\aap},
  \href{https://ui.adsabs.harvard.edu/abs/2016A&A...585A..12B}{585, A12}

\bibitem[{{Bovy} {et~al.}(2012){Bovy}, {Allende Prieto}, {Beers}, {Bizyaev},
  {da Costa}, {Cunha}, {Ebelke}, {Eisenstein}, {Frinchaboy}, {Garc{\'\i}a
  P{\'e}rez}, {Girardi}, {Hearty}, {Hogg}, {Holtzman}, {Maia}, {Majewski},
  {Malanushenko}, {Malanushenko}, {M{\'e}sz{\'a}ros}, {Nidever}, {O'Connell},
  {O'Donnell}, {Oravetz}, {Pan}, {Rocha-Pinto}, {Schiavon}, {Schneider},
  {Schultheis}, {Skrutskie}, {Smith}, {Weinberg}, {Wilson}, \&
  {Zasowski}}]{2012ApJ...759..131B}
{Bovy}, J., {Allende Prieto}, C., {Beers}, T.~C., {et~al.} 2012,
  \href{http://dx.doi.org/10.1088/0004-637X/759/2/131}{\color{magenta}\apj},
  \href{https://ui.adsabs.harvard.edu/abs/2012ApJ...759..131B}{759, 131}

\bibitem[{{Buder} {et~al.}(2021){Buder}, {Sharma}, {Kos}, {Amarsi},
  {Nordlander}, {Lind}, {Martell}, {Asplund}, {Bland-Hawthorn}, {Casey}, {de
  Silva}, {D'Orazi}, {Freeman}, {Hayden}, {Lewis}, {Lin}, {Schlesinger},
  {Simpson}, {Stello}, {Zucker}, {Zwitter}, {Beeson}, {Buck}, {Casagrande},
  {Clark}, {{\v{C}}otar}, {da Costa}, {de Grijs}, {Feuillet}, {Horner},
  {Kafle}, {Khanna}, {Kobayashi}, {Liu}, {Montet}, {Nandakumar}, {Nataf},
  {Ness}, {Spina}, {Tepper-Garc{\'\i}a}, {Ting}, {Traven},
  {Vogrin{\v{c}}i{\v{c}}}, {Wittenmyer}, {Wyse}, {{\v{Z}}erjal}, \& {Galah
  Collaboration}}]{2021MNRAS.506..150B}
{Buder}, S., {Sharma}, S., {Kos}, J., {et~al.} 2021,
  \href{http://dx.doi.org/10.1093/mnras/stab1242}{\color{magenta}\mnras},
  \href{https://ui.adsabs.harvard.edu/abs/2021MNRAS.506..150B}{506, 150}

\bibitem[{{Cami} {et~al.}(1997){Cami}, {Sonnentrucker}, {Ehrenfreund}, \&
  {Foing}}]{1997A&A...326..822C}
{Cami}, J., {Sonnentrucker}, P., {Ehrenfreund}, P., \& {Foing}, B.~H. 1997,
  \aap, \href{https://ui.adsabs.harvard.edu/abs/1997A&A...326..822C}{326, 822}

\bibitem[{{Campbell} {et~al.}(2015){Campbell}, {Holz}, {Gerlich}, \&
  {Maier}}]{2015Natur.523..322C}
{Campbell}, E.~K., {Holz}, M., {Gerlich}, D., \& {Maier}, J.~P. 2015,
  \href{http://dx.doi.org/10.1038/nature14566}{\color{magenta}\nat},
  \href{https://ui.adsabs.harvard.edu/abs/2015Natur.523..322C}{523, 322}

\bibitem[{{Campbell} \& {Maier}(2018)}]{2018ApJ...858...36C}
{Campbell}, E.~K. \& {Maier}, J.~P. 2018,
  \href{http://dx.doi.org/10.3847/1538-4357/aab963}{\color{magenta}\apj},
  \href{https://ui.adsabs.harvard.edu/abs/2018ApJ...858...36C}{858, 36}

\bibitem[{{Castellanos} {et~al.}(2024){Castellanos}, {Najarro}, {Garcia},
  {Patrick}, \& {Geballe}}]{2024MNRAS.532.2065C}
{Castellanos}, R., {Najarro}, F., {Garcia}, M., {Patrick}, L.~R., \& {Geballe},
  T.~R. 2024,
  \href{http://dx.doi.org/10.1093/mnras/stae1472}{\color{magenta}\mnras},
  \href{https://ui.adsabs.harvard.edu/abs/2024MNRAS.532.2065C}{532, 2065}

\bibitem[{{Chen} {et~al.}(2013){Chen}, {Lallement}, {Babusiaux}, {Puspitarini},
  {Bonifacio}, \& {Hill}}]{2013A&A...550A..62C}
{Chen}, H.~C., {Lallement}, R., {Babusiaux}, C., {et~al.} 2013,
  \href{http://dx.doi.org/10.1051/0004-6361/201220413}{\color{magenta}\aap},
  \href{https://ui.adsabs.harvard.edu/abs/2013A&A...550A..62C}{550, A62}

\bibitem[{{Cordiner} {et~al.}(2011){Cordiner}, {Cox}, {Evans}, {Trundle},
  {Smith}, {Sarre}, \& {Gordon}}]{2011ApJ...726...39C}
{Cordiner}, M.~A., {Cox}, N. L.~J., {Evans}, C.~J., {et~al.} 2011,
  \href{http://dx.doi.org/10.1088/0004-637X/726/1/39}{\color{magenta}\apj},
  \href{https://ui.adsabs.harvard.edu/abs/2011ApJ...726...39C}{726, 39}

\bibitem[{{Cordiner} {et~al.}(2008{\natexlab{a}}){Cordiner}, {Cox}, {Trundle},
  {Evans}, {Hunter}, {Przybilla}, {Bresolin}, \&
  {Salama}}]{2008A&A...480L..13C}
{Cordiner}, M.~A., {Cox}, N.~L.~J., {Trundle}, C., {et~al.} 2008{\natexlab{a}},
  \href{http://dx.doi.org/10.1051/0004-6361:20079309}{\color{magenta}\aap},
  \href{https://ui.adsabs.harvard.edu/abs/2008A&A...480L..13C}{480, L13}

\bibitem[{{Cordiner} {et~al.}(2019){Cordiner}, {Linnartz}, {Cox}, {Cami},
  {Najarro}, {Proffitt}, {Lallement}, {Ehrenfreund}, {Foing}, {Gull}, {Sarre},
  \& {Charnley}}]{2019ApJ...875L..28C}
{Cordiner}, M.~A., {Linnartz}, H., {Cox}, N.~L.~J., {et~al.} 2019,
  \href{http://dx.doi.org/10.3847/2041-8213/ab14e5}{\color{magenta}\apjl},
  \href{https://ui.adsabs.harvard.edu/abs/2019ApJ...875L..28C}{875, L28}

\bibitem[{{Cordiner} {et~al.}(2008{\natexlab{b}}){Cordiner}, {Smith}, {Cox},
  {Evans}, {Hunter}, {Przybilla}, {Bresolin}, \& {Sarre}}]{2008A&A...492L...5C}
{Cordiner}, M.~A., {Smith}, K.~T., {Cox}, N.~L.~J., {et~al.}
  2008{\natexlab{b}},
  \href{http://dx.doi.org/10.1051/0004-6361:200810906}{\color{magenta}\aap},
  \href{https://ui.adsabs.harvard.edu/abs/2008A&A...492L...5C}{492, L5}

\bibitem[{{Cox} {et~al.}(2024){Cox}, {Vergely}, \&
  {Lallement}}]{2024A&A...689A..38C}
{Cox}, N.~L.~J., {Vergely}, J.~L., \& {Lallement}, R. 2024,
  \href{http://dx.doi.org/10.1051/0004-6361/202450297}{\color{magenta}\aap},
  \href{https://ui.adsabs.harvard.edu/abs/2024A&A...689A..38C}{689, A38}

\bibitem[{{Cropper} {et~al.}(2018){Cropper}, {Katz}, {Sartoretti}, {Prusti},
  {de Bruijne}, {Chassat}, {Charvet}, {Boyadjian}, {Perryman}, {Sarri}, {Gare},
  {Erdmann}, {Munari}, {Zwitter}, {Wilkinson}, {Arenou}, {Vallenari},
  {G{\'o}mez}, {Panuzzo}, {Seabroke}, {Allende Prieto}, {Benson}, {Marchal},
  {Huckle}, {Smith}, {Dolding}, {Jan{\ss}en}, {Viala}, {Blomme}, {Baker},
  {Boudreault}, {Crifo}, {Soubiran}, {Fr{\'e}mat}, {Jasniewicz}, {Guerrier},
  {Guy}, {Turon}, {Jean-Antoine-Piccolo}, {Th{\'e}venin}, {David}, {Gosset}, \&
  {Damerdji}}]{2018A&A...616A...5C}
{Cropper}, M., {Katz}, D., {Sartoretti}, P., {et~al.} 2018,
  \href{http://dx.doi.org/10.1051/0004-6361/201832763}{\color{magenta}\aap},
  \href{https://ui.adsabs.harvard.edu/abs/2018A&A...616A...5C}{616, A5}

\bibitem[{{Cui} {et~al.}(2012){Cui}, {Zhao}, {Chu}, {Li}, {Li}, {Zhang}, {Su},
  {Yao}, {Wang}, {Xing}, {Li}, {Zhu}, {Wang}, {Gu}, {Luo}, {Xu}, {Zhang},
  {Liu}, {Zhang}, {Yang}, {Cao}, {Chen}, {Chen}, {Chen}, {Chen}, {Chu}, {Feng},
  {Gong}, {Hou}, {Hu}, {Hu}, {Hu}, {Jia}, {Jiang}, {Jiang}, {Jiang}, {Jin},
  {Li}, {Li}, {Li}, {Liu}, {Liu}, {Lu}, {Mao}, {Men}, {Qi}, {Qi}, {Shi},
  {Tang}, {Tao}, {Wang}, {Wang}, {Wang}, {Wang}, {Wang}, {Wang}, {Wang},
  {Wang}, {Wang}, {Wang}, {Wang}, {Wang}, {Xu}, {Xu}, {Yang}, {Yu}, {Yuan},
  {Yuan}, {Zhai}, {Zhang}, {Zhang}, {Zhang}, {Zhao}, {Zhou}, {Zhou}, {Zhu}, \&
  {Zou}}]{2012RAA....12.1197C}
{Cui}, X.-Q., {Zhao}, Y.-H., {Chu}, Y.-Q., {et~al.} 2012,
  \href{http://dx.doi.org/10.1088/1674-4527/12/9/003}{\color{magenta}Research
  in Astronomy and Astrophysics},
  \href{https://ui.adsabs.harvard.edu/abs/2012RAA....12.1197C}{12, 1197}

\bibitem[{{Dame} {et~al.}(2001){Dame}, {Hartmann}, \&
  {Thaddeus}}]{2001ApJ...547..792D}
{Dame}, T.~M., {Hartmann}, D., \& {Thaddeus}, P. 2001,
  \href{http://dx.doi.org/10.1086/318388}{\color{magenta}\apj},
  \href{https://ui.adsabs.harvard.edu/abs/2001ApJ...547..792D}{547, 792}

\bibitem[{{Danks} \& {Lambert}(1976)}]{1976MNRAS.174..571D}
{Danks}, A.~C. \& {Lambert}, D.~L. 1976,
  \href{http://dx.doi.org/10.1093/mnras/174.3.571}{\color{magenta}\mnras},
  \href{https://ui.adsabs.harvard.edu/abs/1976MNRAS.174..571D}{174, 571}

\bibitem[{{Deng} {et~al.}(2012){Deng}, {Newberg}, {Liu}, {Carlin}, {Beers},
  {Chen}, {Chen}, {Christlieb}, {Grillmair}, {Guhathakurta}, {Han}, {Hou},
  {Lee}, {L{\'e}pine}, {Li}, {Liu}, {Pan}, {Sellwood}, {Wang}, {Wang}, {Yang},
  {Yanny}, {Zhang}, {Zhang}, {Zheng}, \& {Zhu}}]{2012RAA....12..735D}
{Deng}, L.-C., {Newberg}, H.~J., {Liu}, C., {et~al.} 2012,
  \href{http://dx.doi.org/10.1088/1674-4527/12/7/003}{\color{magenta}Research
  in Astronomy and Astrophysics},
  \href{https://ui.adsabs.harvard.edu/abs/2012RAA....12..735D}{12, 735}

\bibitem[{{Ebenbichler} {et~al.}(2022){Ebenbichler}, {Postel}, {Przybilla},
  {Seifahrt}, {We{\ss}mayer}, {Kausch}, {Firnstein}, {Butler}, {Kaufer}, \&
  {Linnartz}}]{2022A&A...662A..81E}
{Ebenbichler}, A., {Postel}, A., {Przybilla}, N., {et~al.} 2022,
  \href{http://dx.doi.org/10.1051/0004-6361/202142990}{\color{magenta}\aap},
  \href{https://ui.adsabs.harvard.edu/abs/2022A&A...662A..81E}{662, A81}

\bibitem[{{Fan} {et~al.}(2019){Fan}, {Hobbs}, {Dahlstrom}, {Welty}, {York},
  {Rachford}, {Snow}, {Sonnentrucker}, {Baskes}, \&
  {Zhao}}]{2019ApJ...878..151F}
{Fan}, H., {Hobbs}, L.~M., {Dahlstrom}, J.~A., {et~al.} 2019,
  \href{http://dx.doi.org/10.3847/1538-4357/ab1b74}{\color{magenta}\apj},
  \href{https://ui.adsabs.harvard.edu/abs/2019ApJ...878..151F}{878, 151}

\bibitem[{{Fan} {et~al.}(2022){Fan}, {Schwartz}, {Farhang}, {Cox},
  {Ehrenfreund}, {Monreal-Ibero}, {Foing}, {Salama}, {Kulik}, {MacIsaac}, {van
  Loon}, \& {Cami}}]{2022MNRAS.510.3546F}
{Fan}, H., {Schwartz}, M., {Farhang}, A., {et~al.} 2022,
  \href{http://dx.doi.org/10.1093/mnras/stab3651}{\color{magenta}\mnras},
  \href{https://ui.adsabs.harvard.edu/abs/2022MNRAS.510.3546F}{510, 3546}

\bibitem[{{Fan} {et~al.}(2017){Fan}, {Welty}, {York}, {Sonnentrucker},
  {Dahlstrom}, {Baskes}, {Friedman}, {Hobbs}, {Jiang}, {Rachford}, {Snow},
  {Sherman}, \& {Zhao}}]{2017ApJ...850..194F}
{Fan}, H., {Welty}, D.~E., {York}, D.~G., {et~al.} 2017,
  \href{http://dx.doi.org/10.3847/1538-4357/aa9480}{\color{magenta}\apj},
  \href{https://ui.adsabs.harvard.edu/abs/2017ApJ...850..194F}{850, 194}

\bibitem[{{Farhang} {et~al.}(2015{\natexlab{a}}){Farhang}, {Khosroshahi},
  {Javadi}, \& {van Loon}}]{2015ApJS..216...33F}
{Farhang}, A., {Khosroshahi}, H.~G., {Javadi}, A., \& {van Loon}, J.~T.
  2015{\natexlab{a}},
  \href{http://dx.doi.org/10.1088/0067-0049/216/2/33}{\color{magenta}\apjs},
  \href{https://ui.adsabs.harvard.edu/abs/2015ApJS..216...33F}{216, 33}

\bibitem[{{Farhang} {et~al.}(2015{\natexlab{b}}){Farhang}, {Khosroshahi},
  {Javadi}, {van Loon}, {Bailey}, {Molaeinezhad}, {Tavasoli}, {Habibi},
  {Kourkchi}, {Rezaei}, {Saberi}, \& {Hardy}}]{2015ApJ...800...64F}
{Farhang}, A., {Khosroshahi}, H.~G., {Javadi}, A., {et~al.} 2015{\natexlab{b}},
  \href{http://dx.doi.org/10.1088/0004-637X/800/1/64}{\color{magenta}\apj},
  \href{https://ui.adsabs.harvard.edu/abs/2015ApJ...800...64F}{800, 64}

\bibitem[{{Farhang} {et~al.}(2019){Farhang}, {van Loon}, {Khosroshahi},
  {Javadi}, \& {Bailey}}]{2019NatAs...3..922F}
{Farhang}, A., {van Loon}, J.~T., {Khosroshahi}, H.~G., {Javadi}, A., \&
  {Bailey}, M. 2019,
  \href{http://dx.doi.org/10.1038/s41550-019-0814-z}{\color{magenta}Nature
  Astronomy}, \href{https://ui.adsabs.harvard.edu/abs/2019NatAs...3..922F}{3,
  922}

\bibitem[{{Friedman} {et~al.}(2011){Friedman}, {York}, {McCall}, {Dahlstrom},
  {Sonnentrucker}, {Welty}, {Drosback}, {Hobbs}, {Rachford}, \&
  {Snow}}]{2011ApJ...727...33F}
{Friedman}, S.~D., {York}, D.~G., {McCall}, B.~J., {et~al.} 2011,
  \href{http://dx.doi.org/10.1088/0004-637X/727/1/33}{\color{magenta}\apj},
  \href{https://ui.adsabs.harvard.edu/abs/2011ApJ...727...33F}{727, 33}

\bibitem[{{Gaia Collaboration} {et~al.}(2023{\natexlab{a}}){Gaia
  Collaboration}, {Schultheis}, {Zhao}, {Zwitter}, {Bailer-Jones}, {Carballo},
  {Sordo}, {Drimmel}, {Ordenovic}, {Pailler}, {Fouesneau}, {Creevey}, {Heiter},
  {Recio-Blanco}, {Kordopatis}, {de Laverny}, {Marshall}, {Dharmawardena},
  {Brown}, {Vallenari}, {Prusti}, {de Bruijne}, {Arenou}, {Babusiaux},
  {Barbier}, {Biermann}, {Ducourant}, {Evans}, {Eyer}, {Guerra}, {Hutton},
  {Jordi}, {Klioner}, {Lammers}, {Lindegren}, {Luri}, {Mignard}, {Randich},
  {Sartoretti}, {Smiljanic}, {Tanga}, {Walton}, {Bastian}, {Cropper}, {Katz},
  {Soubiran}, {van Leeuwen}, {Andrae}, {Audard}, {Bakker}, {Blomme},
  {Casta{\~n}eda}, {De Angeli}, {Fabricius}, {Fr{\'e}mat}, {Galluccio},
  {Guerrier}, {Masana}, {Messineo}, {Nicolas}, {Nienartowicz}, {Panuzzo},
  {Riclet}, {Roux}, {Seabroke}, {Th{\'e}venin}, {Gracia-Abril}, {Portell},
  {Teyssier}, {Altmann}, {Benson}, {Berthier}, {Burgess}, {Busonero}, {Busso},
  {C{\'a}novas}, {Carry}, {Cheek}, {Clementini}, {Damerdji}, {Davidson}, {de
  Teodoro}, {Delchambre}, {Dell'Oro}, {Fraile Garcia}, {Garabato},
  {Garc{\'\i}a-Lario}, {Garralda Torres}, {Gavras}, {Haigron}, {Hambly},
  {Harrison}, {Hatzidimitriou}, {Hern{\'a}ndez}, {Hodgkin}, {Holl}, {Jamal},
  {Jordan}, {Krone-Martins}, {Lanzafame}, {L{\"o}ffler}, {Lorca}, {Marchal},
  {Marrese}, {Moitinho}, {Muinonen}, {Nu{\~n}ez Campos}, {Oreshina-Slezak},
  {Osborne}, {Pancino}, {Pauwels}, {Riello}, {Rimoldini}, {Robin}, {Roegiers},
  {Sarro}, {Siopis}, {Smith}, {Sozzetti}, {Utrilla}, {van Leeuwen},
  {Weingrill}, {Abbas}, {{\'A}brah{\'a}m}, {Abreu Aramburu}, {Aerts},
  {Altavilla}, {{\'A}lvarez}, {Alves}, {Anders}, {Anderson}, {Antoja},
  {Baines}, {Baker}, {Balog}, {Barache}, {Barbato}, {Barros}, {Barstow},
  {Bartolom{\'e}}, {Bashi}, {Bauchet}, {Baudeau}, {Becciani}, {Bedin},
  {Bellas-Velidis}, {Bellazzini}, {Beordo}, {Berihuete}, {Bernet},
  {Bertolotto}, {Bertone}, {Bianchi}, {Binnenfeld}, {Blazere}, {Boch},
  {Bombrun}, {Bouquillon}, {Bragaglia}, {Braine}, {Bramante}, {Breedt},
  {Bressan}, {Brouillet}, {Brugaletta}, {Bucciarelli}, {Butkevich}, {Buzzi},
  {Caffau}, {Cancelliere}, {Cannizzo}, {Carlucci}, {Carnerero}, {Carrasco},
  {Carretero}, {Carton}, {Casamiquela}, {Castellani}, {Castro-Ginard},
  {Cesare}, {Charlot}, {Chemin}, {Chiaramida}, {Chiavassa}, {Chornay},
  {Collins}, {Contursi}, {Cooper}, {Cornez}, {Crosta}, {Crowley}, {Dafonte},
  {De Luise}, {De March}, {de Souza}, {de Torres}, {del Peloso}, {Delbo},
  {Delgado}, {Diakite}, {Diener}, {Distefano}, {Dolding}, {Dsilva},
  {Dur{\'a}n}, {Enke}, {Esquej}, {Fabre}, {Fabrizio}, {Faigler}, {Fatovi{\'c}},
  {Fedorets}, {Fern{\'a}ndez-Hern{\'a}ndez}, {Fernique}, {Figueras},
  {Fournier}, {Fouron}, {Gai}, {Galinier}, {Garcia-Gutierrez},
  {Garc{\'\i}a-Torres}, {Garofalo}, {Gerlach}, {Geyer}, {Giacobbe}, {Gilmore},
  {Girona}, {Giuffrida}, {Gomel}, {Gomez}, {Gonz{\'a}lez-N{\'u}{\~n}ez},
  {Gonz{\'a}lez-Santamar{\'\i}a}, {Gosset}, {Granvik}, {Gregori Barrera},
  {Guti{\'e}rrez-S{\'a}nchez}, {Haywood}, {Helmer}, {Helmi}, {Henares},
  {Hidalgo}, {Hilger}, {Hobbs}, {Hottier}, {Huckle}, {Jab{\l}o{\'n}ska},
  {Jansen}, {Jim{\'e}nez-Arranz}, {Juaristi Campillo}, {Khanna}, {Korn},
  {K{\'o}sp{\'a}l}, {Kostrzewa-Rutkowska}, {Kun}, {Lambert}, {Lanza}, {Le
  Campion}, {Lebreton}, {Lebzelter}, {Leccia}, {Lecoeur-Taibi}, {Lecoutre},
  {Liao}, {Liberato}, {Licata}, {Lindstr{\o}m}, {Lister}, {Livanou}, {Lobel},
  {Loup}, {Mahy}, {Mann}, {Manteiga}, {Marchant}, {Marconi}, {Mar{\'\i}n Pina},
  {Marinoni}, {Mart{\'\i}n Lozano}, {Mart{\'\i}n-Fleitas}, {Marton}, {Mary},
  {Masip}, {Massari}, {Mastrobuono-Battisti}, {Mazeh}, {McMillan}, {Meichsner},
  {Messina}, {Michalik}, {Millar}, {Mints}, {Molina}, {Molinaro}, {Moln{\'a}r},
  {Monari}, {Mongui{\'o}}, {Montegriffo}, {Montero}, {Mor}, {Mora},
  {Morbidelli}, {Morel}, {Morris}, {Mowlavi}, {Munoz}, {Muraveva}, {Murphy},
  {Musella}, {Nagy}, {Nieto}, {Noval}, {Ogden}, {Pagani}, {Pagano},
  {Palaversa}, {Palicio}, {Pallas-Quintela}, {Panahi}, {Panem},
  {Payne-Wardenaar}, {Pegoraro}, {Penttil{\"a}}, {Pesciullesi}, {Piersimoni},
  {Pinamonti}, {Pineau}, {Plachy}, {Plum}, {Poggio}, {Pourbaix}, {Pr{\v{s}}a},
  {Pulone}, {Racero}, {Rainer}, {Raiteri}, {Ramos}, {Ramos-Lerate},
  {Ratajczak}, {Re Fiorentin}, {Regibo}, {Reyl{\'e}}, {Ripepi}, {Riva}, {Rix},
  {Rixon}, {Robichon}, {Robin}, {Romero-G{\'o}mez}, {Rowell}, {Royer}, {Ruz
  Mieres}, {Rybicki}, {Sadowski}, {S{\'a}ez N{\'u}{\~n}ez}, {Sagrist{\`a}
  Sell{\'e}s}, {Sahlmann}, {Sanchez Gimenez}, {Sanna}, {Santove{\~n}a},
  {Sarasso}, {Sarrate Riera}, {Sciacca}, {Segovia}, {S{\'e}gransan}, {Shahaf},
  {Siebert}, {Siltala}, {Slezak}, {Smart}, {Snaith}, {Solano}, {Solitro},
  {Souami}, {Souchay}, {Spina}, {Spitoni}, {Spoto}, {Squillante}, {Steele},
  {Steidelm{\"u}ller}, {Surdej}, {Szabados}, {Taris}, {Taylor}, {Teixeira},
  {Tisani{\'c}}, {Tolomei}, {Torra}, {Torralba Elipe}, {Trabucchi}, {Tsantaki},
  {Ulla}, {Unger}, {Vanel}, {Vecchiato}, {Vicente}, {Voutsinas}, {Weiler},
  {Wyrzykowski}, {Zorec}, {Balaguer-N{\'u}{\~n}ez}, {Leclerc}, {Morgenthaler},
  {Robert}, \& {Zucker}}]{2023A&A...680A..38G}
{Gaia Collaboration}, {Schultheis}, M., {Zhao}, H., {et~al.}
  2023{\natexlab{a}},
  \href{http://dx.doi.org/10.1051/0004-6361/202347103}{\color{magenta}\aap},
  \href{https://ui.adsabs.harvard.edu/abs/2023A&A...680A..38G}{680, A38}

\bibitem[{{Gaia Collaboration} {et~al.}(2023{\natexlab{b}}){Gaia
  Collaboration}, {Schultheis}, {Zhao}, {Zwitter}, {Marshall}, {Drimmel},
  {Fr{\'e}mat}, {Bailer-Jones}, {Recio-Blanco}, {Kordopatis}, \&
  et~al.}]{2023A&A...674A..40G}
{Gaia Collaboration}, {Schultheis}, M., {Zhao}, H., {et~al.}
  2023{\natexlab{b}},
  \href{http://dx.doi.org/10.1051/0004-6361/202243283}{\color{magenta}\aap},
  \href{https://ui.adsabs.harvard.edu/abs/2023A&A...674A..40G}{674, A40}

\bibitem[{{Gaia Collaboration} {et~al.}(2023{\natexlab{c}}){Gaia
  Collaboration}, {Vallenari}, {Brown}, {Prusti}, {de Bruijne}, {Arenou},
  {Babusiaux}, {Biermann}, {Creevey}, {Ducourant}, {Evans}, {Eyer}, {Guerra},
  {Hutton}, {Jordi}, {Klioner}, {Lammers}, {Lindegren}, {Luri}, {Mignard},
  {Panem}, {Pourbaix}, {Randich}, {Sartoretti}, {Soubiran}, {Tanga}, {Walton},
  {Bailer-Jones}, {Bastian}, {Drimmel}, {Jansen}, {Katz}, {Lattanzi}, {van
  Leeuwen}, {Bakker}, {Cacciari}, {Casta{\~n}eda}, {De Angeli}, {Fabricius},
  {Fouesneau}, {Fr{\'e}mat}, {Galluccio}, {Guerrier}, {Heiter}, {Masana},
  {Messineo}, {Mowlavi}, {Nicolas}, {Nienartowicz}, {Pailler}, {Panuzzo},
  {Riclet}, {Roux}, {Seabroke}, {Sordo}, {Th{\'e}venin}, {Gracia-Abril},
  {Portell}, {Teyssier}, {Altmann}, {Andrae}, {Audard}, {Bellas-Velidis},
  {Benson}, {Berthier}, {Blomme}, {Burgess}, {Busonero}, {Busso},
  {C{\'a}novas}, {Carry}, {Cellino}, {Cheek}, {Clementini}, {Damerdji},
  {Davidson}, {de Teodoro}, {Nu{\~n}ez Campos}, {Delchambre}, {Dell'Oro},
  {Esquej}, {Fern{\'a}ndez-Hern{\'a}ndez}, {Fraile}, {Garabato},
  {Garc{\'\i}a-Lario}, {Gosset}, {Haigron}, {Halbwachs}, {Hambly}, {Harrison},
  {Hern{\'a}ndez}, {Hestroffer}, {Hodgkin}, {Holl}, {Jan{\ss}en}, {Jevardat de
  Fombelle}, {Jordan}, {Krone-Martins}, {Lanzafame}, {L{\"o}ffler}, {Marchal},
  {Marrese}, {Moitinho}, {Muinonen}, {Osborne}, {Pancino}, {Pauwels},
  {Recio-Blanco}, {Reyl{\'e}}, {Riello}, {Rimoldini}, {Roegiers}, {Rybizki},
  {Sarro}, {Siopis}, {Smith}, {Sozzetti}, {Utrilla}, {van Leeuwen}, {Abbas},
  {{\'A}brah{\'a}m}, {Abreu Aramburu}, {Aerts}, {Aguado}, {Ajaj},
  {Aldea-Montero}, {Altavilla}, {{\'A}lvarez}, {Alves}, {Anders}, {Anderson},
  {Anglada Varela}, {Antoja}, {Baines}, {Baker}, {Balaguer-N{\'u}{\~n}ez},
  {Balbinot}, {Balog}, {Barache}, {Barbato}, {Barros}, {Barstow},
  {Bartolom{\'e}}, {Bassilana}, {Bauchet}, {Becciani}, {Bellazzini},
  {Berihuete}, {Bernet}, {Bertone}, {Bianchi}, {Binnenfeld}, {Blanco-Cuaresma},
  {Blazere}, {Boch}, {Bombrun}, {Bossini}, {Bouquillon}, {Bragaglia},
  {Bramante}, {Breedt}, {Bressan}, {Brouillet}, {Brugaletta}, {Bucciarelli},
  {Burlacu}, {Butkevich}, {Buzzi}, {Caffau}, {Cancelliere}, {Cantat-Gaudin},
  {Carballo}, {Carlucci}, {Carnerero}, {Carrasco}, {Casamiquela}, {Castellani},
  {Castro-Ginard}, {Chaoul}, {Charlot}, {Chemin}, {Chiaramida}, {Chiavassa},
  {Chornay}, {Comoretto}, {Contursi}, {Cooper}, {Cornez}, {Cowell}, {Crifo},
  {Cropper}, {Crosta}, {Crowley}, {Dafonte}, {Dapergolas}, {David}, {David},
  {de Laverny}, {De Luise}, {De March}, {De Ridder}, {de Souza}, {de Torres},
  {del Peloso}, {del Pozo}, {Delbo}, {Delgado}, {Delisle}, {Demouchy},
  {Dharmawardena}, {Di Matteo}, {Diakite}, {Diener}, {Distefano}, {Dolding},
  {Edvardsson}, {Enke}, {Fabre}, {Fabrizio}, {Faigler}, {Fedorets}, {Fernique},
  {Fienga}, {Figueras}, {Fournier}, {Fouron}, {Fragkoudi}, {Gai},
  {Garcia-Gutierrez}, {Garcia-Reinaldos}, {Garc{\'\i}a-Torres}, {Garofalo},
  {Gavel}, {Gavras}, {Gerlach}, {Geyer}, {Giacobbe}, {Gilmore}, {Girona},
  {Giuffrida}, {Gomel}, {Gomez}, {Gonz{\'a}lez-N{\'u}{\~n}ez},
  {Gonz{\'a}lez-Santamar{\'\i}a}, {Gonz{\'a}lez-Vidal}, {Granvik}, {Guillout},
  {Guiraud}, {Guti{\'e}rrez-S{\'a}nchez}, {Guy}, {Hatzidimitriou}, {Hauser},
  {Haywood}, {Helmer}, {Helmi}, {Sarmiento}, {Hidalgo}, {Hilger},
  {H{\l}adczuk}, {Hobbs}, {Holland}, {Huckle}, {Jardine}, {Jasniewicz},
  {Jean-Antoine Piccolo}, {Jim{\'e}nez-Arranz}, {Jorissen}, {Juaristi
  Campillo}, {Julbe}, {Karbevska}, {Kervella}, {Khanna}, {Kontizas},
  {Kordopatis}, {Korn}, {K{\'o}sp{\'a}l}, {Kostrzewa-Rutkowska},
  {Kruszy{\'n}ska}, {Kun}, {Laizeau}, {Lambert}, {Lanza}, {Lasne}, {Le
  Campion}, {Lebreton}, {Lebzelter}, {Leccia}, {Leclerc}, {Lecoeur-Taibi},
  {Liao}, {Licata}, {Lindstr{\o}m}, {Lister}, {Livanou}, {Lobel}, {Lorca},
  {Loup}, {Madrero Pardo}, {Magdaleno Romeo}, {Managau}, {Mann}, {Manteiga},
  {Marchant}, {Marconi}, {Marcos}, {Marcos Santos}, {Mar{\'\i}n Pina},
  {Marinoni}, {Marocco}, {Marshall}, {Martin Polo}, {Mart{\'\i}n-Fleitas},
  {Marton}, {Mary}, {Masip}, {Massari}, {Mastrobuono-Battisti}, {Mazeh},
  {McMillan}, {Messina}, {Michalik}, {Millar}, {Mints}, {Molina}, {Molinaro},
  {Moln{\'a}r}, {Monari}, {Mongui{\'o}}, {Montegriffo}, {Montero}, {Mor},
  {Mora}, {Morbidelli}, {Morel}, {Morris}, {Muraveva}, {Murphy}, {Musella},
  {Nagy}, {Noval}, {Oca{\~n}a}, {Ogden}, {Ordenovic}, {Osinde}, {Pagani},
  {Pagano}, {Palaversa}, {Palicio}, {Pallas-Quintela}, {Panahi},
  {Payne-Wardenaar}, {Pe{\~n}alosa Esteller}, {Penttil{\"a}}, {Pichon},
  {Piersimoni}, {Pineau}, {Plachy}, {Plum}, {Poggio}, {Pr{\v{s}}a}, {Pulone},
  {Racero}, {Ragaini}, {Rainer}, {Raiteri}, {Rambaux}, {Ramos}, {Ramos-Lerate},
  {Re Fiorentin}, {Regibo}, {Richards}, {Rios Diaz}, {Ripepi}, {Riva}, {Rix},
  {Rixon}, {Robichon}, {Robin}, {Robin}, {Roelens}, {Rogues}, {Rohrbasser},
  {Romero-G{\'o}mez}, {Rowell}, {Royer}, {Ruz Mieres}, {Rybicki}, {Sadowski},
  {S{\'a}ez N{\'u}{\~n}ez}, {Sagrist{\`a} Sell{\'e}s}, {Sahlmann}, {Salguero},
  {Samaras}, {Sanchez Gimenez}, {Sanna}, {Santove{\~n}a}, {Sarasso},
  {Schultheis}, {Sciacca}, {Segol}, {Segovia}, {S{\'e}gransan}, {Semeux},
  {Shahaf}, {Siddiqui}, {Siebert}, {Siltala}, {Silvelo}, {Slezak}, {Slezak},
  {Smart}, {Snaith}, {Solano}, {Solitro}, {Souami}, {Souchay}, {Spagna},
  {Spina}, {Spoto}, {Steele}, {Steidelm{\"u}ller}, {Stephenson}, {S{\"u}veges},
  {Surdej}, {Szabados}, {Szegedi-Elek}, {Taris}, {Taylor}, {Teixeira},
  {Tolomei}, {Tonello}, {Torra}, {Torra}, {Torralba Elipe}, {Trabucchi},
  {Tsounis}, {Turon}, {Ulla}, {Unger}, {Vaillant}, {van Dillen}, {van Reeven},
  {Vanel}, {Vecchiato}, {Viala}, {Vicente}, {Voutsinas}, {Weiler}, {Wevers},
  {Wyrzykowski}, {Yoldas}, {Yvard}, {Zhao}, {Zorec}, {Zucker}, \&
  {Zwitter}}]{2023A&A...674A...1G}
{Gaia Collaboration}, {Vallenari}, A., {Brown}, A.~G.~A., {et~al.}
  2023{\natexlab{c}},
  \href{http://dx.doi.org/10.1051/0004-6361/202243940}{\color{magenta}\aap},
  \href{https://ui.adsabs.harvard.edu/abs/2023A&A...674A...1G}{674, A1}

\bibitem[{{Galazutdinov} {et~al.}(2002){Galazutdinov}, {Moutou}, {Musaev}, \&
  {Kre{\l}owski}}]{2002A&A...384..215G}
{Galazutdinov}, G., {Moutou}, C., {Musaev}, F., \& {Kre{\l}owski}, J. 2002,
  \href{http://dx.doi.org/10.1051/0004-6361:20020003}{\color{magenta}\aap},
  \href{https://ui.adsabs.harvard.edu/abs/2002A&A...384..215G}{384, 215}

\bibitem[{{Galazutdinov} {et~al.}(2008){Galazutdinov}, {Lo Curto}, \&
  {Kre{\l}owski}}]{2008MNRAS.386.2003G}
{Galazutdinov}, G.~A., {Lo Curto}, G., \& {Kre{\l}owski}, J. 2008,
  \href{http://dx.doi.org/10.1111/j.1365-2966.2008.13015.x}{\color{magenta}\mnras},
  \href{https://ui.adsabs.harvard.edu/abs/2008MNRAS.386.2003G}{386, 2003}

\bibitem[{{Galazutdinov} {et~al.}(2000){Galazutdinov}, {Musaev},
  {Kre{\l}owski}, \& {Walker}}]{2000PASP..112..648G}
{Galazutdinov}, G.~A., {Musaev}, F.~A., {Kre{\l}owski}, J., \& {Walker},
  G.~A.~H. 2000, \href{http://dx.doi.org/10.1086/316570}{\color{magenta}\pasp},
  \href{https://ui.adsabs.harvard.edu/abs/2000PASP..112..648G}{112, 648}

\bibitem[{{Green} {et~al.}(2019){Green}, {Schlafly}, {Zucker}, {Speagle}, \&
  {Finkbeiner}}]{2019ApJ...887...93G}
{Green}, G.~M., {Schlafly}, E., {Zucker}, C., {Speagle}, J.~S., \&
  {Finkbeiner}, D. 2019,
  \href{http://dx.doi.org/10.3847/1538-4357/ab5362}{\color{magenta}\apj},
  \href{https://ui.adsabs.harvard.edu/abs/2019ApJ...887...93G}{887, 93}

\bibitem[{{Hamano} {et~al.}(2022){Hamano}, {Kobayashi}, {Kawakita}, {Takenaka},
  {Ikeda}, {Matsunaga}, {Kondo}, {Sameshima}, {Fukue}, {Otsubo}, {Arai},
  {Yasui}, {Kobayashi}, {Bono}, \& {Saviane}}]{2022ApJS..262....2H}
{Hamano}, S., {Kobayashi}, N., {Kawakita}, H., {et~al.} 2022,
  \href{http://dx.doi.org/10.3847/1538-4365/ac7567}{\color{magenta}\apjs},
  \href{https://ui.adsabs.harvard.edu/abs/2022ApJS..262....2H}{262, 2}

\bibitem[{{Heger}(1922)}]{1922LicOB..10..141H}
{Heger}, M.~L. 1922,
  \href{http://dx.doi.org/10.5479/ADS/bib/1922LicOB.10.141H}{\color{magenta}Lick
  Observatory Bulletin},
  \href{https://ui.adsabs.harvard.edu/abs/1922LicOB..10..141H}{10, 141}

\bibitem[{{Herbig}(1975)}]{1975ApJ...196..129H}
{Herbig}, G.~H. 1975,
  \href{http://dx.doi.org/10.1086/153400}{\color{magenta}\apj},
  \href{https://ui.adsabs.harvard.edu/abs/1975ApJ...196..129H}{196, 129}

\bibitem[{{Herbig}(1993)}]{1993ApJ...407..142H}
{Herbig}, G.~H. 1993,
  \href{http://dx.doi.org/10.1086/172500}{\color{magenta}\apj},
  \href{https://ui.adsabs.harvard.edu/abs/1993ApJ...407..142H}{407, 142}

\bibitem[{{Herbig}(1995)}]{1995ARA&A..33...19H}
{Herbig}, G.~H. 1995,
  \href{http://dx.doi.org/10.1146/annurev.aa.33.090195.000315}{\color{magenta}\araa},
  \href{https://ui.adsabs.harvard.edu/abs/1995ARA&A..33...19H}{33, 19}

\bibitem[{{Herbig} \& {Soderblom}(1982)}]{1982ApJ...252..610H}
{Herbig}, G.~H. \& {Soderblom}, D.~R. 1982,
  \href{http://dx.doi.org/10.1086/159588}{\color{magenta}\apj},
  \href{https://ui.adsabs.harvard.edu/abs/1982ApJ...252..610H}{252, 610}

\bibitem[{{Ho} {et~al.}(2017){Ho}, {Ness}, {Hogg}, {Rix}, {Liu}, {Yang},
  {Zhang}, {Hou}, \& {Wang}}]{2017ApJ...836....5H}
{Ho}, A. Y.~Q., {Ness}, M.~K., {Hogg}, D.~W., {et~al.} 2017,
  \href{http://dx.doi.org/10.3847/1538-4357/836/1/5}{\color{magenta}\apj},
  \href{https://ui.adsabs.harvard.edu/abs/2017ApJ...836....5H}{836, 5}

\bibitem[{{Hobbs} {et~al.}(2008){Hobbs}, {York}, {Snow}, {Oka}, {Thorburn},
  {Bishof}, {Friedman}, {McCall}, {Rachford}, {Sonnentrucker}, \&
  {Welty}}]{2008ApJ...680.1256H}
{Hobbs}, L.~M., {York}, D.~G., {Snow}, T.~P., {et~al.} 2008,
  \href{http://dx.doi.org/10.1086/587930}{\color{magenta}\apj},
  \href{https://ui.adsabs.harvard.edu/abs/2008ApJ...680.1256H}{680, 1256}

\bibitem[{{Iglesias-Groth}(2007)}]{2007ApJ...661L.167I}
{Iglesias-Groth}, S. 2007,
  \href{http://dx.doi.org/10.1086/518832}{\color{magenta}\apjl},
  \href{https://ui.adsabs.harvard.edu/abs/2007ApJ...661L.167I}{661, L167}

\bibitem[{{Jenniskens} \& {Desert}(1994)}]{1994AandAS..106...39J}
{Jenniskens}, P. \& {Desert}, F.~X. 1994, \aaps,
  \href{https://ui.adsabs.harvard.edu/abs/1994A&AS..106...39J}{106, 39}

\bibitem[{{Kimeswenger} {et~al.}(2015){Kimeswenger}, {Kausch}, {Noll}, \&
  {Jones}}]{2015EPJWC..8901001K}
{Kimeswenger}, S., {Kausch}, W., {Noll}, S., \& {Jones}, A.~M. 2015, in
  European Physical Journal Web of Conferences, Vol.~89, European Physical
  Journal Web of Conferences,
  \href{https://ui.adsabs.harvard.edu/abs/2015EPJWC..8901001K}{01001}

\bibitem[{{Kos} \& {Zwitter}(2013)}]{2013ApJ...774...72K}
{Kos}, J. \& {Zwitter}, T. 2013,
  \href{http://dx.doi.org/10.1088/0004-637X/774/1/72}{\color{magenta}\apj},
  \href{https://ui.adsabs.harvard.edu/abs/2013ApJ...774...72K}{774, 72}

\bibitem[{{Kos} {et~al.}(2013){Kos}, {Zwitter}, {Grebel}, {Bienayme}, {Binney},
  {Bland-Hawthorn}, {Freeman}, {Gibson}, {Gilmore}, {Kordopatis}, {Navarro},
  {Parker}, {Reid}, {Seabroke}, {Siebert}, {Siviero}, {Steinmetz}, {Watson}, \&
  {Wyse}}]{2013ApJ...778...86K}
{Kos}, J., {Zwitter}, T., {Grebel}, E.~K., {et~al.} 2013,
  \href{http://dx.doi.org/10.1088/0004-637X/778/2/86}{\color{magenta}\apj},
  \href{https://ui.adsabs.harvard.edu/abs/2013ApJ...778...86K}{778, 86}

\bibitem[{{Kos} {et~al.}(2014){Kos}, {Zwitter}, {Wyse}, {Bienaym{\'e}},
  {Binney}, {Bland-Hawthorn}, {Freeman}, {Gibson}, {Gilmore}, {Grebel},
  {Helmi}, {Kordopatis}, {Munari}, {Navarro}, {Parker}, {Reid}, {Seabroke},
  {Sharma}, {Siebert}, {Siviero}, {Steinmetz}, {Watson}, \&
  {Williams}}]{2014Sci...345..791K}
{Kos}, J., {Zwitter}, T., {Wyse}, R., {et~al.} 2014,
  \href{http://dx.doi.org/10.1126/science.1253171}{\color{magenta}Science},
  \href{https://ui.adsabs.harvard.edu/abs/2014Sci...345..791K}{345, 791}

\bibitem[{{Kre{\l}owski} {et~al.}(2011){Kre{\l}owski}, {Galazutdinov}, \&
  {Ko{\l}os}}]{2011ApJ...735..124K}
{Kre{\l}owski}, J., {Galazutdinov}, G., \& {Ko{\l}os}, R. 2011,
  \href{http://dx.doi.org/10.1088/0004-637X/735/2/124}{\color{magenta}\apj},
  \href{https://ui.adsabs.harvard.edu/abs/2011ApJ...735..124K}{735, 124}

\bibitem[{{Krelowski} \& {Walker}(1987)}]{1987ApJ...312..860K}
{Krelowski}, J. \& {Walker}, G.~A.~H. 1987,
  \href{http://dx.doi.org/10.1086/164932}{\color{magenta}\apj},
  \href{https://ui.adsabs.harvard.edu/abs/1987ApJ...312..860K}{312, 860}

\bibitem[{{Krelowski} \& {Westerlund}(1988)}]{1988A&A...190..339K}
{Krelowski}, J. \& {Westerlund}, B.~E. 1988, \aap,
  \href{https://ui.adsabs.harvard.edu/abs/1988A&A...190..339K}{190, 339}

\bibitem[{{Lallement} {et~al.}(2018){Lallement}, {Cox}, {Cami}, {Smoker},
  {Farhang}, {Elyajouri}, {Cordiner}, {Linnartz}, {Smith}, {Ehrenfreund}, \&
  {Foing}}]{2018A&A...614A..28L}
{Lallement}, R., {Cox}, N.~L.~J., {Cami}, J., {et~al.} 2018,
  \href{http://dx.doi.org/10.1051/0004-6361/201832647}{\color{magenta}\aap},
  \href{https://ui.adsabs.harvard.edu/abs/2018A&A...614A..28L}{614, A28}

\bibitem[{{Lan} {et~al.}(2015){Lan}, {M{\'e}nard}, \&
  {Zhu}}]{2015MNRAS.452.3629L}
{Lan}, T.-W., {M{\'e}nard}, B., \& {Zhu}, G. 2015,
  \href{http://dx.doi.org/10.1093/mnras/stv1519}{\color{magenta}\mnras},
  \href{https://ui.adsabs.harvard.edu/abs/2015MNRAS.452.3629L}{452, 3629}

\bibitem[{{Linnartz} {et~al.}(2020){Linnartz}, {Cami}, {Cordiner}, {Cox},
  {Ehrenfreund}, {Foing}, {Gatchell}, \& {Scheier}}]{2020JMoSp.36711243L}
{Linnartz}, H., {Cami}, J., {Cordiner}, M., {et~al.} 2020,
  \href{http://dx.doi.org/10.1016/j.jms.2019.111243}{\color{magenta}Journal of
  Molecular Spectroscopy},
  \href{https://ui.adsabs.harvard.edu/abs/2020JMoSp.36711243L}{367, 111243}

\bibitem[{{Lobel}(2011)}]{2011CaJPh..89..395L}
{Lobel}, A. 2011,
  \href{http://dx.doi.org/10.1139/p11-030}{\color{magenta}Canadian Journal of
  Physics}, \href{https://ui.adsabs.harvard.edu/abs/2011CaJPh..89..395L}{89,
  395}

\bibitem[{{Luo} {et~al.}(2012){Luo}, {Zhang}, {Zhao}, {Zhao}, {Cui}, {Li},
  {Chu}, {Shi}, {Wang}, {Zhang}, {Bai}, {Chen}, {Wang}, {Guo}, {Chen}, {Du},
  {Kong}, {Lei}, {Li}, {Song}, {Wu}, {Zhang}, {Zhou}, {Zuo}, {Du}, {He}, {Hou},
  {Dong}, {Li}, {Li}, {Li}, {Song}, {Tian}, {Wang}, {Wu}, {Yang}, {Yuan},
  {Cao}, {Chen}, {Chen}, {Chen}, {Chu}, {Feng}, {Gong}, {Gu}, {Hou}, {Huo},
  {Hu}, {Hu}, {Hu}, {Jia}, {Jiang}, {Jiang}, {Jiang}, {Jin}, {Li}, {Li}, {Li},
  {Li}, {Li}, {Liu}, {Liu}, {Liu}, {Lu}, {Lu}, {Luo}, {Mao}, {Men}, {Ni}, {Qi},
  {Qi}, {Shi}, {Su}, {Sun}, {Su}, {Tang}, {Tao}, {Tu}, {Wang}, {Wang}, {Wang},
  {Wang}, {Wang}, {Wang}, {Wang}, {Wang}, {Wang}, {Wang}, {Wang}, {Wang},
  {Wang}, {Wang}, {Wei}, {Xue}, {Xing}, {Xu}, {Xu}, {Xu}, {Yang}, {Yang},
  {Yao}, {Yu}, {Yuan}, {Zhai}, {Zhang}, {Zhang}, {Zhang}, {Zhang}, {Zhang},
  {Zhang}, {Zhao}, {Zhou}, {Zhu}, {Zhu}, \& {Zou}}]{2012RAA....12.1243L}
{Luo}, A.~L., {Zhang}, H.-T., {Zhao}, Y.-H., {et~al.} 2012,
  \href{http://dx.doi.org/10.1088/1674-4527/12/9/004}{\color{magenta}Research
  in Astronomy and Astrophysics},
  \href{https://ui.adsabs.harvard.edu/abs/2012RAA....12.1243L}{12, 1243}

\bibitem[{{Luo} {et~al.}(2015){Luo}, {Zhao}, {Zhao}, {Deng}, {Liu}, {Jing},
  {Wang}, {Zhang}, {Shi}, {Cui}, {Chu}, {Li}, {Bai}, {Wu}, {Cai}, {Cao}, {Cao},
  {Carlin}, {Chen}, {Chen}, {Chen}, {Chen}, {Chen}, {Chen}, {Chen},
  {Christlieb}, {Chu}, {Cui}, {Dong}, {Du}, {Fan}, {Feng}, {Fu}, {Gao}, {Gong},
  {Gu}, {Guo}, {Han}, {He}, {Hou}, {Hou}, {Hou}, {Hu}, {Hu}, {Hu}, {Huo},
  {Jia}, {Jiang}, {Jiang}, {Jiang}, {Jin}, {Kong}, {Kong}, {Lei}, {Li}, {Li},
  {Li}, {Li}, {Li}, {Li}, {Li}, {Li}, {Li}, {Li}, {Li}, {Li}, {Liang}, {Lin},
  {Liu}, {Liu}, {Liu}, {Liu}, {Lu}, {Luo}, {Mao}, {Newberg}, {Ni}, {Qi}, {Qi},
  {Shen}, {Shi}, {Song}, {Song}, {Su}, {Su}, {Tang}, {Tao}, {Tian}, {Wang},
  {Wang}, {Wang}, {Wang}, {Wang}, {Wang}, {Wang}, {Wang}, {Wang}, {Wang},
  {Wang}, {Wang}, {Wang}, {Wang}, {Wang}, {Wang}, {Wang}, {Wang}, {Wang},
  {Wang}, {Wei}, {Wei}, {Wu}, {Wu}, {Wu}, {Wu}, {Xing}, {Xu}, {Xu}, {Xu},
  {Yan}, {Yang}, {Yang}, {Yang}, {Yang}, {Yao}, {Yu}, {Yuan}, {Yuan}, {Yuan},
  {Yuan}, {Zhai}, {Zhang}, {Zhang}, {Zhang}, {Zhang}, {Zhang}, {Zhang},
  {Zhang}, {Zhang}, {Zhao}, {Zhou}, {Zhou}, {Zhu}, {Zhu}, {Zou}, \&
  {Zuo}}]{2015RAA....15.1095L}
{Luo}, A.~L., {Zhao}, Y.-H., {Zhao}, G., {et~al.} 2015,
  \href{http://dx.doi.org/10.1088/1674-4527/15/8/002}{\color{magenta}Research
  in Astronomy and Astrophysics},
  \href{https://ui.adsabs.harvard.edu/abs/2015RAA....15.1095L}{15, 1095}

\bibitem[{{MacIsaac} {et~al.}(2022){MacIsaac}, {Cami}, {Cox}, {Farhang},
  {Smoker}, {Elyajouri}, {Lallement}, {Sarre}, {Cordiner}, {Fan}, {Kulik},
  {Linnartz}, {Foing}, {van Loon}, {Mulas}, \& {Smith}}]{2022A&A...662A..24M}
{MacIsaac}, H., {Cami}, J., {Cox}, N. L.~J., {et~al.} 2022,
  \href{http://dx.doi.org/10.1051/0004-6361/202142225}{\color{magenta}\aap},
  \href{https://ui.adsabs.harvard.edu/abs/2022A&A...662A..24M}{662, A24}

\bibitem[{{Majewski} {et~al.}(2017){Majewski}, {Schiavon}, {Frinchaboy},
  {Allende Prieto}, {Barkhouser}, {Bizyaev}, {Blank}, {Brunner}, {Burton},
  {Carrera}, {Chojnowski}, {Cunha}, {Epstein}, {Fitzgerald}, {Garc{\'\i}a
  P{\'e}rez}, {Hearty}, {Henderson}, {Holtzman}, {Johnson}, {Lam}, {Lawler},
  {Maseman}, {M{\'e}sz{\'a}ros}, {Nelson}, {Nguyen}, {Nidever}, {Pinsonneault},
  {Shetrone}, {Smee}, {Smith}, {Stolberg}, {Skrutskie}, {Walker}, {Wilson},
  {Zasowski}, {Anders}, {Basu}, {Beland}, {Blanton}, {Bovy}, {Brownstein},
  {Carlberg}, {Chaplin}, {Chiappini}, {Eisenstein}, {Elsworth}, {Feuillet},
  {Fleming}, {Galbraith-Frew}, {Garc{\'\i}a}, {Garc{\'\i}a-Hern{\'a}ndez},
  {Gillespie}, {Girardi}, {Gunn}, {Hasselquist}, {Hayden}, {Hekker}, {Ivans},
  {Kinemuchi}, {Klaene}, {Mahadevan}, {Mathur}, {Mosser}, {Muna}, {Munn},
  {Nichol}, {O'Connell}, {Parejko}, {Robin}, {Rocha-Pinto}, {Schultheis},
  {Serenelli}, {Shane}, {Silva Aguirre}, {Sobeck}, {Thompson}, {Troup},
  {Weinberg}, \& {Zamora}}]{2017AJ....154...94M}
{Majewski}, S.~R., {Schiavon}, R.~P., {Frinchaboy}, P.~M., {et~al.} 2017,
  \href{http://dx.doi.org/10.3847/1538-3881/aa784d}{\color{magenta}\aj},
  \href{https://ui.adsabs.harvard.edu/abs/2017AJ....154...94M}{154, 94}

\bibitem[{{Matheson} {et~al.}(2000){Matheson}, {Filippenko}, {Ho}, {Barth}, \&
  {Leonard}}]{2000AJ....120.1499M}
{Matheson}, T., {Filippenko}, A.~V., {Ho}, L.~C., {Barth}, A.~J., \& {Leonard},
  D.~C. 2000, \href{http://dx.doi.org/10.1086/301519}{\color{magenta}\aj},
  \href{https://ui.adsabs.harvard.edu/abs/2000AJ....120.1499M}{120, 1499}

\bibitem[{{McCall} {et~al.}(2010){McCall}, {Drosback}, {Thorburn}, {York},
  {Friedman}, {Hobbs}, {Rachford}, {Snow}, {Sonnentrucker}, \&
  {Welty}}]{2010ApJ...708.1628M}
{McCall}, B.~J., {Drosback}, M.~M., {Thorburn}, J.~A., {et~al.} 2010,
  \href{http://dx.doi.org/10.1088/0004-637X/708/2/1628}{\color{magenta}\apj},
  \href{https://ui.adsabs.harvard.edu/abs/2010ApJ...708.1628M}{708, 1628}

\bibitem[{Men\'{e}ndez {et~al.}(1997)Men\'{e}ndez, Pardo, Pardo, \&
  Pardo}]{MENENDEZ1997307}
Men\'{e}ndez, M., Pardo, J., Pardo, L., \& Pardo, M. 1997,
  \href{http://dx.doi.org/https://doi.org/10.1016/S0016-0032(96)00063-4}{\color{magenta}Journal
  of the Franklin Institute}, 334, 334

\bibitem[{{Merrill}(1930)}]{1930ApJ....72...98M}
{Merrill}, P.~W. 1930,
  \href{http://dx.doi.org/10.1086/143266}{\color{magenta}\apj},
  \href{https://ui.adsabs.harvard.edu/abs/1930ApJ....72...98M}{72, 98}

\bibitem[{{Moutou} {et~al.}(1999){Moutou}, {Kre{\l}owski}, {D'Hendecourt}, \&
  {Jamroszczak}}]{1999A&A...351..680M}
{Moutou}, C., {Kre{\l}owski}, J., {D'Hendecourt}, L., \& {Jamroszczak}, J.
  1999,
  \href{http://dx.doi.org/10.48550/arXiv.astro-ph/9912560}{\color{magenta}\aap},
  \href{https://ui.adsabs.harvard.edu/abs/1999A&A...351..680M}{351, 680}

\bibitem[{{Ochsenbein} {et~al.}(2000){Ochsenbein}, {Bauer}, \&
  {Marcout}}]{2000A&AS..143...23O}
{Ochsenbein}, F., {Bauer}, P., \& {Marcout}, J. 2000,
  \href{http://dx.doi.org/10.1051/aas:2000169}{\color{magenta}\aaps},
  \href{https://ui.adsabs.harvard.edu/abs/2000A&AS..143...23O}{143, 23}

\bibitem[{{Poznanski} {et~al.}(2011){Poznanski}, {Ganeshalingam}, {Silverman},
  \& {Filippenko}}]{2011MNRAS.415L..81P}
{Poznanski}, D., {Ganeshalingam}, M., {Silverman}, J.~M., \& {Filippenko},
  A.~V. 2011,
  \href{http://dx.doi.org/10.1111/j.1745-3933.2011.01084.x}{\color{magenta}\mnras},
  \href{https://ui.adsabs.harvard.edu/abs/2011MNRAS.415L..81P}{415, L81}

\bibitem[{{Puspitarini} {et~al.}(2015){Puspitarini}, {Lallement}, {Babusiaux},
  {Chen}, {Bonifacio}, {Sbordone}, {Caffau}, {Duffau}, {Hill}, {Monreal-Ibero},
  {Royer}, {Arenou}, {Peralta}, {Drew}, {Bonito}, {Lopez-Santiago}, {Alfaro},
  {Bensby}, {Bragaglia}, {Flaccomio}, {Lanzafame}, {Pancino}, {Recio-Blanco},
  {Smiljanic}, {Costado}, {Lardo}, {de Laverny}, \&
  {Zwitter}}]{2015A&A...573A..35P}
{Puspitarini}, L., {Lallement}, R., {Babusiaux}, C., {et~al.} 2015,
  \href{http://dx.doi.org/10.1051/0004-6361/201424391}{\color{magenta}\aap},
  \href{https://ui.adsabs.harvard.edu/abs/2015A&A...573A..35P}{573, A35}

\bibitem[{{Puspitarini} {et~al.}(2013){Puspitarini}, {Lallement}, \&
  {Chen}}]{2013A&A...555A..25P}
{Puspitarini}, L., {Lallement}, R., \& {Chen}, H.~C. 2013,
  \href{http://dx.doi.org/10.1051/0004-6361/201321173}{\color{magenta}\aap},
  \href{https://ui.adsabs.harvard.edu/abs/2013A&A...555A..25P}{555, A25}

\bibitem[{{Raimond} {et~al.}(2012){Raimond}, {Lallement}, {Vergely},
  {Babusiaux}, \& {Eyer}}]{2012A&A...544A.136R}
{Raimond}, S., {Lallement}, R., {Vergely}, J.~L., {Babusiaux}, C., \& {Eyer},
  L. 2012,
  \href{http://dx.doi.org/10.1051/0004-6361/201219191}{\color{magenta}\aap},
  \href{https://ui.adsabs.harvard.edu/abs/2012A&A...544A.136R}{544, A136}

\bibitem[{{Reid} {et~al.}(2019){Reid}, {Menten}, {Brunthaler}, {Zheng}, {Dame},
  {Xu}, {Li}, {Sakai}, {Wu}, {Immer}, {Zhang}, {Sanna}, {Moscadelli}, {Rygl},
  {Bartkiewicz}, {Hu}, {Quiroga-Nu{\~n}ez}, \& {van
  Langevelde}}]{2019ApJ...885..131R}
{Reid}, M.~J., {Menten}, K.~M., {Brunthaler}, A., {et~al.} 2019,
  \href{http://dx.doi.org/10.3847/1538-4357/ab4a11}{\color{magenta}\apj},
  \href{https://ui.adsabs.harvard.edu/abs/2019ApJ...885..131R}{885, 131}

\bibitem[{{Saydjari} {et~al.}(2023){Saydjari}, {Uzsoy}, {Zucker}, {Peek}, \&
  {Finkbeiner}}]{2023ApJ...954..141S}
{Saydjari}, A.~K., {Uzsoy}, A. S.~M., {Zucker}, C., {Peek}, J.~E.~G., \&
  {Finkbeiner}, D.~P. 2023,
  \href{http://dx.doi.org/10.3847/1538-4357/acd454}{\color{magenta}\apj},
  \href{https://ui.adsabs.harvard.edu/abs/2023ApJ...954..141S}{954, 141}

\bibitem[{{Schlafly} \& {Finkbeiner}(2011)}]{2011ApJ...737..103S}
{Schlafly}, E.~F. \& {Finkbeiner}, D.~P. 2011,
  \href{http://dx.doi.org/10.1088/0004-637X/737/2/103}{\color{magenta}\apj},
  \href{https://ui.adsabs.harvard.edu/abs/2011ApJ...737..103S}{737, 103}

\bibitem[{{Seabroke} {et~al.}(2021){Seabroke}, {Fabricius}, {Teyssier},
  {Sartoretti}, {Katz}, {Cropper}, {Antoja}, {Benson}, {Smith}, {Dolding},
  {Gosset}, {Panuzzo}, {Th{\'e}venin}, {Allende Prieto}, {Blomme}, {Guerrier},
  {Huckle}, {Jean-Antoine}, {Haigron}, {Marchal}, {Baker}, {Damerdji}, {David},
  {Fr{\'e}mat}, {Jan{\ss}en}, {Jasniewicz}, {Lobel}, {Samaras}, {Plum},
  {Soubiran}, {Vanel}, {Zwitter}, {Ajaj}, {Caffau}, {Chemin}, {Royer},
  {Brouillet}, {Crifo}, {Guy}, {Hambly}, {Leclerc}, {Mastrobuono-Battisti}, \&
  {Viala}}]{2021A&A...653A.160S}
{Seabroke}, G.~M., {Fabricius}, C., {Teyssier}, D., {et~al.} 2021,
  \href{http://dx.doi.org/10.1051/0004-6361/202141008}{\color{magenta}\aap},
  \href{https://ui.adsabs.harvard.edu/abs/2021A&A...653A.160S}{653, A160}

\bibitem[{{Smith} {et~al.}(2021){Smith}, {Harriott}, {Majaess}, {Massa}, \&
  {Matta}}]{2021MNRAS.507.5236S}
{Smith}, F.~M., {Harriott}, T.~A., {Majaess}, D., {Massa}, L., \& {Matta},
  C.~F. 2021,
  \href{http://dx.doi.org/10.1093/mnras/stab2444}{\color{magenta}\mnras},
  \href{https://ui.adsabs.harvard.edu/abs/2021MNRAS.507.5236S}{507, 5236}

\bibitem[{{Snell} \& {vanden Bout}(1981)}]{1981ApJ...244..844S}
{Snell}, R.~L. \& {vanden Bout}, P.~A. 1981,
  \href{http://dx.doi.org/10.1086/158759}{\color{magenta}\apj},
  \href{https://ui.adsabs.harvard.edu/abs/1981ApJ...244..844S}{244, 844}

\bibitem[{{Snow} \& {Cohen}(1974)}]{1974ApJ...194..313S}
{Snow}, T.~P., J. \& {Cohen}, J.~G. 1974,
  \href{http://dx.doi.org/10.1086/153247}{\color{magenta}\apj},
  \href{https://ui.adsabs.harvard.edu/abs/1974ApJ...194..313S}{194, 313}

\bibitem[{{Song} {et~al.}(2012){Song}, {Luo}, {Comte}, {Bai}, {Zhang}, {Du},
  {Zhang}, {Chen}, {Zuo}, \& {Zhao}}]{2012RAA....12..453S}
{Song}, Y.-H., {Luo}, A.~L., {Comte}, G., {et~al.} 2012,
  \href{http://dx.doi.org/10.1088/1674-4527/12/4/009}{\color{magenta}Research
  in Astronomy and Astrophysics},
  \href{https://ui.adsabs.harvard.edu/abs/2012RAA....12..453S}{12, 453}

\bibitem[{{Steinmetz} {et~al.}(2006){Steinmetz}, {Zwitter}, {Siebert},
  {Watson}, {Freeman}, {Munari}, {Campbell}, {Williams}, {Seabroke}, {Wyse},
  {Parker}, {Bienaym{\'e}}, {Roeser}, {Gibson}, {Gilmore}, {Grebel}, {Helmi},
  {Navarro}, {Burton}, {Cass}, {Dawe}, {Fiegert}, {Hartley}, {Russell},
  {Saunders}, {Enke}, {Bailin}, {Binney}, {Bland-Hawthorn}, {Boeche}, {Dehnen},
  {Eisenstein}, {Evans}, {Fiorucci}, {Fulbright}, {Gerhard}, {Jauregi}, {Kelz},
  {Mijovi{\'c}}, {Minchev}, {Parmentier}, {Pe{\~n}arrubia}, {Quillen}, {Read},
  {Ruchti}, {Scholz}, {Siviero}, {Smith}, {Sordo}, {Veltz}, {Vidrih}, {von
  Berlepsch}, {Boyle}, \& {Schilbach}}]{2006AJ....132.1645S}
{Steinmetz}, M., {Zwitter}, T., {Siebert}, A., {et~al.} 2006,
  \href{http://dx.doi.org/10.1086/506564}{\color{magenta}\aj},
  \href{https://ui.adsabs.harvard.edu/abs/2006AJ....132.1645S}{132, 1645}

\bibitem[{{Taylor}(2005)}]{2005ASPC..347...29T}
{Taylor}, M.~B. 2005, in Astronomical Society of the Pacific Conference Series,
  Vol. 347, Astronomical Data Analysis Software and Systems XIV, ed.
  P.~{Shopbell}, M.~{Britton}, \& R.~{Ebert},
  \href{https://ui.adsabs.harvard.edu/abs/2005ASPC..347...29T}{29}

\bibitem[{{Vogrin{\v{c}}i{\v{c}}} {et~al.}(2023){Vogrin{\v{c}}i{\v{c}}}, {Kos},
  {Zwitter}, {Traven}, {Beeson}, {{\v{C}}otar}, {Munari}, {Buder}, {Martell},
  {Lewis}, {De Silva}, {Hayden}, {Bland-Hawthorn}, \&
  {D'Orazi}}]{2023MNRAS.521.3727V}
{Vogrin{\v{c}}i{\v{c}}}, R., {Kos}, J., {Zwitter}, T., {et~al.} 2023,
  \href{http://dx.doi.org/10.1093/mnras/stad678}{\color{magenta}\mnras},
  \href{https://ui.adsabs.harvard.edu/abs/2023MNRAS.521.3727V}{521, 3727}

\bibitem[{{Vos} {et~al.}(2011){Vos}, {Cox}, {Kaper}, {Spaans}, \&
  {Ehrenfreund}}]{2011A&A...533A.129V}
{Vos}, D.~A.~I., {Cox}, N.~L.~J., {Kaper}, L., {Spaans}, M., \& {Ehrenfreund},
  P. 2011,
  \href{http://dx.doi.org/10.1051/0004-6361/200809746}{\color{magenta}\aap},
  \href{https://ui.adsabs.harvard.edu/abs/2011A&A...533A.129V}{533, A129}

\bibitem[{{Wenger} {et~al.}(2000){Wenger}, {Ochsenbein}, {Egret}, {Dubois},
  {Bonnarel}, {Borde}, {Genova}, {Jasniewicz}, {Lalo{\"e}}, {Lesteven}, \&
  {Monier}}]{2000A&AS..143....9W}
{Wenger}, M., {Ochsenbein}, F., {Egret}, D., {et~al.} 2000,
  \href{http://dx.doi.org/10.1051/aas:2000332}{\color{magenta}\aaps},
  \href{https://ui.adsabs.harvard.edu/abs/2000A&AS..143....9W}{143, 9}

\bibitem[{{Weselak}(2019)}]{2019A&A...625A..55W}
{Weselak}, T. 2019,
  \href{http://dx.doi.org/10.1051/0004-6361/201834576}{\color{magenta}\aap},
  \href{https://ui.adsabs.harvard.edu/abs/2019A&A...625A..55W}{625, A55}

\bibitem[{{Weselak} {et~al.}(2004){Weselak}, {Galazutdinov}, {Musaev}, \&
  {Kre{\l}owski}}]{2004A&A...414..949W}
{Weselak}, T., {Galazutdinov}, G.~A., {Musaev}, F.~A., \& {Kre{\l}owski}, J.
  2004,
  \href{http://dx.doi.org/10.1051/0004-6361:20031663}{\color{magenta}\aap},
  \href{https://ui.adsabs.harvard.edu/abs/2004A&A...414..949W}{414, 949}

\bibitem[{{Xiang} {et~al.}(2015){Xiang}, {Liu}, {Yuan}, {Huang}, {Huo},
  {Zhang}, {Chen}, {Zhang}, {Sun}, {Wang}, {Zhao}, {Shi}, {Luo}, {Li}, {Wu},
  {Bai}, {Zhang}, {Hou}, {Yuan}, {Li}, \& {Wei}}]{2015MNRAS.448..822X}
{Xiang}, M.~S., {Liu}, X.~W., {Yuan}, H.~B., {et~al.} 2015,
  \href{http://dx.doi.org/10.1093/mnras/stu2692}{\color{magenta}\mnras},
  \href{https://ui.adsabs.harvard.edu/abs/2015MNRAS.448..822X}{448, 822}

\bibitem[{{Yanny} {et~al.}(2009){Yanny}, {Rockosi}, {Newberg}, {Knapp},
  {Adelman-McCarthy}, {Alcorn}, {Allam}, {Allende Prieto}, {An}, {Anderson},
  {Anderson}, {Bailer-Jones}, {Bastian}, {Beers}, {Bell}, {Belokurov},
  {Bizyaev}, {Blythe}, {Bochanski}, {Boroski}, {Brinchmann}, {Brinkmann},
  {Brewington}, {Carey}, {Cudworth}, {Evans}, {Evans}, {Gates}, {G{\"a}nsicke},
  {Gillespie}, {Gilmore}, {Nebot Gomez-Moran}, {Grebel}, {Greenwell}, {Gunn},
  {Jordan}, {Jordan}, {Harding}, {Harris}, {Hendry}, {Holder}, {Ivans},
  {Ivezi{\v{c}}}, {Jester}, {Johnson}, {Kent}, {Kleinman}, {Kniazev},
  {Krzesinski}, {Kron}, {Kuropatkin}, {Lebedeva}, {Lee}, {French Leger},
  {L{\'e}pine}, {Levine}, {Lin}, {Long}, {Loomis}, {Lupton}, {Malanushenko},
  {Malanushenko}, {Margon}, {Martinez-Delgado}, {McGehee}, {Monet}, {Morrison},
  {Munn}, {Neilsen}, {Nitta}, {Norris}, {Oravetz}, {Owen}, {Padmanabhan},
  {Pan}, {Peterson}, {Pier}, {Platson}, {Re Fiorentin}, {Richards}, {Rix},
  {Schlegel}, {Schneider}, {Schreiber}, {Schwope}, {Sibley}, {Simmons},
  {Snedden}, {Allyn Smith}, {Stark}, {Stauffer}, {Steinmetz}, {Stoughton},
  {SubbaRao}, {Szalay}, {Szkody}, {Thakar}, {Sivarani}, {Tucker}, {Uomoto},
  {Vanden Berk}, {Vidrih}, {Wadadekar}, {Watters}, {Wilhelm}, {Wyse}, {Yarger},
  \& {Zucker}}]{2009AJ....137.4377Y}
{Yanny}, B., {Rockosi}, C., {Newberg}, H.~J., {et~al.} 2009,
  \href{http://dx.doi.org/10.1088/0004-6256/137/5/4377}{\color{magenta}\aj},
  \href{https://ui.adsabs.harvard.edu/abs/2009AJ....137.4377Y}{137, 4377}

\bibitem[{{Yuan} \& {Liu}(2012)}]{2012MNRAS.425.1763Y}
{Yuan}, H.~B. \& {Liu}, X.~W. 2012,
  \href{http://dx.doi.org/10.1111/j.1365-2966.2012.21674.x}{\color{magenta}\mnras},
  \href{https://ui.adsabs.harvard.edu/abs/2012MNRAS.425.1763Y}{425, 1763}

\bibitem[{{Zanolli} {et~al.}(2023){Zanolli}, {Malc{\i}o{\u{g}}lu}, \&
  {Charlier}}]{2023A&A...675L...9Z}
{Zanolli}, Z., {Malc{\i}o{\u{g}}lu}, O.~B., \& {Charlier}, J.-C. 2023,
  \href{http://dx.doi.org/10.1051/0004-6361/202245721}{\color{magenta}\aap},
  \href{https://ui.adsabs.harvard.edu/abs/2023A&A...675L...9Z}{675, L9}

\bibitem[{{Zasowski} {et~al.}(2015){Zasowski}, {M{\'e}nard}, {Bizyaev},
  {Garc{\'\i}a-Hern{\'a}ndez}, {Garc{\'\i}a P{\'e}rez}, {Hayden}, {Holtzman},
  {Johnson}, {Kinemuchi}, {Majewski}, {Nidever}, {Shetrone}, \&
  {Wilson}}]{2015ApJ...798...35Z}
{Zasowski}, G., {M{\'e}nard}, B., {Bizyaev}, D., {et~al.} 2015,
  \href{http://dx.doi.org/10.1088/0004-637X/798/1/35}{\color{magenta}\apj},
  \href{https://ui.adsabs.harvard.edu/abs/2015ApJ...798...35Z}{798, 35}

\bibitem[{{Zhang} {et~al.}(2020){Zhang}, {Liu}, \&
  {Deng}}]{2020ApJS..246....9Z}
{Zhang}, B., {Liu}, C., \& {Deng}, L.-C. 2020,
  \href{http://dx.doi.org/10.3847/1538-4365/ab55ef}{\color{magenta}\apjs},
  \href{https://ui.adsabs.harvard.edu/abs/2020ApJS..246....9Z}{246, 9}

\bibitem[{{Zhao} {et~al.}(2012){Zhao}, {Zhao}, {Chu}, {Jing}, \&
  {Deng}}]{2012RAA....12..723Z}
{Zhao}, G., {Zhao}, Y.-H., {Chu}, Y.-Q., {Jing}, Y.-P., \& {Deng}, L.-C. 2012,
  \href{http://dx.doi.org/10.1088/1674-4527/12/7/002}{\color{magenta}Research
  in Astronomy and Astrophysics},
  \href{https://ui.adsabs.harvard.edu/abs/2012RAA....12..723Z}{12, 723}

\bibitem[{{Zhao} {et~al.}(2024){Zhao}, {Schultheis}, {Qu}, \&
  {Zwitter}}]{2024A&A...683A.199Z}
{Zhao}, H., {Schultheis}, M., {Qu}, C., \& {Zwitter}, T. 2024,
  \href{http://dx.doi.org/10.1051/0004-6361/202348671}{\color{magenta}\aap},
  \href{https://ui.adsabs.harvard.edu/abs/2024A&A...683A.199Z}{683, A199}

\bibitem[{{Zhao} {et~al.}(2021{\natexlab{a}}){Zhao}, {Schultheis},
  {Recio-Blanco}, {Kordopatis}, {de Laverny}, {Rojas-Arriagada}, {Zoccali},
  {Surot}, \& {Valenti}}]{2021A&A...645A..14Z}
{Zhao}, H., {Schultheis}, M., {Recio-Blanco}, A., {et~al.} 2021{\natexlab{a}},
  \href{http://dx.doi.org/10.1051/0004-6361/202039736}{\color{magenta}\aap},
  \href{https://ui.adsabs.harvard.edu/abs/2021A&A...645A..14Z}{645, A14}

\bibitem[{{Zhao} {et~al.}(2021{\natexlab{b}}){Zhao}, {Schultheis},
  {Rojas-Arriagada}, {Recio-Blanco}, {de Laverny}, {Kordopatis}, \&
  {Surot}}]{2021A&A...654A.116Z}
{Zhao}, H., {Schultheis}, M., {Rojas-Arriagada}, A., {et~al.}
  2021{\natexlab{b}},
  \href{http://dx.doi.org/10.1051/0004-6361/202141128}{\color{magenta}\aap},
  \href{https://ui.adsabs.harvard.edu/abs/2021A&A...654A.116Z}{654, A116}

\end{thebibliography}

\begin{appendix}
\section{Instrumental broadening in LAMOST LRS} \label{appd:ib}

\renewcommand{\arraystretch}{1.1}
\begin{table}[htpb]
\caption{Broadened Gaussian width of DIB [\AA] \ \label{tab:ib}}
\centering
\setlength\tabcolsep{7pt}
\begin{tabular}{cccccc}
\hline\hline
DIB & $\rm FWHM_{ins}$ & $\rm \sigma_{ins}$ & $\rm FWHM_{ref}$ & $\rm \sigma_{ref}$ & $\rm \sigma_{DIB}$ \\
\hline
$\lambda 5780$ & 3.06 & 1.30 & 2.13 & 0.90 & 1.58 \\
$\lambda 5797$ & 3.06 & 1.30 & 0.99 & 0.42 & 1.37 \\
$\lambda 6614$ & 4.17 & 1.77 & 1.02 & 0.43 & 1.82 \\
\hline
\end{tabular}
\end{table}

Compared with the high-resolution spectrographs, the low-resolution spectrographs have a more significant instrumental broadening effect. For the LAMOST LRS, the instrumental broadening effect is mainly caused by the slit width, $\rm \Delta \lambda$. Although \cite{2012RAA....12.1197C} and \cite{2012RAA....12.1243L} have reported the design of LAMOST including $\rm \Delta \lambda$, it was an experimental parameter at that time. The actual resolution of LAMOST LRS is about 1800 (private communication, Yonghui Hou, 2023) at 5500 \AA \ in red end (from 3300 \AA \ to 5900 \AA) and at 7500 \AA \ in blue end (from 5700 \AA \ to 9000 \AA). The resolution R is defined as $\rm \lambda / \Delta \lambda$, where $\rm \lambda$ is the corresponding central wavelength. Hence, $\rm \Delta \lambda$, namely the instrumental FWHM, at 5500 \AA \ and 7500 \AA \ are roughly 3.06 \AA \ and 4.17 \AA.

We refer to the canonical FWHM of DIB (see the column $\rm FWHM_{ref}$ in Table \ref{tab:ib}) from \cite{2023MNRAS.521.3727V} as our benchmark to estimate the practical Gaussian width of DIB against the LAMOST LRS. The broadened Gaussian width of DIB can be approximated by Eq. (\ref{eq:ib}).
\begin{equation} \label{eq:ib}
    \sigma_{\text{DIB}} = \sqrt{\sigma_{\text{ins}}^2 + \sigma_{\text{ref}}^2} \ ,
\end{equation}
where $\rm \sigma = FWHM \ / \ 2\sqrt{2\ln2}$, and the subscripts $\rm ins$ and $\rm ref$ stand for the instrumental broadening and the reference. Table \ref{tab:ib} provides the practical Gaussian widths of DIBs $\lambda 5780$, $\lambda 5797$, and $\lambda 6614$ against the LAMOST LRS, which are annotated in Fig. \ref{fig:lambda_sigma}.

\end{appendix}

\clearpage

\end{CJK*}
\end{document}